\documentstyle[12pt]{article}

\catcode`\@=11

\newcount\hour
\newcount\minute
\newtoks\amorpm \hour=\time\divide\hour by 60\minute
=\time{\multiply\hour by 60 \global\advance\minute by-\hour}
\edef\standardtime{{\ifnum\hour<12 \global\amorpm={am}%
        \else\global\amorpm={pm}\advance\hour by-12 \fi
        \ifnum\hour=0 \hour=12 \fi
        \number\hour:\ifnum\minute<10
        0\fi\number\minute\the\amorpm}}
\edef\militarytime{\number\hour:\ifnum\minute<10
0\fi\number\minute}

\def\draftlabel#1{{\@bsphack\if@filesw {\let\thepage\relax
   \xdef\@gtempa{\write\@auxout{\string
      \newlabel{#1}{{\@currentlabel}{\thepage}}}}}\@gtempa
   \if@nobreak \ifvmode\nobreak\fi\fi\fi\@esphack}
        \gdef\@eqnlabel{#1}}
\def\@eqnlabel{}
\def\@vacuum{}
\def\marginnote#1{}
\def\draftmarginnote#1{\marginpar{\raggedright\scriptsize\tt#1}}
\def\draft{
        \pagestyle{plain}
        \overfullrule=2pt
        \oddsidemargin -.5truein
        \def\@oddhead{\sl \phantom{\today\quad\militarytime} \hfil
        \smash{\Large\sl DRAFT} \hfil \today\quad\militarytime}
        \let\@evenhead\@oddhead
        \let\label=\draftlabel
        \let\marginnote=\draftmarginnote
        \def\ps@empty{\let\@mkboth\@gobbletwo
        \def\@oddfoot{\hfil \smash{\Large\sl DRAFT} \hfil}
        \let\@evenfoot\@oddhead}
        \def\@eqnnum{(\theequation)\rlap{\kern\marginparsep\tt\@eqnlabel}%
        \global\let\@eqnlabel\@vacuum}  }

\newcommand{\rf}[1]{(\ref{#1})}
\renewcommand{\theequation}{\thesection.\arabic{equation}}
\renewcommand{\thefootnote}{\fnsymbol{footnote}}
\newcommand{\newsection}{    
\setcounter{equation}{0}\section}
\def\appendix#1{\addtocounter{section}{1}\setcounter{equation}{0}
\renewcommand{\thesection}{\Alph{section}}
\section*{Appendix \thesection\protect\indent \parbox[t]{11.15cm}{#1}}
\addcontentsline{toc}{section}{Appendix \thesection\ \ \ #1}}

\def\DL#1{{D_{#1}^{{\scriptscriptstyle\rm L}}}}

\def\ipr{{i^\prime}}
\def\jpr{{j^\prime}}
\def\kpr{{k^\prime}}
\def\lpr{{l^\prime}}

\def\cI{{\cal I}}
\def\cJ{{\cal J}}

\def\bQ{{\bf Q}}
\def\bS{{\bf S}}
\def\bP{{\bf P}}
\def\bK{{\bf K}}
\def\bJ{{\bf J}}
\def\bD{{\bf D}}

\def\hi{{\hat{i}}}
\def\hj{{\hat{j}}}
\def\hk{{\hat{k}}}

\def\mun{{\underline{m}}}
\def\nun{{\underline{n}}}
\def\kun{{\underline{k}}}

\def\muun{{\underline{\mu}}}
\def\nuun{{\underline{\nu\!\!\!\!\phantom{\mu}}}}
\def\rhoun{{\underline{\rho}}}

\def\f{{\rm f}}
\def \we {\wedge}
\def\rQ{{\rm Q}}

\def \bi{\bibitem}

\def\PP{{\cal P}}

\def\cN{{\cal N}}

\def \L{{\widehat{L}}}

\def\rw{{\rm w}}
\def\nline{\,\nabla\kern -0.7em\raise0.2ex\hbox{/}\,\,}
\def\yline{\,y\kern -0.47em /}
\def\aline{\,a\kern -0.49em /}
\def\parline{\,\partial\kern -0.55em /\,\,}

\def\del{\partial}

\def\FF{{\rm F}}
\def\rTr{{\rm Tr}}
\def\rTrpr{\hbox{\'Tr}}

\def\det{\hbox{det}}

\def\be{\begin{equation}}
\def\ee{\end{equation}}

\begin{document}


\begin{flushright}
FIAN/TD/02-05   \\
hep-th/0211178
\end{flushright}

\vspace{1cm}

\begin{center}

{\Large \bf Supersymmetric $D3$ brane  and $\cN=4$ SYM actions

\bigskip
in plane wave backgrounds
}\\[.2cm]

\vspace{2.5cm}

R.R. Metsaev\footnote{
E-mail: metsaev@lpi.ru }

\vspace{1cm}

{\it Department of Theoretical Physics, P.N. Lebedev Physical
Institute, \\ Leninsky prospect 53,  Moscow 119991, Russia }

\vspace{3.5cm}

{\bf Abstract}

\end{center}

The explicit (all-order in fermions) form of the kappa-symmetric
$D3$ brane probe action was previously found in the two maximally
supersymmetric type IIB vacua: flat space and $AdS_5 \times S^5$.
Here we present the form of the action in the third maximally
supersymmetric  type IIB  background: gravitational plane wave
supported by constant null 5-form strength. We study $D3$ brane
action in both  covariant and light cone kappa symmetry gauges.
Like the  fundamental string action, the $D3$ brane  action takes a
simple form once written in the light cone  kappa-symmetry gauge.
We also consider the  $\cN=4$ SYM  theory in $4d$ plane wave
background. Since some (super)symmetries of plane wave SYM action
are friendly to (super)symmetries of the type IIB superstring in
plane wave Ramond-Ramond background we suggest this SYM  model may
be useful in the context of AdS/CFT duality. We develop  the
Hamiltonian light cone gauge formulation for this theory.

\newpage
\setcounter{page}{1}
\renewcommand{\thefootnote}{\arabic{footnote}}
\setcounter{footnote}{0}

\newsection{Introduction}

Recently,  new maximally supersymmetric solution of IIB
supergravity with Ramond-Ramond flux was found \cite{blau}. It
turns out that the light cone gauge Green-Schwarz superstring
action on this background is quadratic in both  bosonic and
fermionic superstring $2d$ fields, and therefore, this model can
be explicitly quantized \cite{rrm0112}. On the other hand,  in
\cite{beren} it was proposed that this superstring in plane wave
background corresponds to a certain (large R-charge)  sector of
the $\cN=4$ SYM theory. Given that the plane wave superstring model
can be quantized explicitly \cite{rrm0112,rrm0202} one can study
the duality correspondence between string states and gauge theory
operators at the string-mode level \cite{beren}. This  new duality
\cite{beren} (which can be understood also from the point of view
of  semiclassical approximation to  the original AdS/CFT setting
\cite{gkp,frt}, see for review \cite{Tseytlin:2002ny}) turned out
to be very fruitful. It renewed interest in various aspects of
string/gauge theory correspondence and triggered many
investigations  both on the gauge theory and string theory sides.
On the string theory side many new interesting more general
solutions to IIA and IIB supergravities were found and corresponding
world-sheet superstring actions were constructed
\cite{Cvetic:2002hi}. Remarkably simple structure of plane wave
background admits also to construct string superfield theory
\cite{Spradlin:2002ar,Schwarz:2002bc,Pankiewicz:2002gs} in plane
wave background hence giving first example of string field theory
on curved background\footnote{Surprisingly the plane wave
superstring field theory can be successfully used for a study of
new string/gauge theory duality \cite{Kiem:2002xn}. Discussion of
Penrose limit and plane wave duality for $AdS_5\times T^{1,1}$ and
various orbifolds like $AdS_5\times S^5/Z_N$ may be found in
\cite{Itzhaki:2002kh},\cite{Alishahiha:2002ev}. Non-supersymmetric
$0B$ theories and theories with $N=1,2$ supersymmetries were
studied \cite{Bigazzi:2002gw} and \cite{Brecher:2002ar}. Plane
wave duality for $ads_3\times s^3$ and for various gauge theories
in six dimensions were investigated in
\cite{Parnachev:2002qs},\cite{Oz:2002ku}. Study of strings in RR
plane wave background at finite temperature is given in
\cite{PandoZayas:2002hh}.}. On the gauge theory side various matrix
models techniques were developed and applied to the study of new
duality
\cite{Kristjansen:2002bb,Gross:2002su,Santambrogio:2002sb}.

According to the  idea of string/gauge theory duality each state
of string theory can be associated  to some operator of gauge
theory \cite{polyakov2001}. For the case of the type IIB
superstring on $AdS_5\times S^5$ Ramond-Ramond background  the
dual theory should be $\cN=4$ supersymmetric Yang-Mill theory (SYM)
``living'' at the boundary of $AdS_5$ \cite{malda97}. This
boundary can be taken to be flat Minkowski space time or any
space-time which is obtainable from the Minkowski space-time by a
conformal transformation. All such theories  have the same global
conformal supersymmetries generated by the $psu(2,2|4)$
superalgebra which on  the string theory side is realized as the
algebra of super-isometries of the AdS+RR background. Thus in the
case of the AdS string we have  the same superalgebra in the bulk
and at the boundary.

Now let us turn to the  plane-wave version (or ``limit'') of the
superstring - gauge theory duality. Here  we still deal with (a
sector of) the $\cN=4$ SYM, (as it is this theory that we want to
study), i.e. on  the gauge theory side we have  again $psu(2,2|4)$
as defining superconformal algebra. In contrast to  the original
AdS string case,  however, this superalgebra no longer  coincides
with the algebra of global symmetries of the plane wave
background, i.e. the  plane wave superstring and the $\cN=4$ SYM
have different algebras of global symmetries.\footnote{Indeed, the
symmetry superalgebra of plane wave superstring is obtainable from
$psu(2,2|4)$ superalgebra via  a contraction procedure
\cite{Hatsuda:2002xp}. Interesting discussion of this  contraction
procedure at the level of oscillator construction may be found in
\cite{Fernando:2002wv}.} One may   argue that this is not
surprising since we should restrict our attention only to a
particular sector of (large R-charge) states of SYM theory on
which a smaller symmetry may be acting.

Still,  it is reasonable, and this is what we are going to do in
this paper, to look for a subset of symmetries in the plane wave
superstring and SYM theories which can be matched. Here by
matching the symmetries we simply mean    a coincidence of
commutation relations of the symmetry generators on the string
theory side and the appropriate generators on the SYM side. It
seems very  likely that in order to increase the number of
matching symmetries as much as  possible we should consider the
SYM theory not in flat but in a $4d$ plane wave background.
Another attractive feature of plane wave SYM is a discrete
spectrum of {\it the light cone energy operator}. As is well known
spectrum of {\it the light cone energy operator} of plane wave
superstring is also discrete \cite{rrm0202}. Therefore is it
natural to expect that it is the plane wave SYM theory that is
most appropriate for establishing a precise correspondence with
plane wave superstring\footnote{$4d$ plane wave background can be
obtained via Penrose limit from the space-time $R\times S^3$ which
is used for a study of $AdS$ holography. Therefore one can expect
that the $4d$ plane wave background is most appropriate for the study
holographical issues of the new duality. Various discussions of
plane wave holography may be found in \cite{Das:2002cw}.}. This is
our motivation for the study of $\cN=4$ SYM theory in plane wave
background.

Another closely related  theme of  this paper is the structure of
$D3$ brane action in plane wave Ramond-Ramond background. Our
interest in the form of the $D3$ brane action is due to the
following  reasons. First, as is well known, the action for a
$D3$ brane probe  in plane wave background respects the same
symmetries as the plane wave superstring, and after imposing
appropriate kappa symmetry gauge and static gauges to be discussed
below the quadratic part of the $D3$ brane action describes the
abelian  $\cN=4$ SYM multiplet propagating  in  $4d$ plane wave
background. In this  quadratic approximation some of the original
symmetries of the  $10d$ plane wave become broken. In other words,
the study of the $D3$ brane action suggests  a natural way how to
formulate the plane wave SYM theory.

The second reason for our interest in  $D3$ brane action is
related to  the desire to apply the  supercoset method developed
in \cite{rrm9805,rrm9806} (see also \cite{kr}) to the study of the
$D3$ brane dynamics. The low-energy action of a probe D-brane in
a curved type II supergravity background is given by a superspace
generalization of the sum of the Born-Infeld  action and a
Wess-Zumino type term (similar in form to the curved superspace
version of the Green-Schwarz string action \cite{howe}). While the
formal expression for such an action in a generic on-shell type II
supergravity  background is known \cite{ced,aps,bt},  its explicit
component form is hard to determine (explicit solution of
superspace constraints is not known in general). Resorting  to
expansion in powers of fermions it is cumbersome to go  much far
beyond terms quadratic in the fermions (see \cite{mil,grana}).
There are, however, a few exceptional cases where there is a lot
of symmetry allowing one to explicitly determine the full
structure of the supervielbeins and field strengths, and thus the
form of the D-brane actions  to all orders in the fermions. The
two previously studied  cases are the  maximally supersymmetric
flat space \cite{aps}  and  the $AdS_5 \times S^5$ background
\cite{rrm9806}. Here we shall find the explicit form of the
$D3$ brane action\footnote{ We shall consider primarily the $D3$ brane
action though similar actions can be found for all other $Dp$ branes
of type IIB theory.} in the third maximally supersymmetric type
IIB background - the symmetric gravitational  wave  supported by
the constant null 5-form  strength  \cite{blau}. Note that
alternative way to derive plane wave $D3$ brane action is to
consider AdS supersymmetric $D3$ brane action \cite{rrm9806} in
the Penrose limit. Discussion of this approach for the bosonic
part of various $Dp$ brane actions may be found in
\cite{Blau:2002mw}. We do not use this approach for the study of
supersymmetric $D3$ brane action because it does not make explicit
use of the basic (super)symmetries of the problem.

As in the  case of the $D3$ brane in $AdS_5 \times S^5$
\cite{rrm9806}, we shall find the explicit form of the $D3$ brane
action using  the supercoset approach \cite{rrm9805}. It was
already used  in \cite{rrm0112} in order to find the complete form
of  the fundamental GS string in the plane wave
background.\footnote{ The supercoset construction turned out to be
very effective and fruitful and was used to construct $AdS(3)$ and
$AdS(2)$ superstring actions \cite{pes2}, \cite{zho}, $D1$ and $D5$
brane actions in various super $AdS$+RR backgrounds \cite{daw},
and various $AdS$ supermembrane actions \cite{wpps}-\cite{dik}.
Dual $D3$ $AdS$ brane action in the framework of supercoset
construction were discussed in \cite{oda}.} We shall then see that
like the fundamental string action  \cite{rrm0112}, the $D3$ brane
action takes quite  simple form  when  written in the light cone
kappa symmetry gauge.\footnote{Our expression for the $D3$ brane
action in the fermionic light cone gauge  may be of interest also
in the flat-space limit, as it  has simpler structure than the
corresponding covariant kappa symmetry gauge  action in
\cite{aps}.}

We shall first determine the form of the action for a generic
embedding of a $D3$ brane  in the plane wave background (i.e. we
shall not make a particular choice of a  static gauge). Then we
will be  able  to see  explicitly in principle which $D3$ brane
orientations preserve some parts of supersymmetry (following,
e.g., the   general approach  of \cite{kal} and the analysis
\cite{bil}  of  the BPS  brane states in the $AdS_5 \times S^5$
case). A  related discussion of supersymmetry of particular
$Dp$ brane probe orientations in plane wave background appeared in
\cite{Chu:2002in,skenderis}.\footnote{ Microscopic open-string
approaches to construction of supersymmetric $Dp$ branes in the
plane wave background   were developed in \cite{Billo:2002ff} (see
also \cite{Bak:2002rq} for the corresponding discussion of branes
in $11d$ case). Study of $Dp$ brane interactions in the framework of
microscopic approach may be found in \cite{Bergman:2002hv}.
Derivation of conformal operators which are dual to open strings
states ending on $D5$ branes may be found in \cite{Lee:2002cu}.
Interesting discussion  of realization of $Dp$ brane dynamics in
plane wave background is given in \cite{Ganor:2002ju}.}

In Section \ref{2} we find supersymmetric and $\kappa$ invariant
action of $D3$ brane in plane wave Ramond-Ramond background. Most
of this section follows closely the same strategy as was used in
the $AdS_5 \times S^5$ case in \cite{rrm9806}.

In Section \ref{GFACT} we study covariant $\kappa$ symmetry gauge
and static gauge fixed $D3$ brane action. We discuss broken and
unbroken symmetries of such brane action.

In Section \ref{LCD3} we discuss light cone kappa symmetry gauge
fixed $D3$ brane action.

In Section \ref{PSUVARBAS} we study $psu(2,2|4)$ superalgebra in
various bases. We introduce notion of plane wave basis of this
superalgebra and discuss interrelation of this basis with the
conventional Lorentz basis.

In Section \ref{PWSYM} we discuss the covariant and gauge
invariant formulation of $\cN=4$ SYM in plane wave background. We
study realization of 32 plane wave supersymmetries of the
$psu(2,2|4)$ superalgebra in the covariant formulation of SYM
theory.

In Section \ref{HAMFOR} we study the Hamiltonian light cone gauge
formulation of plane wave SYM. We demonstrate that in contrast to
covariant and gauge invariant formulation the 3-point and 4-point
light cone gauge vertices of plane wave SYM theory takes exactly
the same form as the ones of SYM in flat Minkowski space-time.

Section \ref{GLOSYM} is devoted to a study of realization of
global symmetries of $psu(2,2|4)$ superalgebra on the physical
fields. We find field theoretical realization of Noether
(super)charges as generators of the $psu(2,2|4)$ superalgebra.

In Section \ref{TRARUL} we discuss transformation rules of
physical fields in the framework of light cone formulation.

Section \ref{CONCLU} summarizes our conclusions and suggests
directions for future research.

Our notation  and conventions are explained in Appendices A and B.
Appendix C contains some basic relations for Cartan 1-forms
on the coset superspace.
Appendix D contains some additional details about transformation
rules of physical fields of plane wave SYM and generalization to
plane wave massless higher spin fields.

\newsection{General form of $D3$ brane action  }\label{2}

The $D3$ brane action  depends on the coset superspace coordinates
$X=(x^{\underline{\nu}},\Theta)$ and vector field strength
$F_{ab}=\del_a A_b -\del_b A_a$. As in \cite{ced,aps,bt}, it is
given by the  sum of the BI and WZ  terms
\begin{equation}\label{action}
S=
\int d^4 \sigma {\cal L}\,,\qquad {\cal L} ={\cal L}_{\rm BI}+
{\cal L}_{\rm WZ}\,,
\end{equation}
where (we set the 3-brane tension to  be 1)

\be {\cal L}_{\rm BI}=-\sqrt{-\det(G_{ab}+{\FF}_{ab})}\ ,
\label{hhh} \ee \be {\cal L}_{\rm WZ} = d^{-1} {H}_5 \ .
\label{hdo} \ee The induced world-volume metric $G_{ab}$ is
($a,b=0,1,2,9$)

\be\label{Gab} G_{ab}=L_a^\mu L_b^\mu \,, \qquad\qquad L^\mu
(X(\sigma)) = d\sigma^a L_a^\mu\,, \ee where $L^\mu$ are Cartan
1-forms (see Appendix C for definition of Cartan 1-forms). The
supersymmetric extension
${\FF}=\frac{1}{2}{\FF}_{ab}d\sigma^a\wedge d\sigma^b$ of the
world-volume gauge field strength  2-form $dA$  is  found to be
\footnote{ For fermionic coordinates we assume the convention for
hermitian conjugation and permutations rules $(\theta_1
\theta_2)^\dagger = \theta_2^\dagger \theta_1^\dagger$,
$\theta_1\theta_2 =-\theta_2\theta_1$, while for fermionic Cartan
1-form we adopt $(L_1 \wedge L_2)^\dagger = -L_2^\dagger \wedge
L_1^\dagger$, \ $L_1\wedge L_2  = L_2\wedge L_1$. For bosonic
Cartan 1-forms we adopt the convention $B_1\wedge B_2=-B_2\wedge
B_1$.} \be\label{fexp} {\FF}=dA +2{\rm i}\int_0^1 dt\  L_t^\mu\we
\Theta\bar{\gamma}^\mu \tau_3{\bf L}_t \  , \ee where
$L_t^\mu(x,\Theta)\equiv L^\mu(x,t\Theta), \ \ {\bf
L}^\alpha_t(x,\Theta)\equiv {\bf L}^\alpha (x,t\Theta) $ (we
suppressed the
 spinor indices $\alpha$ in \rf{fexp}).
The $\Theta$-dependent correction term  in \rf{fexp} given by the
integral over the auxiliary parameter $t$ is exactly  the same
2-form as in the string action \cite{rrm0112} (see also
\cite{rrm9805,rrm9806}\footnote{Discussion of alternative
representation for WZ part of GS action may be found in
\cite{Hatsuda:2002iu}. Various alternative covariant formulations
of plane wave superstring action are given in
\cite{Berkovits:2002vn}.}). This representation corresponds to the
specific choice of coset representative made above. Note that
while ${\FF}$ is not expressible in terms of Cartan forms only,
its exterior derivative is \be d{\FF}={\rm i} {\bf L}\we \L
\we\tau_3  {\bf L} \,, \ \ \ \ \ \ \ \ \ \ \L_{\alpha\beta}\equiv
L^\mu \gamma_{\alpha\beta}^\mu\,. \label{dee} \ee This important
formula can be proved my making use of the Maurer-Cartan equations
\rf{mc1} -- \rf{mc2} and equations \rf{t1}--\rf{t6} from Appendix
C. As a result, $d{\FF}$ is manifestly invariant under
supersymmetry, and then  so is ${\FF}$, provided  one defines
appropriately   the transformation of $A$ to cancel the  exact
variation of the second (string WZ) term in \rf{fexp}  (cf.
\cite{ced,aps}).\footnote{Note that this is  the same
transformation that is needed  to make  the superstring action
\cite{rrm0112} defined on a { disc}  and coupled to $A$ at the
boundary invariant under supersymmetry.}

As in flat space \cite{ced,aps}, the  super-invariance of $S_{\rm
WZ}$ follows  from supersymmetry of the closed 5-form  $H_5 $. We
shall determine the supersymmetric $H_5$ from the requirement of
$\kappa$-symmetry of the full action $S$  which    fixes  this
5-form uniquely.  The $\kappa$-transformations are defined by (see
\rf{hatdel}) \be\label{kaptra} \widehat{\delta_\kappa} x^\mu=0\,,\
\ \qquad \widehat{\delta_\kappa \Theta} = {\bf K}\,, \ee where the
transformation parameter satisfies the constraint
\begin{equation}\label{kappa}
\Gamma {\bf K} ={\bf K}\,, \qquad \ \
\Gamma^2=1\,.
\end{equation}
Here  $\Gamma$  is  given   by \be\label{Gammat}
\Gamma=\frac{\epsilon^{a_1\ldots
a_4}}{\sqrt{-\det(G_{ab}+{\FF}_{ab})}} \bigg(\frac{1}{4!}
\gamma_{a_1\ldots a_4} \tau_2
+\frac{1}{4}\gamma_{a_1a_2}{\FF}_{a_3a_4} \tau_1
+\frac{1}{8}{\FF}_{a_1a_2}{\FF}_{a_3a_4} \tau_2\bigg) \,, \ee and
we use the notation \be \gamma_{a_1\ldots a_n}\equiv
\L_{[a_1}\ldots \L_{a_n]}\,,\ \ \ \ \  \ \ \ \qquad \L_a\equiv
L_a^\mu \gamma^\mu \,.\ee The corresponding variation of the
metric $G_{ab}$ is
\begin{equation}\label{kvg}
\delta_\kappa G_{ab}
=-2{\rm i}\widehat{\delta_\kappa \Theta}
(\L_a {\bf L}_b+\L_b {\bf L}_a) \,,
\end{equation}
while the variation of ${\FF}$  is given by
\be\label{kvf}
\delta_\kappa {\FF}
=2{\rm i}\widehat{\delta_\kappa \Theta} \L\we \tau_3 {\bf L}\,,
\ \ \ {\rm
i.e.} \ \ \ \
\delta_\kappa
{\FF}_{ab}=2{\rm i}\widehat{\delta_\kappa  \Theta}
(\L_a \tau_3 {\bf L}_b-\L_b \tau_3 {\bf L}_a)\,.
\end{equation}
Then the $D3$ brane
action $S$  in (3.1) is $\kappa$-invariant provided
the   5-form ${H}_5$ is given by
$$
H_5
={\rm i}{\bf L}\we \bigg(\frac{1}{6}\L\we\L\we\L \tau_2
+ {\FF}\we \L \tau_1
\bigg)   \we  {\bf L}
$$
\begin{equation}\label{h5}
+\ \frac{\f}{6}\bigg(\epsilon^{i_1\ldots i_4}L^+ \we L^{i_1}\we
\ldots\we L^{i_4} + \epsilon^{i_1'\ldots i_4'} L^+\we L^{i'_1}\we
\ldots\we  L^{i'_4}\bigg) \,.\end{equation} It is possible to
check  using Maurer-Cartan equations (see Appendix C) and
well-known Fierz identity\footnote{Throughout this paper
symmetrization and anti-symmetrization rules are defined with
normalization $(ab)=\frac{1}{2}(ab+ba)$,
$[ab]=\frac{1}{2}(ab-ba)$.} \be {\bf
L}\wedge \bar{\gamma}^{\mu\nu\rho}\tau_2{\bf  L}\wedge\,{\bf L}\wedge
\bar{\gamma}^\rho {\bf L} = 2{\bf L}\wedge\bar{\gamma}^{[\mu} \tau_1
{\bf L}\,\wedge {\bf L}\wedge\bar{\gamma}^{\nu]} \tau_3{\bf L}\ee
 that ${H}_5$ is closed, i.e. the equation \rf{hdo} is consistent
and thus determines  ${\cal L}_{\rm WZ}$.

The important fact is that  ${H}_5$ is expressed  in terms of the
Cartan 1-forms and super-invariant  ${\FF}$ only. This implies
that ${H}_5$ is invariant under space-time supersymmetry. Then
from \rf{hdo} we conclude that $\delta_{susy} (d^{-1} H_5)$ is
exact, so  that the WZ term \rf{hdo},  like the BI  term \rf{hhh},
is supersymmetry-invariant.

To put the fermionic part of the WZ term in the action in a more
explicit form let us make a rescaling $\Theta \to t\Theta$ and
define \be H_{5t} \equiv H_5|_{\Theta \rightarrow t \Theta}\ , \ \
\ \ \ \ \ \ {\rm F}_t\equiv {\rm F}|_{\Theta\rightarrow t\Theta}
\, .  \ee Since $ L(x,tt'\Theta)=
L_t(x,t'\Theta)=L_{tt'}(x,\Theta) $ one can show that (cf.
\rf{fexp},\rf{dee}) \be {\rm F}_t=dA+2{\rm i}\int_0^t dt' \,
\Theta \widehat{L}_{t'} \wedge \tau_3 {\bf L}_{t'} \ , \ \ \ \ \ \
\
\partial_t {\rm F}_t=2{\rm i} \Theta\widehat{L}_t\wedge \tau_3{\bf L}_t \ .
\label{iim} \ee Then using the  defining equations for the Cartan
1-forms \rf{t1}-\rf{t6} one finds from \rf{h5} the following
differential equation \be
\partial_t H_{5t}
=d\bigg[ 2{\rm i}(\frac{1}{6} \Theta \widehat{L}_t\wedge
\widehat{L}_t\wedge \widehat{L}_t \wedge \tau_2 {\bf L}_t +\Theta
\widehat{L}_t\wedge {\rm F}_t\wedge \tau_1 {\bf L}_t)\bigg] \,,\ee
which determines the $\Theta$-dependence of $H_5$. With the
initial condition \be (H_{5t})_{t=0} = H_5|_{\Theta=0} =
H_5^{(bose)} = \frac{\f}{6}e^+\we (\epsilon^{i_1\ldots i_4}
e^{i_1}\wedge \ldots \wedge e^{i_4} + \epsilon^{i_1'\ldots i_4'}
e^{i_1'}\wedge \ldots \wedge e^{i_4'}) \ , \ee where $e^\mu$ are
(pull-backs of) the vielbein forms  of plane wave background. The
explicit form of the $\Theta$-independent part ${\cal L}_{\rm
WZ}^{(bose)}= d^{-1} H_5^{(bose)}$ depends on a particular choice
of coordinates on plane wave background. Thus  the ${\cal L}_{\rm
WZ}$  term in \rf{hdo} can be written as \be \label{exx} {\cal
L}_{\rm WZ} = 2{\rm i} \int_0^1 dt \, \bigg(\frac{1}{6}
\Theta\widehat{L}_t\wedge \widehat{L}_t\wedge \widehat{L}_t
\wedge \tau_2{\bf L}_t +\Theta\widehat{L}_t\wedge{\rm F}_t\wedge
\tau_1{\bf L}_t\bigg) + {\cal L}_{\rm WZ}^{(bose)} \,. \ee Using
\rf{iim} and expansion of Cartan 1-forms in terms of $\Theta$ (see
Appendix C and \cite{rrm0112}) one can then find the expansion of
${\cal L}_{\rm WZ}$ in powers of $\Theta$.

The only non-trivial
background fields  in  plane wave  vacuum  are
the space-time metric and the self-dual RR 5-form.
The bosonic parts of the last two terms in
${H}_5$ \rf{h5}    represent, indeed, the
standard bosonic couplings  of the $D3$ brane
 to the   5-form background.
The action we have obtained  contains
also the fermionic terms required to make this
coupling supersymmetric and
$\kappa$-invariant.

We started with the BI action expressed in terms of the Cartan
1-forms and the  2-form in \rf{fexp}  \cite{rrm0112} as implied by
the structure of the plane wave  space or the basic symmetry
superalgebra.  We then fixed the form of ${H}_5$ from the
requirement of $\kappa$-symmetry of the full action. As in the
$AdS_5 \times S^5$ case \cite{rrm9806}, the  fact that we have
reproduced the bosonic part of the self-dual 5-form is in
agreement   with the result of \cite{ced,bt}  that  the $D3$ brane
action is $\kappa$-symmetric  only in a background  which is a
solution of type IIB supergravity.

Let us briefly discuss global plane wave supersymmetries and
$\kappa$ symmetries of $D3$ brane action. The supersymmetry
transformations of (super)coordinates brane fields
$x^{\underline{\nu}}$, $\Theta$, $A=A_ad\sigma^a$ can be presented
as Taylor series expansion in fermionic field $\Theta$ which
terminates at terms $\Theta^{32}$ in general. The leading terms of
this expansion we will need below are fixed to be
\begin{eqnarray}
&& \label{susy1}\delta_{susy} x^{\underline{\nu}} = {\rm i}
e_\mu^{\underline{\nu}}\, (\delta_{susy}\Theta)\bar{\gamma}^\mu
\Theta+O(\Theta^3)\,,
\\
&& \label{susy2} \delta_{susy} \Theta =\epsilon(x) +O(\Theta^2)\,,
\\
&&  \label{susy3}\delta_{susy} A = {\rm i} e^\mu (\delta_{susy}
\Theta) \bar{\gamma}^\mu \tau_3 \Theta +O(\Theta^3)\,,
\end{eqnarray}
where the Killing spinor $\epsilon(x)$ is given by \be\label{epsU}
\epsilon(x) = U\epsilon_0\,,\qquad U = \exp(-\frac{\f}{2}
x^I\Pi\gamma^+\bar{\gamma}^I \tau_2) \exp(-\frac{\f}{2}
x^+\Pi\gamma^+\bar{\gamma}^- \tau_2)\,. \ee The supersymmetry
transformations for (super)coordinates $x^{\underline{\nu}}$ and
$\Theta$ \rf{susy1},\rf{susy2} are fixed via standard coset
construction. The supersymmetry transformations for brane
world-volume field $A$ \rf{susy3} is chosen then so that the generalized
field strength $\FF$ \rf{fexp} be invariant with respect to
supersymmetry transformations \be \delta_{susy} \FF =0\,. \ee The
$\kappa$ transformations given in \rf{kaptra} can be expressed in
terms of conventional variation of super(coordinates)
$\delta_\kappa x^{\underline{\nu}}$ and $\delta_\kappa \Theta$ by
using formulas \rf{hatdel} and the expressions for Cartan 1-forms
given in \rf{exprep1}
\begin{eqnarray}
&&\label{kappa1} \delta_\kappa x^{\underline{\nu}} = -{\rm
i}e^{\underline{\nu}}_\mu (\delta_\kappa\Theta)\bar{\gamma}^\mu
\Theta +O(\Theta^3)\,,
\\
&&\label{kappa2} \delta_\kappa \Theta = {\bf K} +O(\Theta^2)\,,
\\
&&\label{kappa3} \delta_\kappa A = - {\rm i}e^\mu (\delta_\kappa
\Theta)\bar{\gamma}^\mu\tau_3\Theta + O(\Theta^3)\,.
\end{eqnarray}
Here the kappa-transformation of world-volume field $A$ is chosen
so that the kappa-transformation of generalized field strength
$\FF$ \rf{fexp} takes the form given in \rf{kvf}.

\newsection{Gauge fixed $D3$ brane action}\label{GFACT}

$D3$ brane action can be simplified by fixing local fermionic
kappa and world-volume diffeomorphism symmetries. Various
possibilities to fix these symmetries can be divided into two
classes - covariant and noncovariant gauges. In this Section we
find the form of the $D3$ brane action in the plane wave
background with R-R 5-form flux in the covariant gauge. Our
discussion  of the covariant gauge fixing closely repeats the same
steps as in ref.\cite{aps}, where the case of flat space-time was
treated.

In flat space  the $D3$ brane gauge  fixing procedure consists of
the two  stages:

(I) fermionic covariant gauge choice, i.e., fixing the
$\kappa$-symmetry by  $\theta^1 = 0$

(II) bosonic covariant gauge choice, i.e., fixing  the
world-volume diffeomorphism symmetry by $x^a(\sigma)  =
\sigma^a$\footnote{Discussions of alternative covariant bosonic
gauge choices may be found in \cite{Ryang:1998eu}.}.

\subsection{$\kappa$ symmetry covariant gauge fixed action}

Our fermionic  $\kappa$-symmetry covariant gauge is the same as in
flat $D3$ brane   $\theta^1 =0$. One usually imposes the
$\kappa$-symmetry covariant gauge by starting with the explicit
representation for the $D3$ brane Lagrangian in terms of
$\theta'$s. However it is convenient to first impose the covariant
gauge at the level of the Cartan forms $L^\mu$, ${\bf L}^\alpha$ and
then to use them in \rf{action}. In what follows we adopt this
strategy.\footnote{This strategy was first successfully used in
\cite{pes},\cite{kr} while deriving $\kappa$-symmetry fixed action
for long superstring in $AdS_5\times S^5$ and for
$\kappa$-symmetry fixed light cone $AdS$ superstring in
\cite{rrm0202}. Application of this method to plane wave
superstring may be found in \cite{rrm0112}.}

By applying argument similar to the ones in \cite{aps} we impose
the following covariant kappa symmetry gauge \be\label{covg}
\theta^1=0,\quad  \theta^2 =\lambda\,, \qquad {i.e.} \qquad \Theta
= \left( \begin{array}{c}
0 \\
\lambda \end{array} \right) \ee Because the $D3$ brane action
\rf{action} is expressible in terms of Cartan 1-forms we should
simply to evaluate the Cartan 1-forms in this gauge. This can be
done straightforwardly by using representation for the Cartan
1-forms given in Appendix C (see \rf{exprep1}-\rf{comM}) and
plugging there the expression for $\Theta$ given in \rf{covg}.
This leads to the following Cartan 1-forms \be\label{carggg1} {\bf
L} = \frac{\hbox{sinh}  {\bf m}}{\bf  m }{\cal D}\Theta\,,\qquad
L^\mu = e^\mu -2{\rm i}\Theta \,\bar{\gamma}^\mu
\frac{\hbox{cosh}{\bf m} -1}{{\bf m}^2}{\cal D}\Theta\,, \ee where
covariant derivative is simplified as compared with \rf{comder}
and is given by \be {\cal D}\Theta = \left( \begin{array}{c}
 \frac{\f}{2}e^\mu \Pi\gamma^+\bar{\gamma}^\mu \lambda
\\[7pt]
\DL{}\lambda
\end{array}\right)\,,\qquad
\DL{}\equiv d +\frac{1}{4}\omega^{\mu\nu}\gamma^{\mu\nu}\,, \ee
while the matrix ${\bf m}$ is \be {\bf m}^2
=\left(\begin{array}{cc} 0 & -{\rm i}\f
(\Pi\gamma^+\bar{\gamma}^\mu\lambda)^\alpha (\lambda
\bar{\gamma}^\mu)_\beta
\\
{\rm i}\f(\gamma^+\bar{\gamma}^\mu\lambda)^\alpha
(\lambda\bar{\gamma}^\mu\Pi)_\beta
& 0
\end{array}\right)\,.
\ee In contrast to the matrix ${\cal M}$ \rf{comM} which enters
definition of general Cartan 1-form the gauge fixed matrix ${\bf
m}$ turns out to be off-diagonal. This considerably simplify
structure of gauge fixed action. In order to enter definition of
action we evaluate gauge fixed 2-form $\FF$ which is given by
$\FF = \FF_{t=1}$ \be {\rm F}_t=dA - 2{\rm i}\int_0^t dt' \,
\lambda \widehat{L}_{t'} \wedge  L_{t'}^2 \,, \ \ \ \ \ \ \ \ee
where $L^2$ is the second component of ${\bf L}$ (see
\rf{carggg1},\rf{vec2com}). The expressions above given define BI
part the action \rf{hhh}. Gauge fixed  WZ part of the action can
be obtained from \rf{exx} \be \label{exxcov} {\cal L}_{\rm WZ} =
2{\rm i} \int_0^1 dt \, \bigg( - \frac{1}{6}
\lambda\widehat{L}_t\wedge \widehat{L}_t \wedge \widehat{L}_t
L_t^1 +\lambda\widehat{L}_t\wedge{\rm F}_t\wedge L_t^1\bigg) +
{\cal L}_{\rm WZ}^{(bose)} \,. \ee Above given expression provides
possibility to find expansion of $D3$ brane action in terms of
fermionic field. Note however that though the gauge fixed action
is much more simpler than the covariant one there is no natural
way to integrate out parameter $t$ in expressions for $F_t$ and
${\cal L}_{WZ}$ to bring action to completely explicit
form\footnote{This minor problem can be by passed in light cone
gauge.}. Total Lagrangian can presented as Taylor series in field
strength $F_{ab}$ and the fermionic field $\lambda$ \be {\cal L}=
{\cal L}_0 + {\cal L}_2 + {\cal L}_3+ {\cal L}_4 +O(F^3\lambda^2,
F\lambda^4, \lambda^6)\,, \ee where
\begin{eqnarray}
\label{L0}{\cal L}_0 & = &  -\sqrt{g} +{\cal L}_{WZ}^{(bose)}\,,
\\
\label{L2}
{\cal L}_2 & = & \sqrt{g}\Bigl( -\frac{1}{4}F^{ab}F_{ab}
+ {\rm i}e^{\mu a}\lambda\bar{\gamma}^\mu \DL{a} \lambda\Bigr) -
\frac{{\rm i}\f}{12}\epsilon^{abcd} e_a^\mu e_b^\nu e_c^\rho
e_d^\sigma (\lambda \bar{\gamma}^{\mu\nu\rho}\Pi\gamma^+
\bar{\gamma}^\sigma \lambda)\,,
\\
{\cal L}_3 & = & {\rm i}\,\sqrt{g} F^{ab}e_a^\mu
\lambda\bar{\gamma}^\mu \DL{b} \lambda +\frac{{\rm
i}\f}{4}\epsilon^{abcd} F_{ab}e_c^\mu e_d^\nu (\lambda
\bar{\gamma}^\mu\Pi\gamma^+ \bar{\gamma}^\nu \lambda)\,,
\end{eqnarray}
\begin{eqnarray}\label{L4}
{\cal L}_4 & = & \sqrt{g}\Bigl(\frac{1}{8} F^{ab}
F_{bc}F^{cd}F_{da} -\frac{1}{32} (F^{ab}F_{ab})^2
\nonumber\\
&-&{\rm i}F^{ac}F^b{}_c e_a^\mu \lambda\bar{\gamma}^\mu \DL{b}\lambda
+\frac{\rm i}{4} F^{bc}F_{bc} e^{\mu a} \lambda\bar{\gamma}^\mu \DL{a}\lambda
\nonumber\\
& + &  \frac{1}{2}\lambda\bar{\gamma}^\mu \DL{}^a \lambda
\lambda\bar{\gamma}^\mu \DL{a} \lambda +\frac{1}{2}(e^{\mu
a}\lambda\bar{\gamma}^\mu \DL{a} \lambda)^2 -e_a^\mu e_b^\nu
\lambda\bar{\gamma}^\mu \DL{b} \lambda \lambda\bar{\gamma}^\nu
\DL{a} \lambda\Bigr) \nonumber
\nonumber\\
&+&\f\epsilon^{abcd}e_a^\mu e_b^\nu
\Bigl(\frac{1}{8}
e_c^\sigma
\lambda\bar{\gamma}^\rho \DL{d}\lambda
-\frac{1}{72}
e_c^\rho
\lambda\bar{\gamma}^\sigma \DL{d}\lambda\Bigr)
(\lambda \bar{\gamma}^{\mu\nu\rho}\Pi\gamma^+ \bar{\gamma}^\sigma \lambda)
\nonumber\\
&+& \frac{\f}{4}\epsilon^{abcd} e_a^\mu e_b^\nu e_c^\rho
\lambda\bar{\gamma}^\rho \DL{d}\lambda (\lambda
\bar{\gamma}^\mu\Pi\gamma^+ \bar{\gamma}^\nu \lambda) +\sqrt{g}
\frac{\f^2}{24}\lambda \bar{\gamma}^{+\mu\nu}\lambda \lambda
\bar{\gamma}^{+\mu\nu}\lambda\,.
\end{eqnarray}
In these formulas $g_{ab}$ is a bosonic body of the induced
world-volume metric $G_{ab}$ \rf{Gab} \be\label{gab0} g_{ab}=
e_a^\mu e_b^\mu, \qquad e^\mu =d\sigma^a e_a^\mu\,, \qquad g\equiv
-\hbox{det}\, g_{ab}\,.\ee The above given representation for $D3$
brane action is valid for arbitrary coordinate on plane wave
background. The explicit form of the bosonic body of WZ part of
the action, ${\cal L}_{\rm WZ}^{(bose)}$, depends on a particular
choice of coordinates on plane wave background. We chose the
coordinate frame in which vielbeins $e^\mu$ take the form
\be\label{bosvie} e^+ =dx^+\,, \qquad e^I =dx^I\,,\qquad e^- =
dx^- -\frac{\f^2}{2}x^Ix^I dx^+\,,\ee For this particular choice
of coordinates we get the following representation for ${\cal
L}_{\rm WZ}^{(bose)}$ \be\label{LWZB} {\cal L}_{WZ}^{(bose)} = -
\frac{\f }{6}\epsilon^{abcd}\partial_a x^+
(\epsilon^{ijkl}x^i\partial_b x^j \partial_c x^k \partial_d x^l +
\epsilon^{\ipr\jpr\kpr\lpr}x^\ipr \partial_b x^\jpr
\partial_c x^\kpr \partial_d x^\lpr)\,,
\ee while the induced metric tensor takes the form \be\label{gab1}
g_{ab} \equiv 2\partial_{(a} x^+\partial_{b)} x^- -\f^2x^Ix^I
\partial_a x^+\partial_b x^+ +\partial_ax^I\partial_b x^I\,. \ee
In what follows we assume this particular choice of the
coordinates. We normalize the Levi-Civita symbols to be
$\epsilon^{0129}=\epsilon^{1234}=\epsilon^{5678}=1$.
Note that as $\f \rightarrow 0$ the expansion
for ${\cal L}$ shown in \rf{L0}-\rf{L4} reduces to the one in the
flat space time (see \cite{aps},\cite{yon}).

\subsection{Static gauge fixed $D3$ brane action and its symmetries}

Our static gauge being the same as in flat space leads to the
action of self-interacting abelian $\cN=4$ SYM in $4d$ plane wave
background. In this section we study realization of bosonic
symmetries of static gauge fixed brane action. We demonstrate that
only isometry symmetries of $4d$ plane wave background plus
R-symmetries $SO(2)\times SO'(4)$ are realized linearly while the
remaining bosonic symmetries are realized non-linearly. In this
respect the situation is similar to the one in flat space. Because
number of isometry symmetries in $4d$ plane wave background is
equal to seven we note that number of linearly realized space-time
symmetries is equal to number of linearly realized
R-symmetries\footnote{This is not the case in flat space. After
imposing the static gauge in the $D3$ brane action the Poincar\'e
symmetries of ten-dimensional Minkowski space-time reduces to the
10 linearly realized Poincar\'e symmetries of $4d$ Minkowski
space,  15 linearly realized R-symmetries of $SO(6)$ plus certain
non-linearly realized symmetries. Thus in the flat space the
number of linearly realized space-time symmetries is not equal to
the number of linearly realized R-symmetries.}.

Making field redefinitions and introducing conventional notation
for six scalar fields

\be\label{fiered} x^\mun\, (\sigma) \rightarrow -
x^\mun\,(\sigma)\,, \qquad \phi^M(\sigma) \equiv -
x^M(\sigma)\,,\quad \mun=0,1,2,9\,;\quad M=3,\ldots,8\,, \ee we
impose the standard static gauge

\be\label{stagau} x^\mun\, (\sigma) =
\delta_a^\mun\,\sigma^a\equiv \sigma^\mun\,,\qquad a=0,1,2,9\,,\ee
where $\delta_a^\mun =1(0)$ for $\mun=a$($\mun=\!\!\!\!\!/\,\,
a$). Introducing light cone coordinates on $D3$ brane

\be \sigma^\pm =\frac{1}{\sqrt{2}}(\sigma^9 \pm \sigma^0)\,,
\qquad \sigma^\hi\,,\qquad  \hi =1,2\,, \ee and inserting the
static gauge into \rf{gab1} we get the gauge fixed induced
metric\footnote{After choice of static gauge the indices
$\mun\,,\nun$ are used for target space vectors while indices
$a,b$ are used for tangent space vectors. These vectors are
related as $A^\mun= e^\mun_a A^a$, where $e^\mun_a$ is inverse to
the basis of one forms $e^a=e_\mun^adx^\mun\,$, $e^\mun_a\,
e_\mun^b =\delta_a^b$. The basis of $e^a$ is specified to be
$e^+=dx^+$, $e^\hi=dx^\hi$, $e^-=dx^- - (\f^2/2)\sigma^\hi
\sigma^\hi dx^+$.}

\be\label{gabstagau} g_{\mun\,\nun} = g_{\mun\,\nun}^{(pw)} -\f^2
\phi^M\phi^M \delta_\mun^+\delta_\nun^+
+\partial_\mun\phi^M\partial_\nun\phi^M\,,\ee where
$g_{\mun\,\nun}^{(pw)}$ is a metric tensor of $4d$ plane wave
background

\be g_{\mun\,\nun}^{(pw)}d\sigma^\mun\, d\sigma^\nun
=2d\sigma^+d\sigma^- - \f^2 \sigma^\hi\sigma^\hi
d\sigma^+d\sigma^+ + d\sigma^\hi d\sigma^\hi\,. \ee Taking into
account the formula \rf{gabstagau} we see that if we will treat
$g^{(pw)}$ as background field and make Taylor series expansion
with respect to remaining terms, $\f^2\phi^2$ and
$(\partial\phi)^2$, then the static gauge fixed $D3$ brane action
will be manifestly invariant with respect to isometry symmetries
of $4d$ plane wave background. It is natural to expect then that
these isometry symmetries are realized linearly. To demonstrate
this point explicitly we consider transformations of brane fields
with respect to original plane wave symmetries in ten dimensions.
This is to say we start with global plane wave transformations
supplemented with local diffeomorphism transformation
\be\label{tottra1} \delta x^\muun =  \xi^{G\muun} +
\epsilon^a\partial_a x^\muun\,, \ee where $\xi^{G\muun}$ is
Killing vector corresponding to plane wave global transformation
generated by element of algebra denoted by $G$. Representation of
the Killing vectors in term of differential operators
$G=\xi^{G\underline{\nu}}\partial_{x^{\underline{\nu}}}$ is given
in \rf{PWK1}-\rf{PWK4}. From the requirement that the transformation
\rf{tottra1} maintains the static gauge \rf{stagau}, $\delta
x^\mun =0$, we fix parameter of compensating transformation \be
\epsilon^\mun=-\xi^{G\mun}\,. \ee Plugging this into \rf{tottra1}
we get transformation rules of six scalar fields \be \delta_G x^M
= \xi^{G\, M}  -\xi^{G\,\mun}\partial_\mun x^M\,. \ee Making use
of concrete representation for Killing vectors plane wave
background \rf{PWK1}-\rf{PWK4} we get the following
transformations with respect to transverse translations and
Lorentz boosts

\begin{eqnarray}
\label{tottra2}&&
 \delta_{(a^IP^I)} x^M =\cos \f x^+ a^M -(\cos\f x^+
a^\hi\partial_\hi + \f\sin\f x^+ (a^Ix^I)\partial^+ )x^M\,,
\\
\label{tottra3}&&\delta_{(b^IJ^{+I})} x^M =\frac{\sin \f x^+}{\f}
b^M -(\frac{\sin\f x^+}{\f} b^\hi\partial_\hi -\cos\f x^+
(b^Ix^I)\partial^+ )x^M\,,
\end{eqnarray}
where $a^I$ and $b^I$ are parameters of the appropriate
transformations. Plugging the static gauge \rf{stagau} into
transformation rules \rf{tottra2},\rf{tottra3} we see that the
transformations generated by $P^\hi$ and $J^{+\hi}$

\begin{eqnarray}
&& \delta_{P^\hi}\phi^M  = (\cos\f \sigma^+ \partial_\hi +
\f\sin\f \sigma^+ \sigma^\hi \partial^+ )\phi^M\,,
\\
&& \delta_{J^{+\hi}}\phi^M = (\frac{\sin\f \sigma^+}{\f}
\partial_\hi -\cos\f \sigma^+ \sigma^\hi\partial^+ )\phi^M\,,
\end{eqnarray}
are realized linearly and coincide with transformation of isometry
algebra of $4d$ plane wave background. It is easy to check that
transformations generated by translations in light cone directions
$P^\pm$ and $SO(2)\times SO(2)\times SO'(4)$ rotations generated
by $J^{12}$, $J^{34}$, $J^{\i'j'}$ which are obtainable from
\rf{tottra1}, are also realized linearly. The generators $P^\pm$,
$P^\hi$, $J^{+\hi}$ and $J^{12}$ form algebra of isometry
symmetries of $4d$ plane wave space-time while the generators
$J^{34}$, $J^{i'j'}$ are responsible for $R$-symmetries, which are
$SO(2)\times SO'(4)$ rotations.

The transformations of remaining  six translations $P^N$ and six
Lorentz boosts $J^{+N}$ take the form
\begin{eqnarray}
&& \delta_{a^N P^N}\phi^M =\cos \f \sigma^+ a^M - \f\sin\f
\sigma^+ a^N\phi^N\partial^+ \phi^M\,,
\\
&& \delta_{b^N J^{+N}}\phi^M =\frac{\sin \f \sigma^+}{\f} b^M +
\cos\f \sigma^+ b^N \phi^N\partial^+ \phi^M\,,
\end{eqnarray}
and these transformations are obviously broken and realized
non-linearly.

Broken and unbroken (super)symmetries of kappa symmetry and static
gauge fixed $D3$ brane action are collected in Table 1.

\bigskip\noindent
TABLE 1:  {\sf Broken and unbroken (super)symmetries of kappa
symmetry and static gauge} \\  \phantom{a} \hspace{1.5cm} {\sf
fixed $D3$ brane action}

\begin{center}
\begin{tabular}{|c|c|c|}
\hline        &&
\\ [-3mm]Generators  of & Generators of & Generators of
\\
plane wave  & unbroken & broken
\\
superalgebra  & symmetries             & symmetries
\\[2mm] \hline
& &
\\[-3mm]
$P^+$, $P^-$       &  $P^+$, $P^-$             &
\\[2mm] \hline
 & &
 \\[-3mm]
$ P^I$,\quad $I=1,\ldots 8$ & $P^\hi$, \  \ $\hi=1,2 $ & $P^M$,\ \
\ $M=3, \ldots, 8$
\\ [2mm] \hline
 & &
 \\[-3mm]
$ J^{+I}$,\quad $I=1,\ldots, 8$             & $J^{+\hi}$, \  \
$\hi=1,2 $ & $J^{+M}$,\ \ \ $M=3, \ldots, 8$
\\ [2mm] \hline
& &
\\ [-3mm]
$J^{ij}$,\quad $i,j=1,\ldots,4 $        &  $J^{12}$, $J^{34}$ \ &
$J^{13}$, $J^{14}$, $ J^{23} $, $ J^{24} $
\\[2mm] \hline
& &
\\ [-3mm]
$J^{i'j'}$,\quad $i',j'=5,\ldots,8 $     &  $J^{i'j'}$,\quad
$i',j'=5,\ldots,8 $    &
\\[2mm] \hline
 & &
 \\ [-3mm]
\ \ \  $Q^1_\alpha$, $Q_\alpha^2$,\quad $\alpha=1,\ldots,16$   & $
Q^1\zeta_0 +Q^2$ & $ Q^1\zeta_0 -Q^2$
 \\[2mm]\hline
 \end{tabular}
\end{center}

{}~
\bigskip

It is instructive to find commutation relations of generators of
unbroken symmetries. These commutation relations can be obtained
from the ones of the plane wave superalgebra given in
\rf{pmpi1}-\rf{qq}. Bosonic generators form symmetries of $4d$
plane wave space-time and R-symmetries. All that is required is to
find (anti)commutators involving unbroken supercharge
\be\label{unbsup} \rQ = \frac{1}{\sqrt{2}}(Q^1\zeta_0 +
Q^2)\,,\qquad \zeta_0\equiv \gamma^{-+12}\,.\ee Commutation
relations of bosonic generators with this supercharge take the
form $$ [J^{ij},\rQ_\alpha^\pm] = \frac{1}{2}\rQ_\beta^\pm
(\gamma^{ij})^\beta{}_\alpha \,, \quad
[J^{\ipr\jpr},\rQ_\alpha^\pm] =
\frac{1}{2}\rQ_\beta^\pm(\gamma^{\ipr\jpr})^\beta{}_\alpha\,,
\quad [J^{+\hi },\rQ_\alpha^-] =
\frac{1}{2}\rQ_\beta^+(\gamma^{+\hi})^\beta{}_\alpha\,, \ \ \ $$
\be\label{parq}
[P^a,\rQ_\alpha]
=-\frac{\f}{2}\rQ_\beta(\gamma^{+34}\bar{\gamma}^a)^\beta{}_\alpha\,,
\ee where
we have to keep just the $so(2)\oplus so(2)$ part of $J^{ij}$
given by $J^{12}$, $J^{34}$. The anticommutator of supercharges
corresponding to unbroken supersymmetries is given by
\begin{eqnarray}
\{\rQ_\alpha,\rQ_\beta\} & = & -2{\rm i}\gamma^a_{\alpha\beta}P^a
+2{\rm i}\f\bar\gamma_{\alpha\beta}^{+-34\hi }J^{+\hi}
\nonumber\\
&+&2{\rm i}\f (\bar{\gamma}^+\Pi)_{\alpha\beta}J^{12} - 2{\rm
i}\f\gamma^+_{\alpha\beta}J^{34} -{\rm
i}\f\bar\gamma_{\alpha\beta}^{+34i'j'}J^{i'j'}\,.
\end{eqnarray}
These (anti)commutation relations tell us that generators of
unbroken symmetries form some subsuperalgebra of the original
plane wave superalgebra. We demonstrated that the unbroken bosonic
symmetries are realized linearly. As to the unbroken
supersymmetries it is not obvious that they can also be realized
linearly by appropriate choice of parametrization of fermionic
fields. For the case of static gauge fixed $D9$ brane in flat
space-time it is known that these supersymmetries are realized
non-linearly \cite{Bergshoeff:1986jm,rrm87}. Because $D3$ brane has
the same amount of unbroken supersymmetries it seems highly likely
that the $D3$ brane unbroken supersymmetries are also realized
non-linearly.

\subsection{$D3$ brane friendly form of abelian $\cN=4$ SYM
in plane wave background}

Now we study the $D3$ brane action in quadratic approximation in
fields. Looking at the part of $D3$ brane action given by ${\cal
L}_0$ and ${\cal L}_2$ (see \rf{L0},\rf{L2}) one can expect
appearance of some mass like terms for fermionic and bosonic
scalar fields. Therefore in quadratic approximation in fields the
$D3$ brane action reduces to the standard action of free abelian
$\cN=4$ SYM in plane wave background plus some mass like terms for
both the fermionic and bosonic scalar fields. It is instructive to
understand structure of these mass-like terms. This is what we are
doing in this section.

In static gauge this part of $D3$ brane action is obtainable from
${\cal L}_0$, ${\cal L}_2$ given in \rf{L0}, \rf{L2}. Introducing
the notation \be\label{psilam} \qquad \psi\equiv
\sqrt{2}\lambda\,,\quad  \quad Z \equiv \frac{1}{\sqrt{2}}(\phi^3
+{\rm i} \phi^4)\,,\quad \bar{Z} \equiv \frac{1}{\sqrt{2}}(\phi^3
-{\rm i} \phi^4)\,, \ee we get the following Lagrangian for
abelian spin one field $A_\mun\equiv\delta_\mun^a\,A_a$, four
real-valued scalars $\phi^\ipr$, one complex-valued scalar $Z$,
and the sixteen component real-valued one-half spin fermionic
field $\psi$ \be\label{D3SYM}{\cal L}= {\cal L}_{st} + \Delta
{\cal L}\,, \ee where ${\cal L}_{st}$ stands for the standard
Lagrangian of abelian $\cN=4$ SYM theory in $4d$ plane wave
background \be\label{D3SYM1} {\cal L}_{st}  =
-\frac{1}{4}F^{\mun\,\nun}F_{\mun\,\nun}
-\frac{1}{2}g^{\mun\,\nun}\partial_\mun \phi^\ipr \partial_\nun
\phi^\ipr   -  g^{\mun\,\nun}\partial_\mun \bar{Z}
\partial_\nun Z  -  \frac{\rm i}{2}\psi\bar{\gamma}^\mun
\DL{\mun}\psi\,,\ee \be \DL{\mun}=\partial_\mun
-\frac{\f^2}{2}\sigma^\hi\gamma^{+\hi}\delta_\mun^+\,,\qquad
\gamma^\mun =e_a^\mun\, \gamma^a\,, \ee while $\Delta{\cal L}$
describes unusual mass-like terms \be\label{D3SYM2} \Delta {\cal
L}= - 2{\rm i}\f (\bar{Z}\partial^+ Z - Z\partial^+
\bar{Z})-\frac{\rm i}{2}\f \psi \bar{\gamma}^{+34}\psi\,. \ee Note
that ${\cal L}_{st}$ is obtainable from the kinetic part of $D3$
brane action \rf{hhh} while the mass-like terms $\Delta{\cal L}$
are coming from WZ part of brane action \rf{hdo}. Thus modulo
unusual mass-like terms for the complex-valued scalar fields $Z$
and fermionic field $\psi$ we get the Lagrangian for abelian $\cN=4$
SYM theory in $4d$ plane wave background. It turns out that these
unusual mass-like terms depend on field redefinitions. Indeed
making use of the field redefinition \be\label{fiered1} Z
\rightarrow e^{{\rm i}(\rw + 2)\f \sigma^+}Z\,, \qquad\qquad \psi
\rightarrow e^{-\frac{\rw + 2}{2}\f\gamma^{34}\sigma^+}\psi\,,\ee
where $\rw$ is a constant parameter we get the Lagrangian
\rf{D3SYM} with the following mass terms \be\label{lagw}
\Delta{\cal L}  = {\rm i}\rw\f (\bar{Z}\partial^+ Z -Z\partial^+
\bar{Z})+\frac{{\rm i}\rw\f}{4} \psi
\bar{\gamma}^{+34}\psi\,,\qquad
\partial^+ \equiv \partial/\partial\sigma^-\,. \ee
Thus we see that mass-like
terms depend on the parameter of fields redefinition $\rw$. A
scheme in which $\Delta{\cal L}$ takes the form given in \rf{lagw}
we shall refer to as $\rw$-scheme.

The Lagrangian given in \rf{D3SYM1},\rf{D3SYM2} was derived
directly from the Lagrangian of the $D3$ brane and it corresponds
to $\rw=-2$ scheme. Therefore this scheme we shall refer to as $D3$
brane friendly scheme. Another scheme, which we shall refer to as
conventional scheme, does not involve unusual mass-like terms.
This conventional scheme is achieved by setting $\rw=0$. From the
transformation given in \rf{fiered1} it is clear that Hamiltonian
of abelian plane wave SYM in arbitrary $\rw$ scheme is given by
\be\label{pmw} P_\rw^- = P^- + \f(\rw+2) J^{34}\,, \ee where $P^-$
is the Hamiltonian of plane wave SYM taken to be in $D3$ friendly
$\rw=-2$ scheme (see \rf{D3SYM1},\rf{D3SYM2}). Peculiar properties
of various schemes, $D3$ brane friendly and conventional ones, can
be understood by study of supersymmetry. Let us discuss therefore
supersymmetry transformations of abelian plane wave SYM.

Because the $4d$ plane wave metric is conformally flat the abelian
$\cN=4$ plane wave SYM is invariant with respect to 30 bosonic and
32 supersymmetries. Only fourteen bosonic and sixteen
supersymmetry transformations of plane wave abelian SYM can be
derived from unbroken symmetries of $D3$ brane action. The
fourteen unbroken $D3$ brane symmetries which are visible in plane
wave SYM are shown in Table 1\footnote{Remaining $D3$ brane 16
bosonic and 16 super symmetries being broken already in static
gauge fixed $D3$ brane action become to be contracted when we
restrict $D3$ brane action to the action of abelian SYM.}.
According this Table the sixteen supersymmetries of abelian SYM
related with unbroken symmetries of $D3$ brane are generated by
supercharge $\rQ$ \rf{unbsup}. Taking into account representation
for the plane wave SYM Hamiltonian $P_\rw^-$ \rf{pmw} and the
commutators \rf{parq} we get the following commutators between
$\rw$-scheme Hamiltonian $P_\rw^-$ and kinematical and dynamical
supercharges \be\label{pmrq1} [P_\rw^-,\rQ_\alpha^+] =
\frac{\rw}{2}\f \rQ_\beta^+(\gamma^{34})^\beta{}_\alpha\,, \qquad
[P_\rw^-,\rQ_\alpha^-] =  \frac{2+\rw}{2}\f
\rQ_\beta^-(\gamma^{34})^\beta{}_\alpha\,, \ee where kinematical,
$\rQ^+$, and dynamical, $\rQ^-$,  supercharges are defined in
standard way \be \rQ^+ =\frac{1}{2}\bar{\gamma}^-\gamma^+
\rQ\,,\qquad \rQ^- =\frac{1}{2}\bar{\gamma}^+ \gamma^- \rQ\,. \ee
{}From the commutators given \rf{pmrq1} we see that in $D3$ brane
friendly scheme, $\rw=-2$, the Hamiltonian of plane wave SYM is
commuting with dynamical charges and does not commute with
kinematical supercharges. Because supermultiplet of SYM theory is
built out with the help of kinematical supercharges this implies
that lowest energy values for fields of SYM, $A_\mun\,$, $Z$ and
$\psi$, are different\footnote{Alternative way to achieve this
conclusion is simply to solve equations of motions.}. On the other
hand in conventional scheme, $\rw=0$, the Hamiltonian is commuting
with kinematical supercharge and does not commute with dynamical
supercharge. This implies that in conventional scheme all fields
of SYM have the same lowest energy values.

To finish discussion of plane wave SYM we write down an explicit
form of sixteen supersymmetry transformations of abelian SYM which
are obtainable from the unbroken supersymmetries of $D3$
brane\footnote{Note that the supersymmetry transformations
obtainable from $D3$ brane correspond to $\rw=-2$ scheme.
Transformations rules \rf{supsym1},\rf{supsym2} can be derived
then by going from this $\rw=-2$ scheme to the arbitrary scheme
via transformations \rf{fiered1}. }
\begin{eqnarray}
\label{supsym1}\delta A_\mun & =& {\rm i} e_\mun^a  \psi
\bar{\gamma}^a \hat{\epsilon}\,, \qquad \delta \phi^M  =   {\rm i}
\psi \bar{\gamma}^M \hat{\epsilon}\,,
\\
\label{supsym2} \delta \psi \ \ &=&
(\frac{1}{2}\gamma^{\mun\,\nun}F_{\mun\,\nun} -\gamma^M
\bar{\gamma}^\mun \partial_\mun \phi^M)\hat{\epsilon} -
\f\phi^\ipr \gamma^\ipr\bar{\gamma}^{+34}\hat{\epsilon}
\nonumber\\
& - &\f(\rw + 1
)(Z\gamma^{\bar{Z}}+\bar{Z}\gamma^Z)\bar{\gamma}^{+34}\hat{\epsilon}\,,
\quad
\gamma^Z\equiv \frac{1}{\sqrt{2}}(\gamma^3+{\rm i}\gamma^4)\,,\quad
\gamma^{\bar{Z}}=(\gamma^Z)^*\,,
\end{eqnarray}
where parameter of transformation $\hat{\epsilon}$, which is the
Killing spinor, satisfies the equation \be \DL{}\hat{\epsilon} =
\frac{\f}{2}(e^a \gamma^a\bar{\gamma}^{+34} +\rw
e^+\gamma^{34})\hat{\epsilon} \,, \qquad \DL{} =
d\sigma^\mun\,\DL{\mun}\,. \ee Explicit solution to this equation
is fixed to be \be \hat{\epsilon} =
\exp(\frac{\rw+2}{2}\f\gamma^{34}x^+)
\exp(-\frac{\f}{2}x^\hi\gamma^{+34}\bar{\gamma}^\hi)
\exp(-\frac{\f}{2}x^+\gamma^{+34}\bar{\gamma}^-)\epsilon_0\,, \ee
where $\epsilon_0$ is a sixteen component real-valued constant
fermionic parameter. The $\rw$-scheme Lagrangian which is
invariant with respect to the transformations \rf{supsym1},
\rf{supsym2} is given by  \rf{D3SYM},\rf{lagw}.

\subsection{Supersymmetries of gauge fixed brane action}

In this section we investigate realization of supersymmetries of
kappa gauge and static gauge fixed $D3$ brane action. We would
like to learn which supersymmetries are unbroken and which ones
become broken. To investigate this problem we study supersymmetry
transformations of brane fields. In general all original
supersymmetry transformations of brane fermionic fields (see
\rf{susy2}) look like Goldstone type, i.e. they shift the
fermionic field by constant parameter. It turns out however that
it is possible to construct some combination of original brane
transformations which do not shift the fermionic field by constant
parameter and these transformations correspond to unbroken
symmetries. All the remaining transformations are Goldstone type
and they are associated with broken symmetries.

The plane wave supersymmetries of covariant kappa symmetry gauge
and static gauge fixed brane action can be found following
standard procedure \cite{aps}. Introducing notation $\Xi$ for
brane fields $(x^{\underline{\nu}}\,, \Theta, A_a)$ we start with
supersymmetry transformations supplemented by local
$\kappa$-symmetry transformations \be\label{comtra1} \delta \Xi
=\delta_{susy}\Xi + \delta_\kappa \Xi\,. \ee The parameter of
$\kappa$ transformations is fixed by the requirement that the complete
transformation \rf{comtra1} maintains the $\kappa$-symmetry gauge
\rf{covg} \be\label{delthe0} \delta_{susy} \theta^1 +\delta_\kappa
\theta^1 = 0\,.\ee Representing the parameters of
$\kappa$-transformation \rf{kaptra}, supersymmetry transformation
$\epsilon(x)$ \rf{epsU} and the matrix $\Gamma$ \rf{Gammat} as \be
{\bf K}^\alpha =\left(\begin{array}{c} \kappa^{1\alpha}
\\[5pt]
\kappa^{2\alpha}
\end{array}\right)\,,\qquad
\epsilon =\left(\begin{array}{c} \epsilon^{1\alpha}
\\[5pt]
\epsilon^{2\alpha}
\end{array}\right)\,,\qquad
\Gamma =\left(\begin{array}{cc}
 0 & \zeta
 \\[5pt]
\tilde{\zeta} & 0
\end{array}\right)\,,
\ee where \be \zeta,\tilde{\zeta}  =  \frac{\epsilon^{a_1\ldots
a_4}}{\sqrt{-\det(G_{ab}+{\FF}_{ab})}} \Bigl( \pm \frac{1}{4!}
\gamma_{a_1\ldots a_4} +\frac{1}{4}\gamma_{a_1a_2}{\FF}_{a_3a_4}
\pm \frac{1}{8}{\FF}_{a_1a_2}{\FF}_{a_3a_4}\Bigr)\,, \ee we find
that the relations \rf{kappa} imply \be\label{varrel}
\zeta\tilde{\zeta}= \tilde{\zeta}\zeta = 1\,, \qquad \kappa^1
=\zeta \kappa^2\,,\qquad \kappa^2 =\tilde{\zeta} \kappa^1\,. \ee
Taking into account supersymmetry and $\kappa$ transformations for
$\Theta$ \rf{susy2},\rf{kappa2} we get from \rf{delthe0} solution
to $\kappa^1$: $\kappa^1 =-\epsilon^1$. In view of \rf{varrel} we
conclude that both the local $\kappa$ parameters, $\kappa^1$ and
$\kappa^2$, are expressible in terms of global parameter of
supersymmetry transformations \be \kappa^1 =-\epsilon^1\,,\quad
\qquad \kappa^2 =-\tilde{\zeta}\epsilon^1\,. \ee Plugging solution
to $\kappa$ parameters into \rf{comtra1} and taking into account
transformations \rf{susy1}-\rf{susy3},\rf{kappa1}-\rf{kappa3} and
notation \rf{covg} we get the following supersymmetry
transformations of brane fields
\begin{eqnarray}
\delta A & = & -{\rm i}e^\mu (\epsilon^2 +
\tilde{\zeta}\epsilon^1)\bar{\gamma}^\mu \lambda\,,
\\
\delta x^{\underline{\nu}} &= & {\rm
i}e^{\underline{\nu}}_\mu(\epsilon^2 + \tilde{\zeta}\epsilon^1)
\bar{\gamma}^\mu \lambda\,,
\\
\label{bratra3} \delta \lambda &= & \epsilon^2
-\tilde{\zeta}\epsilon^1\,.
\end{eqnarray}
From the transformation rule given in \rf{bratra3} it is clear
that the brane fermionic field transforms like Goldstone field in
general. Note however that this does not imply that all
supersymmetries are broken. It turns out that sixteen
supersymmetries are still to be unbroken. Let us demonstrate this
point explicitly. Representing the constant parameter $\epsilon_0$
\rf{epsU} \be \epsilon_0 =\left(\begin{array}{c} \epsilon_0^1
\\[5pt]
\epsilon_0^2
\end{array}\right)\,,
\ee we get from \rf{epsU} the following representation for the
components of the Killing spinor $\epsilon(x)$
\begin{eqnarray}
\label{eps1} \epsilon^1(x) &=& \frac{1}{2}(\gamma^-\bar{\gamma}^+
+ \gamma^+\bar{\gamma}^-\cos\f x^+)\epsilon_0^1 -
\frac{1}{2}\Pi\gamma^+(\f x^I \bar{\gamma}^I +
\bar{\gamma}^-\sin\f x^+) \epsilon_0^2\,,
\\
\label{eps2} \epsilon^2(x) &= & \frac{1}{2}(\gamma^-\bar{\gamma}^+
+ \gamma^+\bar{\gamma}^-\cos\f x^+)\epsilon_0^2 +
\frac{1}{2}\Pi\gamma^+(\f x^I \bar{\gamma}^I +
\bar{\gamma}^-\sin\f x^+) \epsilon_0^1\,.
\end{eqnarray}
Now we restrict the 32 supersymmetries associated with parameters
$\epsilon_0^1$, $\epsilon_0^2$  to the 16 supersymmetries by
imposing the constraint on the parameters of transformations
\be\label{respar} \epsilon_{0r}^1 = \frac{1}{2}\zeta_0 \eta_0\,,
\qquad \epsilon_{0r}^2 =\frac{1}{2}\eta_0\,,\qquad \zeta_0\equiv
\gamma^{-+12}\,, \ee where $\eta_0$ is a real-valued sixteen
component constant spinor and suffix `$r$' is used to indicate the
fact we restrict 32 parameters $\epsilon_0^1$, $\epsilon_0^2$
\rf{eps1},\rf{eps2} to 16 independent parameters $\eta_0$ . Now if
we plug these $\epsilon_{0r}^1$, $\epsilon_{0r}^2$ in expressions
for $\epsilon^1(x)$, $\epsilon^2(x)$ \rf{eps1},\rf{eps2} and
insert such restricted values of $\epsilon^1(x)$, $\epsilon^2(x)$
on r.h.s of \rf{bratra3} then we learn that linear field
independent term proportional to $\eta_0$ cancels out. This means
that 16 supersymmetries associated with parameter $\eta_0$ are
unbroken. Evaluating \be \epsilon_{0r}^1 Q^1 + \epsilon_{0r}^2 Q^2
= \frac{1}{2}\eta_0 (Q^1\zeta_0 +Q^2) \ee we conclude that sixteen
supersymmetries generated by the supercharges $Q^1\zeta_0 +Q^2$
are unbroken while the remaining supersymmetries sixteen generated
by the supercharges $Q^1\zeta_0-Q^2$ are broken and realized
non-linearly.

\subsection{Supersymmetries of static gauge fixed $D3$ brane
action \\ via supercurrents}

In this section we investigate realization of $D3$ brane
supersymmetries by exploring formalism of supercurrent. Concerning
the broken and unbroken supersymmetries we arrive at the same
conclusions of previous section, i.e. a study of this section is
supplementary to the one above given. We evaluate the
supercurrents and divide them into two class. The supercurrents
which do not involve terms linear in fermionic fields are
responsible for unbroken supersymmetries. Remaining supercurrents
involve terms linear in fermionic fields and these supercurrents
are responsible for broken supersymmetries (Goldstone type
supersymmetries). In order to investigate these issues it is
sufficient to restrict an attention to the terms up to the second
order in the fermionic field $\Theta$. The corresponding part of
gauge invariant $D3$ brane Lagrangian is given by (see
\rf{action}) \be {\cal L}^{(2)}= {\cal L}_1 + {\cal L}_2 + {\cal
L}_3 +{\cal L}_4\,, \ee
\begin{eqnarray}
&& {\cal L}_1  =  -\sqrt{g} +{\rm i}\sqrt{g}\Theta
\hat{\gamma}{}^a {\cal D}_a \Theta\,,
\\
&& {\cal L}_2
 =  \sqrt{g}(-\frac{1}{4}F^{ab}F_{ab}
 -{\rm i} F^{ab}\Theta \hat\gamma_a\tau_3 {\cal D}_b\Theta)\,,
\\
&& {\cal L}_3 = \frac{\rm
i}{6}\epsilon^{abcd}\Theta\hat\gamma_{abc}\tau_2
 {\cal D}_d\Theta
+ {\cal L}_{WZ}^{(bose)}\,,
\\
&& {\cal L}_4
 =  \frac{\rm i}{2}\epsilon^{abcd} F_{ab}\Theta \hat\gamma_c\tau_1
 {\cal D}_d\Theta\,,
 \qquad \hat\gamma_{\alpha\beta}^a \equiv
 \gamma_{\alpha\beta}^\mu e^a_\mu\,,
\end{eqnarray}
where the ${\cal L}_{WZ}^{(bose)}$ is given in \rf{LWZB} and  the
covariant Killing spinor derivative ${\cal D}=d\sigma^a {\cal
D}_a$ is given in \rf{comder}. The remaining notation is given in
\rf{gab0},\rf{bosvie}. Supersymmetry transformations given in
\rf{susy1}-\rf{susy3} lead to a conserved current\footnote{The
current \rf{brasupcur} can be found by using standard Noether
method based on localization of the parameters of associated
global transformation $\epsilon_0$ \rf{epsU}. Replacing
$\epsilon_0$ by function of world-volume coordinates $\sigma^a$
the variation of Lagrangian by module of total derivatives is
found to be $\delta {\cal L} = -2{\rm i}(\partial_a \epsilon_0)
Q^a$.} $Q^a$, $\partial_a Q^a =0$, \be\label{brasupcur} Q^a =  V
\Bigl(\sqrt{g}(\hat\gamma^a + F^{ab}\hat\gamma_b \tau_3)
-\frac{1}{6}\epsilon^{abcd}\hat\gamma_{bcd}\tau_2
-\frac{1}{2}\epsilon^{abcd}F_{bc}\hat\gamma_d\tau_1\Bigr)\Theta\,,
\ee where the matrix $V=V_\alpha{}^\beta$ is given by \be V =
\exp(\frac{\f}{2}\tau_2 x^+\bar\gamma^-\Pi\gamma^+)
\exp(\frac{\f}{2} \tau_2 x^I\bar\gamma^I\Pi\gamma^+ )\,. \ee Note
that $ V_\alpha{}^\beta \equiv U^\beta{}_\alpha$ (see  \rf{epsU}).
Now we should impose both the bosonic static gauge and fermionic
covariant kappa symmetry gauge. To this end we represent the
supercurrent $Q^a$ \rf{brasupcur} as follows \be\label{QVQ} Q^a =
\left(\begin{array}{c} Q^{1a}
\\[7pt]
Q^{2a}
\end{array}\right)\,,
\qquad Q^a = V\tilde{Q}^a\,, \ee and evaluate the supercurrents
$\tilde{Q}^a$ taken to be in covariant kappa symmetry gauge
\rf{covg} \be \tilde{Q}^{1a} =
-\frac{1}{6}\epsilon^{abcd}\hat\gamma_{bcd}\lambda
-\frac{1}{2}\epsilon^{abcd}\hat\gamma_b F_{cd}\lambda\,,
\qquad\quad \tilde{Q}^{2a} = \hat\gamma^a\lambda
-F^{ab}\hat\gamma_b\lambda\,. \ee Plugging these expressions into
\rf{QVQ} and by acting with matrix $V$ we find the following
components of supercurrent $Q^a$:
\begin{eqnarray}
Q^{1a} & = &\frac{1}{2}(\gamma^+\gamma^-+\gamma^-\gamma^+\cos\f
x^+)( -\frac{1}{6}\epsilon^{abcd}\hat\gamma_{bcd}
-\frac{1}{2}\epsilon^{abcd}\hat\gamma_b F_{cd})\lambda
\nonumber\\
& + & \frac{1}{2}(\f\gamma^I x^I + \gamma^-\sin\f x^+)\Pi\gamma^+
(\hat\gamma^a - F^{ab}\hat\gamma_b)\lambda\,,
\\
Q^{2a} & = &\frac{1}{2}(\gamma^+\gamma^-+\gamma^-\gamma^+\cos\f
x^+)(\hat\gamma^a -F^{ab}\hat\gamma_b)\lambda
\nonumber\\
& + & \frac{1}{2}(\f\gamma^I x^I + \gamma^-\sin\f x^+)\Pi\gamma^+
(\frac{1}{6}\epsilon^{abcd}\hat\gamma_{bcd} +
\frac{1}{2}\epsilon^{abcd}\hat\gamma_b F_{cd})\lambda\,.
\end{eqnarray}
These supercurrents $Q^{1a}$ and $Q^{2a}$ taken to be in static
gauge involve unwanted terms which are linear in fermionic field
and do not depend on the remaining fields. These are terms that
generate Goldstone transformation. Motivated by desire to cancel
these unwanted terms we look for the linear combination of the
supercurrents $Q^{1a}$ and $Q^{2a}$ which cancels out
unwanted terms. As expected from analysis of previous Section such
a combination does exist and is given by $Q^{1a}\zeta_0+Q^{2a}$.
Making use of field redefinitions \rf{fiered} this desired
combination of supercurrents takes the form\footnote{Here we
introduce $Q^{\mun}=\delta_a^\mun\, Q^{a}$ and collect $A^\mun\,$
and $\phi^M$ into $10d$ vector $A^\muun\,$. Details of $10d$
notation we use may be found in Sect. \ref{PWSYM}.} \be
-(Q^{1\mun}\zeta_0 + Q^{2\mun}) =
\frac{1}{2}\exp(\frac{\f}{2}\sigma^+\gamma^-\gamma^{+34})
\exp(\frac{\f}{2}\sigma^\hi\gamma^\hi\gamma^{+34})
F_{\mu\nu}\gamma^{\mu\nu}\gamma^\mun\,\lambda
 + \f \phi^M \gamma^M \gamma^{+34}\gamma^\mun\, \lambda\,.
\ee To transform this supercurrent into conventional $\rw=0$
scheme which is friendly to $\cN=4$ SYM theory to be studied below
we use the transformation \rf{fiered1} with $\rw=0$ (see also
\rf{psilam}) and we get finally \be -(Q^{1\mun}\,\zeta_0 +
Q^{2\mun}) = \exp(-\frac{\f}{2}\sigma^+\gamma^{+34}\gamma^-)
\exp(-\frac{\f}{2}\sigma^\hi\gamma^{+34}\gamma^\hi)
(\frac{1}{2}F_{\mu\nu}\gamma^{\mu\nu}-\f \phi^M \gamma^{+34}
\gamma^M) \gamma^\mun \lambda\,. \ee

\newsection{$D3$ brane action in the
kappa symmetry light cone gauge}\label{LCD3}

Fixing fermionic kappa symmetry by using light cone gauge
simplifies considerably the structure of $D3$ brane  action. In
order to discuss light cone gauge fixed action we find it
convenient to represent covariant action it terms of complex
spinors. Complex frame fermionic coordinates $\theta^\alpha$,
$\bar{\theta}^\alpha = (\theta^{\alpha})^\dagger$ and  Cartan
1-forms $L^\alpha$, $\bar{L}^\alpha = (L^\alpha)^\dagger$ are
related with 2-vector notation $\Theta$ and ${\bf L}$ used in
previous sections \be\label{comrea0} \theta
=\frac{1}{\sqrt{2}}(\theta^1+{\rm i}\theta^2)\,,\qquad
L=\frac{1}{\sqrt{2}}(L^1 +{\rm i}L^2)\,, \ee where the components
$\theta^1$, $\theta^2$ and $L^1$, $L^2$ are related with $\Theta$
and ${\bf L}$ as in \rf{vec2com}. In this notation the 2-form
field $\FF$ takes the form\footnote{Complex notation counterpart
of relation \rf{iim} is $d{\FF} ={\rm i}L \wedge \L \wedge L
+h.c.$} \be \FF= \FF_{t=1}\,, \qquad
 {\rm F}_t=dA+(2{\rm i}\int_0^t dt' \, \theta\widehat{L}_{t'}
 \wedge L_{t'} + h.c.)\,,
\label{iimcom} \ee while the defining $H_5$ form can be cast into
the form
$$
{H}_5
= \frac{1}{3}\bar{L}\wedge \L\wedge\L\wedge\L\wedge L
+ {\FF}\wedge (L \wedge \L
\wedge L
- \bar{L} \wedge \L \wedge \bar{L})
$$
\begin{equation}\label{h5com}
+\ \frac{\f}{6}\Bigl(\epsilon^{i_1\ldots i_4}L^+ \wedge
L^{i_1}\wedge \ldots \wedge L^{i_4} + \epsilon^{i_1'\ldots i_4'}
L^+\wedge L^{i'_1}\wedge \ldots\wedge  L^{i'_4}\Bigr)
\,,\end{equation} which leads to the following WZ part of the
Lagrangian ${\cal L}_{WZ}=d^{-1}H_5$ \be\label{lwzcom} {\cal
L}_{\rm WZ} =    \int_0^1 dt
 \,
\Bigl(\frac{1}{3} \bar{\theta}\widehat{L}_t\wedge \widehat{L}_t
\wedge \widehat{L}_t \wedge   L_t
+ 2{\rm F}_t\wedge \theta\widehat{L}_t\wedge  L_t +h.c.\Bigr)
+{\cal L}_{\rm WZ}^{(bose)} \ .
\ee

Fermionic kappa symmetry light cone gauge is defined as \be
\bar{\gamma}^+ \theta = \bar{\gamma}^+ \bar{\theta} = 0\ . \ee In
this gauge the Cartan 1-forms can easily be evaluated (by
transforming result of Appendix C into complex form or by using
Ref. \cite{rrm0112}) and they are given by \be L^+ = e^+\,, \quad
L^I = e^I\,, \quad L^- = e^-  - {\rm
i}(\bar{\theta}\bar{\gamma}^-d\theta+ \theta\bar{\gamma}^-
d\bar{\theta}) -2\f e^+ \bar{\theta}\bar{\gamma}^- \Pi \theta\,,
\ee \be L = d\theta - {\rm i}\f e^+ \Pi \theta\,, \ee where we use
coordinates of plane wave background in which vielbeins $e^\mu$
take the form given in \rf{bosvie}. By using these kappa symmetry
gauge fixed Cartan 1-form we find immediately from \rf{Gab}
\rf{iimcom} the corresponding $G_{ab}$ and $\FF_{ab}$ terms
\begin{eqnarray}
&& G_{ab} = g_{ab} - 2{\rm i}\partial_{(a}
x^+(\bar{\theta}\bar{\gamma}^-\partial_{b)}\theta +
\theta\bar{\gamma}^-\partial_{b)}\bar{\theta}) -4 \f
\bar{\theta}\bar{\gamma}^- \Pi\theta \partial_a x^+\partial_b
x^+\,,
\\
&&{\rm F}_{ab}= F_{ab} +  2{\rm i} \partial_{[a} x^+ (\theta\bar
{\gamma}^-\partial_{b]} \theta +
\bar{\theta}\bar{\gamma}\partial_{b]}\bar{\theta})\,, \qquad
F_{ab} = \partial_a A_b -\partial_b A_a\,, \end{eqnarray} where
the bosonic body $g_{ab}$ is given by \rf{gab1}. As compared to
covariant gauge the light cone gauge fixed Cartan 1-forms, the
metric $G_{ab}$, and 2-form $\FF$ do not involve terms higher than
second order in the fermionic coordinates $\theta$ and due to this
we are able to write down explicit representations for both the
kinetic and WZ parts of gauge fixed $D3$ brane Lagrangian.
Plugging these expressions into the Lagrangian (and making a
redefinition $ x^\mu \rightarrow - x^\mu $) we get (after simple
integration over $t$ in ${\cal L}_{WZ})$ the corresponding BI and
WZ parts
\begin{eqnarray}
&& {\cal L}_{\rm BI} = -\sqrt{- \hbox{ det} (g_{ab} +
F_{ab}+M_{ab})}\,,
\\
&&
{\cal L}_{\rm WZ}   = - \epsilon^{abcd}\partial_a x^+\Bigl(
\partial_b x^I \partial_c x^J\bar{\theta} \gamma^{-IJ}\partial_d \theta+
\frac{1}{2}F_{bc}(\theta \bar{\gamma}^-\partial_d\theta
-\bar{\theta}\bar{\gamma}^- \partial_d\bar{\theta})\Bigr)
\nonumber\\
&&\hspace{1cm} +\, \frac{\f }{6}\epsilon^{abcd}\partial_a x^+
(\epsilon^{ijkl} x^i\partial_b x^j \partial_c x^k \partial_d x^l +
\epsilon^{\ipr\jpr\kpr\lpr} x^\ipr \partial_b x^\jpr
\partial_c x^\kpr \partial_d  x^\lpr)\,,
\end{eqnarray} where the matrix $M_{ab}$ describes
the $\theta$-dependent part of the BI Lagrangian \be M_{ab} \equiv
2{\rm i}\partial_b x^+(\bar{\theta}\bar{\gamma}^-\partial_a\theta
+ \theta\bar{\gamma}^-\partial_a\bar{\theta}) -4 \f
\bar{\theta}\bar{\gamma}^- \Pi\theta \partial_a x^+\partial_b
x^+\,.\ee First terms in expansion of ${\cal L}$ are (modulo
cosmological term) \be \frac{1}{\sqrt{g}}{\cal L}_{\rm BI}
=-\frac{1}{4}F_{ab}F^{ab} -{\rm i}(g^{ab}-F^{ab})\partial_a  x^+
(\bar{\theta}\bar{\gamma}^-\partial_b \theta +
\theta\bar{\gamma}^-\partial_b\bar{\theta})+2\f
g^{ab}\partial_ax^+\partial_bx^+
\bar{\theta}\bar{\gamma}^-\Pi\theta\,, \ee where $g$ is defined as
in \rf{gab0}.

Light cone gauge fixing the world-volume diffeomorphism symmetries
can be made by following standard procedure (see for instance
\cite{Manvelian:1997py}). This subject is beyond scope of this
paper.

\newsection{$psu(2,2|4)$ superalgebra
in various bases}\label{PSUVARBAS}

Now we turn to a study of plane wave SYM theory. Since our
approach is based essentially on realization of supersymmetries
generated by $psu(2,2|4)$ superalgebra we start with discussion of
(anti)commutation relations of this superalgebra in various bases.

$psu(2,2|4)$ superalgebra is more familiar in $so(3,1)\oplus
su(4)\simeq sl(2,C)\oplus su(4)$ or $so(4,1)\oplus so(5)$ basis.
The former basis is convenient in study of  $\cN=4$, $4d$
superconformal symmetry of SYM theory while the latter basis
introduced in \cite{rrm9805} is preferable for discussion of the
covariant GS action in $AdS_5\times S^5$ Ramond-Ramond background
\cite{rrm9805}\footnote{Study of covariant $AdS_5\times S^5$
Green-Schwarz superstring action in $su(2,2)\oplus su(4)$ and
$sl(2,C)\times su(4)$ bases may be found in \cite{roiban} and
\cite{rrm0012} respectively.}. Note that all these bases respect
Lorentz symmetries generated by $so(3,1)$ algebra. The Lorentz
symmetries however are broken in plane wave background and
therefore these bases are not convenient for discussion of  the
$4d$ plane wave $\cN=4$ SYM. It turns out that more suitable and
convenient basis for study plane wave SYM is the one in which a
proper plane wave generator of time translation $P^-$ and
dilatation generator $D$ are realized as diagonal elements of
$psu(2,2|4)$ superalgebra. This basis will be referred to as plane
wave basis.

\subsection{$psu(2,2|4)$ superalgebra in Lorentz basis}

Since we use extensively interrelations between Lorentz and plane
wave bases let us start with discussion of $psu(2,2|4)$
superalgebra in Lorentz basis. As is well known bosonic part of
$psu(2,2|4)$ superalgebra is the algebra of conformal
transformations $so(4,2)$ plus the algebra of $R$-symmetries $so(6)$.
The $so(4,2)$ algebra being realized in flat space-time consists
of translation generators $\bP^m$, dilatation generator $\bD$,
generators of $so(3,1)$ rotations $\bJ^{mn}$, generators of so(6)
rotations $\bJ^{MN}$ and conformal boosts generators $\bK^m$. The
fermionic part of $psu(2,2|4)$ superalgebra consists of sixteen
generators of Poincar\'e super-translations $\bQ_\alpha$ and
sixteen generators of superconformal translations
$\bS^\alpha$\footnote{Here and below we use bold face notation for
generators of $psu(2,24)$ superalgebra taken in Lorentz notation.
This notation should not be confused with the one for 2-vectors we
used in previous Sections.}. Commutators between bosonic
generators of superalgebra are given by\footnote{$m,n,m',n'=0,1,2,9$;
$M,N,M',N'=3,\ldots,8$. In light cone frame $m,n,m',n'=+,-,1,2$ and the only
non-vanishing components of flat metric tensor $\eta^{mn}$
are given by $\eta^{+-}=1$, $\eta^{\hi\hj}=\delta^{\hi\hj}$.}
\be\label{comrel1}
[\bD,\bP^m]=-\bP^m\,,\qquad [\bD,\bK^m]=\bK^m\,, \qquad
[\bP^m,\bK^n]=\eta^{mn}\bD - \bJ^{mn}\,, \ee \be\label{comrel2}
[\bP^m,\bJ^{nn'}]=\eta^{mn}\bP^{n'} - \eta^{mn'}\bP^n\,, \qquad
[\bK^m,\bJ^{nn'}]=\eta^{mn}\bK^{n'} - \eta^{mn'}\bK^n\,, \ee
\be\label{comrel3} [\bJ^{mn},\bJ^{m'n'}]=\delta^{nm'}\bJ^{mn'}+
3\hbox{ terms }\,, \qquad [\bJ^{MN},\bJ^{M'N'}]=\delta^{NM'}\bJ^{MN'}+
3\hbox{ terms }\,. \ee Commutators between bosonic generators and
the fermionic ones are \be\label{comrel4} [\bS,\bP^m]=-\gamma^m
\bQ\,,\qquad [\bQ,\bK^m]= \frac{1}{2}\bar{\gamma}^m \bS\,, \ee
\be\label{comrel5} [\bD,\bQ]=-\frac{1}{2}\bQ\,, \qquad
[\bQ,\bJ^{mn}]=\frac{1}{2}\bar{\gamma}^{mn}\bQ\,, \qquad
[\bQ,\bJ^{MN}]=\frac{1}{2}\bar{\gamma}^{MN} \bQ\,, \ee
\be\label{comrel6} [\bD,\bS]=\frac{1}{2}\bS\,, \qquad [\bS,
\bJ^{mn}]=\frac{1}{2}\gamma^{mn}\bS\,, \qquad [\bS,
\bJ^{MN}]=\frac{1}{2}\gamma^{MN}\bS\,. \ee Anticommutators are
fixed to be \be\label{comrel7} \{\bQ,\bQ\}=  - 2{\rm
i}\bar{\gamma}^m \bP^m\,, \qquad \{\bS,\bS\}=  - 4{\rm i}\gamma^m
\bK^m\,, \ee \be\label{comrel8} \{\bS^\alpha,\bQ_\beta\}= - 2{\rm
i}\delta^\alpha_\beta \bD -{\rm i}(\gamma^{mn})^\alpha{}_\beta
\bJ^{mn} + {\rm i}(\gamma^{MN})^\alpha{}_\beta \bJ^{MN}\,. \ee All
bosonic generators $G=(\bP^m,\bK^m,\bD,\bJ^{mn},\bJ^{MN})$ are
taken to be anti-hermitian $G^\dagger = - G$, while fermionic
generators are considered to be hermitian \be\label{qsherrul}
\bQ_\alpha^\dagger = \bQ_\alpha\,, \qquad \bS^{\alpha\dagger}
=\bS^\alpha\,. \ee In flat space the symmetries generated by
translations $\bP^m$, and Lorentz rotations $\bJ^{mn}$ are
realized as isometry symmetries of Minkowski space time while
$\bD$ and $\bK^m$ are responsible for proper conformal
transformations.

\subsection{ $psu(2,2|4)$ superalgebra in plane wave basis}

First of all let us following \cite{rrm9701} to discuss a manifest
representation for generators of conformal symmetries of $4d$
plane wave geometry.  This representation will be used throughout
this paper. Since for plane wave background the Weyl tensor
vanishes, the conformal algebra in question is isomorphic to the
$so(4,2)$ algebra. Below we establish the isomorphism explicitly,
i.e., we map generators of $so(4,2)$ taken to be in the plane wave
basis to generators of $so(4,2)$ taken to be in the Lorentz basis.
Note that because Lorentz symmetries are broken in plane wave
background the symmetries of $so(4,2)$ algebra are realized in a
different way in plane wave space-time as compared to ones in the
Minkowski space-time. In plane wave background only seven
generators of the $so(4,2)$ algebra are realized as isometry
symmetries while the remaining eight generators of the $so(4,2)$
algebra are realized as proper conformal symmetries. This is to be
compared with Minkowski space where the 10 generators of $so(4,2)$
algebra, four translations $\bP^m$ and six Lorentz boosts
$\bJ^{mn}$, are realized as isometry symmetries while the
remaining 5 generators of the $so(4,2)$ algebra, dilatation $\bD$
and conformal boosts $\bK^m$, are realized as proper conformal
transformations.

Our study is based on usage of concrete parametrization of $4d$
plane wave background in which the line element takes the form
\be\label{4dpw} ds^2 = 2dx^+ dx^- -\f^2 x^\hi x^\hi  dx^+dx^+ +
dx^\hi dx^\hi\,, \qquad \hi=1,2. \ee We start with discussion of
the algebra of isometry symmetries which is subalgebra of
conformal algebra. The algebra of isometry symmetries leaves the
line element \rf{4dpw} to be form-invariant and is generated by
translations $P^+$, $P^-$, $P^\hi$, by Lorentz boosts $J^{+\hi}$
and by $so(2)$ rotations $J^{\hi\hj} =\epsilon^{\hi\hj}J^{12}$,
i.e. the dimension of this algebra is equal to seven. These
generators satisfy the well known commutation relations
\be\label{4dpwalg1} [P^-,P^\hi] = -\f^2J^{+\hi}\,,\qquad
[P^\hi,J^{+\hj}] = -\delta^{\hi\hj}P^+\,, \qquad
[P^-,J^{+\hi}]=P^\hi\,,\ee \be\label{4dpwalg2} [P^\hi,J^{\hj\hk}]
= \delta^{\hi\hj}P^\hk -\delta^{\hi\hk}P^\hj\,, \qquad
[J^{+\hi},J^{\hj\hk}] = \delta^{\hi\hj}J^{+\hk}
-\delta^{\hi\hk}J^{+\hj}\,. \ee Sometimes in what follows instead
of generators $P^\hi$, $J^{+\hi}$ we prefer to use complex frame
generators $T^\hi$, $\bar{T}^\hi$ defined by \be\label{ticomnot}
T^\hi = P^\hi - {\rm i}\f J^{+\hi}\,,\qquad \bar{T}^\hi = P^\hi +
{\rm i}\f J^{+\hi}\,, \ee which are related by hermitian
conjugation rule $\bar{T}^\hi =-T^{\hi\dagger}$. Thus the isometry
symmetry algebra is representable by the following seven
generators \be P^-,\quad P^+,\quad T^\hi,\quad \bar{T}^\hi,\quad
J^{12}\,,\qquad \hbox{isometry generators}\,. \ee To find concrete
representation for these generators we solve the equations for the
Killing vectors of isometry transformations \be\label{isokil}
D_\mun \xi_\nun + D_\nun\xi_\mun =0\,. \ee Well known
representation for these Killing vectors in terms of differential
operators $G=\xi^\mun\partial_\mun$ is given by \be\label{isokil1}
P^\pm=\partial^\pm\,,\qquad T^\hi = e^{-{\rm i}\f x^+}
(\partial^\hi + {\rm i}\f x^\hi\partial^+)\,, \qquad J^{\hi\hj} =
x^\hi\partial^\hj - x^\hj\partial^\hi\,. \ee Because these seven
generators form the isometry algebra the remaining eight
generators of the conformal algebra $so(4,2)$ are responsible for
proper conformal transformations, i.e. they scale the line element
\rf{4dpw}. To find these proper conformal generators we are
solving the equations for conformal Killing vectors in $4d$ plane
wave background \rf{4dpw} \be\label{conkil} D_\mun \xi_\nun +
D_\nun\xi_\mun =\frac{1}{2}g_{\mun\,\nun} D_\kun \xi^\kun\,. \ee
Solution to these equations can be described by eight Killing
vectors denoted by \be\label{prconkil} D,\quad C,\quad
\bar{C},\quad C^\hi,\quad \bar{C}^\hi,\quad K^-,\qquad
\hbox{proper conformal generators}\,, \ee where the complex-valued
generators $C^\hi$, $\bar{C}^\hi$, $C$, $\bar{C}$ are related by
hermitian conjugation rule $\bar{C}^\hi =-C^{\hi \dagger}$,
$\bar{C} =- C^\dagger$. Representation of these proper conformal
generators in terms of differential operators $G=
\xi^\mun\partial_\mun$ is fixed to be \cite{rrm9701}
\begin{eqnarray}
\label{dkil}
&&D =2x^-\partial^+ + x\partial\,,
\\
&&
\label{ckil}
C =e^{-2{\rm i}\f x^+}(\partial^-
- {\rm i}\f x\partial +\f^2 x^2\partial^+)\,,
\\
&&
\label{cikil}
C^\hi =e^{-{\rm i}\f x^+}\Bigl(x^-\partial^\hi - x^\hi\partial^-
+{\rm i}\f(
-\frac{1}{2}x^2\partial^\hi +x^\hi( x^-\partial^+ + x\partial))
-\frac{\f^2}{2}x^2 x^\hi \partial^+ \Bigr)\,,\qquad
\\
&&
\label{kmkil}
K^- = -\frac{1}{2}x^2\partial^- +x^-( x^-\partial^+ + x\partial)
-\frac{\f^2}{4}(x^2)^2\partial^+\,.
\end{eqnarray}
Here and below we use the conventions \be\label{squcon} x^2 \equiv
x^\hi x^\hi\,, \qquad x\partial \equiv x^\hi
\partial^\hi\,,\qquad
\partial^2 \equiv \partial^\hi\partial^\hi\,.
\ee Note that generators $P^+$, $J^{\hi\hj}$ \rf{isokil1}, $D$
\rf{dkil} and $K^-$ \rf{kmkil} do not depend on evolution
parameter $x^+$, i.e. they are commuting with plane wave time
translation generator $P^-$ \rf{isokil}. It is easy to see also
that generators $T^\hi$, $C$ \rf{ckil} and $C^\hi$ \rf{cikil} are
eigenvectors of $P^- =\partial^-$. Straightforward inspection of
generators \rf{isokil},\rf{dkil}-\rf{kmkil} demonstrates that all
these generators are also eigenvectors of $D$ and transform in
scalar or vector representations of $J^{12}$.

As mentioned before the isometry generators \rf{isokil} and proper
conformal generators \rf{prconkil} form plane wave basis of the
$so(4,2)$ algebra. Their commutation relations are given in
\rf{pdcha}-\rf{pwcomrel5}. It is instructive to relate generators
\rf{isokil}-\rf{prconkil} with generators of $so(4,2)$ algebra
taken in Lorentz basis $\bP^m$, $\bK^m$, $\bD$, $\bJ^{mn}$ which
satisfy well known recognizable commutation relations
\rf{comrel1}-\rf{comrel3}. This can be done in a rather
straightforward way by comparing commutation relations of
generators in Lorentz basis \rf{comrel1}-\rf{comrel3} with the
commutation relations in plane wave basis
\rf{pdcha}-\rf{pwcomrel5}. Doing this we find an interrelation
between generators of the conformal algebra in Lorentz basis and
the ones in plane wave basis \be\label{pwlor1} P^- = \bP^- + \f^2
\bK^+\,, \qquad P^+ = \bP^+\,, \qquad J^{12} = \bJ^{12}\,, \ee
\be\label{pwlor2} T^\hi = \bP^\hi -{\rm i}\f \bJ^{+\hi}\,, \qquad
\bar{T}^\hi = \bP^\hi  + {\rm i}\f \bJ^{+\hi}\,, \ee
\be\label{pwlor3} D = \bD -\bJ^{+-}\,, \ee \be C = \bP^- -\f^2
\bK^+ - {\rm i}\f (\bD + \bJ^{+-})\,, \qquad \bar{C} = \bP^- -
\f^2 \bK^+ + {\rm i}\f (\bD + \bJ^{+-})\,, \ee \be C^\hi       =
\bJ^{-\hi}  + {\rm i}\f \bK^\hi\,, \qquad \bar{C}^\hi = \bJ^{-\hi}
- {\rm i}\f \bK^\hi\,, \ee \be K^- = \bK^-\,. \ee Note that the
relations given in \rf{pwlor2} and definition of $T^\hi$
\rf{ticomnot} imply matching the transverse translations $P^\hi$
and $+\hi$ Lorentz boosts $J^{+\hi}$: \be\label{pwlor7} P^\hi =
\bP^\hi\,,\qquad J^{+\hi} = \bJ^{+\hi}.\ee As for $R$-symmetries
$so(6)$, these symmetries are obviously unbroken and therefore one
has the simple correspondence \be\label{pwlorso} J^{MN} =
\bJ^{MN}\,. \ee Now let us turn to supercharges. What is required
is to find the interrelation of supercharges in plane wave and
Lorentz bases. As in bosonic case the supercharges in plane wave
basis should, by definition, be eigenvectors of plane wave time
translation generator $P^-$ and plane wave dilatation generator
$D$. Because we have already established interrelation of two
bases in bosonic body of superalgebra the most straightforward way
to establish of interrelation of supercharges in two bases is to
find linear combinations of Lorentz basis supercharges $\bQ$,
$\bS$ so that this combinations be eigenvectors of plane wave time
translation generator $P^-$ and plane wave dilatation generator
$D$. Because the interrelation between bosonic generators is known
we can establish similar interrelation between supercharges in a
rather straightforward way. Doing this we found that plane wave
basis supercharges can be collected into two sorts of supercharges
which we denote by $\Omega$ and $S^-$. This is to say that the
requirement these supercharges be eigenvectors of generators $P^-$
and $D$ fixes the following representation for sixteen component
complex-valued supercharge $\Omega$ and sixteen component
real-valued supercharge $S^-$ \be\label{omqs} \Omega = \bQ
-\frac{{\rm i}\f}{2}\bar{\gamma}^+\bS^+\,, \qquad S^- = \bS^-\,.
\ee Here and below we use decomposition of $\bS^\alpha$ into plus
and minus parts \be\label{sdecpm} \bS = \bS^+ + \bS^-\,, \qquad
\bS^+ = \frac{1}{2}\gamma^-\bar{\gamma}^+\bS\,, \qquad \bS^- =
\frac{1}{2}\gamma^+\bar{\gamma}^-\bS\,, \ee while for $\bQ_\alpha$
similar decomposition takes the form \be\label{qdecpm} \bQ = \bQ^+
+ \bQ^-\,, \qquad \bQ^+ = \frac{1}{2}\bar{\gamma}^-\gamma^+\bQ\,,
\qquad \bQ^- = \frac{1}{2}\bar{\gamma}^+\gamma^-\bQ\,. \ee The
supercharges $\Omega$ and $S^-$ satisfy the algebraic constraints.
Decomposing the supercharge $\Omega$ as in \rf{qdecpm} \be \Omega
= \Omega^+  +  \Omega^-\,, \ee we have the constraints
\be\label{supherrul} \Omega^{+\dagger} = \Omega^+\,, \qquad
\bar{\gamma}^+ S^- =0\,. \ee {}These constraints are derivable
from hermitian properties for generators $\bQ$, $\bS$ given in
\rf{qsherrul} and definitions given in \rf{sdecpm}. The first
constraint in \rf{supherrul} tells us that though the supercharge
$\Omega$ is complex-valued its $\Omega^+$ part is real-valued.
This implies that the supercharge $\Omega$ has 24 real-valued
components. The second constraint in \rf{supherrul} tells us that
there are only 8 real-valued components of $S^-$. Thus the total
number of real-valued supercharges is equal to 32 as it should be.

Now we can write down commutation relations of the $psu(2,2|4)$
superalgebra in plane wave basis. Because in this basis all
generators are eigenvectors of $P^-$ and $D$ it is convenient to
introduce a notion of $q(P^-)$ and $q(D)$ charges. Given generator
$G$ its $q_G(P^-)$ and $q_G(D)$ charges are defined by commutators
\be\label{pdcha} [P^-, G] = -{\rm i}\f q_G(P^-)G\,, \qquad [D,G] =
q_G(D)G\,. \ee These charges for generators of the $psu(2,2|4)$
superalgebra are given in Table 2.

\bigskip
TABLE 2:  {\sf $q(P^-)$ and $q(D)$ charges of generators of
$psu(2,2|4)$ superalgebra}

\begin{center}
\begin{tabular}{|c|c|c|c|c|c|c|c|c|c|c|c|c|}
\hline        && && && && && &&
\\ [-3mm]\ \ \  Generators  && && && && && &&
\\
& $P^+$ & $T^\hi$ & $\bar{T}^\hi$  & $\!K^-$&  $C^\hi$ &
$\bar{C}^\hi$ & $\!C$ & $\bar{C}$ & $\Omega^- $ & $\bar{\Omega}^-$
& $\Omega^+$ &  $\!S^-$
\\
$ q $ - charges \ \ \ \ \ \  &&  && &&   && && &&
\\ \hline
&& && &&    && &&  &&
\\[-3mm]
$q_G(P^-)$ charge&0& $1$ &$-1$&   $0$&1&$-1$&2& $-2$ & 1& $-1$& 0&
0
\\[2mm]\hline
&&&&&&&&&& &&
\\[-3mm]
$q_G(D)$ charge   & $-2$ & $-1$ & $- 1$ & 2 & 1& 1 & 0& 0 & 0& 0 &
$-1$ &  1
\\[2mm]\hline
\end{tabular}
\end{center}
Note that the $q$-charges of the generators $J^{\hi\hj}$ and
$J^{MN}$ are equal to zero. Remaining (anti)commutators can be
collected in several groups.

Commutators between elements of isometry algebra are
\be\label{pwcomrel1} [\bar{T}^\hi,T^\hj]=2{\rm i}\f
\delta^{\hi\hj} P^+\,,\qquad
[T^\hi,J^{\hj\hk}]=\delta^{\hi\hj}T^\hk
-\delta^{\hi\hk}T^\hj\,.\ee

Commutators between proper conformal generators are given by
\be\label{pwcomrel2} [\bar{C}^\hi,C^\hj]=2{\rm
i}\delta^{\hi\hj}K^-\,, \qquad [\bar{C},C]=-4{\rm i}\f P^-\,,
\qquad [\bar{C}^i,C]=-2{\rm i}\f C^i\,. \ee

Commutators between generators of isometry algebra and proper
conformal generators take the form \be\label{pwcomrel3}
[P^+,K^-]=D\,, \qquad [P^+,C^\hi]= T^\hi\,, \qquad
[T^\hi,K^-]=C^\hi\,, \ee \be\label{pwcomrel4}
[T^\hi,C^\hj]=-\delta^{\hi\hj}C\,, \qquad [T^\hi,\bar{C}]= 2{\rm
i}\f \bar{T}^\hi\,, \ee \be\label{pwcomrel5} [\bar{C}^\hi,T^\hj] =
\delta^{\hi\hj}(P^-+{\rm i}\f D)+2{\rm i}\f J^{\hi\hj}\,. \ee
Commutators between isometry algebra and supercharges are \be
[T^\hi,\bar\Omega^-] = {\rm i}\f \bar\gamma^{+\hi}\Omega^+\,,
\qquad [T^\hi,S^-]=\gamma^\hi \Omega^-\,, \qquad [P^+,S^-] =
\gamma^+ \Omega^+\,, \ee
$$
[\Omega^\pm,
J^{\hi\hj}]=\frac{1}{2}\bar\gamma^{\hi\hj}\Omega^\pm\,, \qquad
[S^-, J^{\hi\hj}]=\frac{1}{2}\gamma^{\hi\hj}S^-\,.
$$
Commutators between proper conformal generators and supercharges
are \be [K^-,\Omega^+] = -\frac{1}{2}\bar\gamma^- S^-\,, \qquad
[C^\hi,\Omega^+] = -\frac{1}{2}\bar\gamma^{-i} \Omega^-\,, \qquad
[C^\hi,\bar{\Omega}^-] = -{\rm i}\f \bar\gamma^\hi S^-\,, \ee \be
[C,\bar{\Omega}^-] = 2{\rm i}\f\, \Omega^-\,. \ee

In view of relations \rf{pwlorso} commutators of $so(6)$ algebra
and commutators between generators of the $so(6)$ and supercharges
in plane wave basis take the same form as in the Lorentz basis
(see last commutators in \rf{comrel3}, \rf{comrel5},\rf{comrel6}).

Anticommutators between supercharges are \be
\{\Omega^+,\Omega^+\}=-2{\rm i}\bar{\gamma}^-P^+\,,\quad
\{\Omega^-,\Omega^-\}=-2{\rm i}\bar{\gamma}^+C\,, \quad
\{S^-,S^-\}=-4{\rm i}\gamma^+K^-\,, \ee \be
\{\bar{\Omega}^-,\Omega^-\}= -2{\rm i}\bar\gamma^+ P^-
+\f\bar\gamma^{+\hi\hj}J^{\hi\hj} - \f \bar\gamma^{+MN}J^{MN}\,,
\ee \be \{\Omega^-,\Omega^+\} = -{\rm
i}\bar\gamma^+\gamma^-\bar\gamma^\hi T^\hi\,, \qquad
\{\Omega^-,S^-\} = 2{\rm i}  \bar\gamma^{+\hi} C^\hi\,, \ee \be
\{S^-,\Omega^+\} =-{\rm i}\gamma^+\bar{\gamma}^-D -\frac{\rm
i}{2}\gamma^+\bar{\gamma}^-\gamma^{\hi\hj}J^{\hi\hj} +\frac{\rm
i}{2}\gamma^+\bar{\gamma}^-\gamma^{MN}J^{MN}\,. \ee Modulo
(anti)commutators obtainable from the ones above-given by applying
hermitian conjugation, the remaining (anti)commutators are equal
to zero.

Now let us match the symmetries generated by the $psu(2,2|4)$
superalgebra and the ones generated by superalgebra of plane wave
superstring (see Appendix B). Because these superalgebras do not
coincide we can match only some subset of generators of these
superalgebras. Let $G_{str}$ be generators of plane wave
superstring superalgebra with (anti)commutators given in Appendix
B. Making use of Lorentz basis of the $psu(2,2|4)$ superalgebra we
find the following interrelations of the generators of the
$psu(2,2|4)$ and plane wave superstring superalgebra \be P_{str}^-
+ 2\f J^{34}_{str}= \bP^- + \f^2 \bK^+\,,\qquad P_{str}^\hi
=\bP^\hi\,,\qquad P_{str}^+ = \bP^+\,, \ee \be J_{str}^{+\hi}=
\bJ^{+\hi}\,,\qquad J_{str}^{12}=\bJ^{12}\,, \qquad
J_{str}^{34}=\bJ^{34}\,,\qquad J_{str}^{i'j'}= \bJ^{i'j'}\,, \ee \be
\frac{1}{\sqrt{2}}(Q_{str}^1\gamma^{-+12} + Q_{str}^2) =\bQ
-\frac{\f}{2}\gamma^{+34}\bS^+\,.\ee Thus one can match explicitly
$14$ bosonic and $16$ super symmetries of plane wave SYM and
corresponding 14 bosonic and 16 super- symmetries of plane wave
superstring theory\footnote{Interesting discussion of realization
of conformal and Heisenberg algebras in plane wave CFT
correspondence may be found in \cite{Das:2002ij}.}.

\newsection{\cN=4 SYM in plane wave background}\label{PWSYM}

Our goal is to develop light cone formulation of $\cN=4$ super YM
theory in $4d$ plane wave background. The most convenient way to
do this is to start with covariant and gauge invariant
formulation. As compared to flat space the covariant formulation
of plane wave SYM becomes to be more complicated as in this case
we deal with theory in curved background\footnote{Discussion of
SYM theory in various curved backgrounds may be found in
\cite{Blau:2000xg}. Plane wave SYM does not fall in the cases
considered in \cite{Blau:2000xg}. Detailed study of the
Hamiltonian formulation of SYM in $R\times S^3$ background may be
found in \cite{oku}. Relevance of a curved background
for SYM was discussed (in the original AdS/CFT context) in
\cite{Park:2000du}.}. Light cone gauge formulation leads to
dramatic simplification. A remarkable fact we demonstrate below is
that the 3-point and 4-point interaction vertices of light cone
gauge action of plane wave SYM take the same form as the ones in
the flat space. The only difference of plane wave SYM as compared
to the one in flat space is an appearance of mass-like terms in
the  quadratic part of the action.

Following our strategy we start with covariant and gauge invariant
action of SYM theory defined on  the $4d$ plane wave background
with line element given in \rf{4dpw} To simplify our presentation
of the $4d$ SYM theory we use the standard trick of $10d$
notation. We use $10d$ space-time which is direct product of $4d$
plane wave background \rf{4dpw} and flat $R^6$ Euclidean space. In
this $10d$ space-time we introduce  the $10d$ target space vector
$A_\muun$, $\muun=0,1,\ldots, 9$. This vector is splitted into
$4d$ plane wave target space vector $A_\mun$, $\mun=0,1,2,9$, and
$SO(6)$ vector $A_M=\phi^M$, $M=3,\ldots,8$. The fermionic partner
of $A_\muun$ is a sixteen component spinor field $\psi^\alpha$
which in the $32$ component notation corresponds to positive
chirality Majorana-Weyl spinor. Both the $A_\muun$ and $\psi$ are
transforming in adjoint representation of certain gauge group.
They can be decomposed into generators of Lie algebra and are
assumed to be anti-hermitian \be\label{sunnot} A_\muun =
A_\muun{}^{\sf a} t_{\sf a}\,,\qquad \psi = \psi^{\sf a}{}t_{\sf
a}\,,\qquad A_\muun{}^\dagger = - A_\muun{}\,,\qquad \psi^\dagger
= - \psi\,.\ee The Lagrangian of SYM in plane wave background is
given by\footnote{ Note that because Richi scalar of $4d$ plane
wave metric \rf{4dpw} is equal to zero the action \rf{covlag} is
indeed describes conformal scalar fields. The target space
$\gamma^{\underline{\nu}}$-matrices are defined in terms of
tangent space ones as
$\gamma^{\underline{\nu}}=e^{\underline{\nu}}_\nu \, \gamma^\nu$,
where $e^{\underline{\nu}}_\nu$ is inverse to the basis of one
forms $e^\mu=e_{\underline{\nu}}^\mu\,dx^{\underline{\nu}}\,$,
$e^{\underline{\nu}}_\nu\, e_{\underline{\nu}}^\mu
=\delta_\nu^\mu$. The basis of $e^\mu$ is specified in
\rf{basonefor}. Notation $\gamma^+$, $\gamma^-$, $\gamma^I$ is
used only for tangent space $\gamma$-matrices defined in
\rf{gammaspl0}.} \be\label{covlag} {\cal L} =\rTrpr\,\,
-\frac{1}{4}F^{\muun\,\nuun} F_{\muun\,\nuun} -\frac{\rm
i}{2}\psi\gamma^\muun  D_\muun \psi\,, \ee where the gauge and
Lorentz covariant derivative $D_\muun$  and the field strength
$F_{\muun\,\nuun}$ are defined to be
\begin{eqnarray}
&& D_\muun\psi = \DL{\muun} \psi  + [A_\muun,\psi]\,,\qquad
\DL{\mun}=\partial_\mun
-\frac{\f^2}{2}x^\hi\gamma^{+\hi}\delta_\mun^+\,,\quad
\DL{M}=\partial_M\,,
\\
&& \label{fiestr} F_{\muun\,\nuun} =\partial_\muun A_\nuun
-\partial_\nuun A_\muun + [A_\muun,A_\nuun]\,. \end{eqnarray} In
\rf{covlag} and below the $\rTrpr$ denotes minus trace \be \rTrpr
Y \equiv  - \rTr Y\,. \ee All fields are assumed to be independent
of six coordinates $x^M$ \be\label{redrul}
\partial^M A^\muun =0\,, \qquad \partial^M \psi =0\,.
\ee $10d$ target space indices are contracted by metric tensor
$g_{\muun\,\nuun}$, $A_\muun = g_{\muun\,\nuun}A^\nuun\,$, which
describes $4d$ plane wave space-time \rf{4dpw} $\times$ flat $R^6$
Euclidean space. $10d$ tangent space vectors $A^\mu$, $\mu=0,1,\ldots,9$,
are defined in terms the target space ones as $A^\mu =e^\mu_\muun
A^\muun$ where basis of one-forms $e^\mu=e_\muun^\mu dx^\muun$ is
specified to be \be\label{basonefor} e^+ = dx^+\,, \qquad e^- =
dx^- -\frac{\f^2}{2}x^\hi x^\hi dx^+\,, \qquad e^I =dx^I\,,\qquad
I=1,\ldots,8, \ee so that we have \be g_{\muun\,\nuun}
=\left(\begin{array}{cc} g_{\mun\,\nun} & 0
\\[3pt]
0 & \delta_{MN}
\end{array}\right)\,,
\quad \qquad g_{\muun\,\nuun} = e_\muun^\mu e_{\nuun\,
\mu}^{\vphantom{5pt}}\,, \quad g_{\mun\,\nun} = e_\mun^m e_{\nun\,
m}^{\vphantom{5pt}}\,, \ee $\muun=(\mun,M)$; $\mu=(m,M)$;
$\mun,\nun=0,1,2,9$; $m,n=0,1,2,9$; $M,N=3,\ldots, 8$\footnote{In
light cone frame $\muun,\nuun=+,-,1,\ldots,8$;
$\mu,\nu=+,-,1,\ldots,8$; $\mun,\nun=+,-,1,2$; $m,n=+,-,1,2$.}.

The Lagrangian \rf{covlag} is invariant with respect to the local
gauge transformations and the global $psu(2,2|4)$ superalgebra
transformations.  All these transformations are realized on the
space of covariant fields $A_\muun, \psi$. Below we will be
interested in transformations of physical fields under action of
$psu(2,2|4)$ superalgebra. These transformations can be derived in
a most straightforward way by starting with transformations of the
covariant fields $A_\muun$, $\psi$. To fix our notation let us
briefly discuss the latter transformations.

(i) {\it Local gauge transformations} take the standard well known
form \be \delta A_\muun = \partial_\muun \alpha
+[A_\muun,\alpha]\,, \qquad \delta \psi =[\psi,\alpha]\,. \ee

(ii) {\it Global $so(4,2)$ conformal transformations}. The
transformations of fields under action of the $so(4,2)$ algebra
are given by the relations
\begin{eqnarray}\label{delamu1}
&& \delta^{cov} A^\mun = \xi^\nun\partial_\nun A^\mun -
A^\nun\partial_\nun \xi^\mun
+\frac{1}{2\sqrt{g}}(\partial_\nun\sqrt{g} \xi^\nun) A^\mun\,,
\\
&& \delta^{cov} \phi^M = \xi^\nun\partial_\nun\phi^M
 +\frac{1}{4\sqrt{g}}(\partial_\nun \sqrt{g} \xi^\nun)\phi^M\,,
\\
\label{delamu3} && \delta^{cov} \psi = (\xi^\mun \DL{\mun}
+\frac{1}{4}\gamma^{\mun\,\nun}\partial_\mun\xi_\nun
+\frac{3}{8\sqrt{g}}(\partial_\nun\sqrt{g}\xi^\nun))\psi\,,
\end{eqnarray}
where $\xi^\mun$ are Killing vectors of the $so(4,2)$ algebra
transformations. In plane wave basis these vectors are given by
relations \rf{isokil1}, \rf{dkil}-\rf{kmkil}.

(iii) {\it Global R-symmetries generated by $so(6)$ algebra} take
the standard form.

(iv) {\it Global plane wave supersymmetry transformations of the
$psu(2,2|4)$ superalgebra}. In order to find these transformations
it is convenient to start with transformation of fields under
action of supercharges taken to be in the Lorentz basis, i.e.
basis of $\bQ$ and $\bS$ supercharges. Transformation of fields in
such a basis are well known and are given by
\begin{eqnarray}
\label{lorsup1}&& \delta^{cov} A^\mu ={\rm i}\psi \bar{\gamma}^\mu
\epsilon_\bQ\,, \qquad \quad \delta^{cov} \psi =
\frac{1}{2}F^{\mu\nu}\gamma^{\mu\nu}\epsilon_\bQ
-2\phi^M\gamma^M\epsilon_\bS\,,
\end{eqnarray}
where $\epsilon_\bQ$ and $\epsilon_\bS$ are appropriate
real-valued sixteen component Killing spinors which are still to
fixed. These Killing spinors can be found from the general formula
\be\label{kilspifor} \epsilon_{_\bQ}{}^\alpha \bQ_\alpha +
\epsilon_{_\bS}{}_\alpha \bS^\alpha = g_x (\epsilon_{_{0\bQ}} \bQ
+ \epsilon_{_{0\bS}}\bS)g_x^{-1}\,, \ee where bosonic coset
representative $g_x$ is defined by the relation \be g_x
=\exp(x^\hi P^\hi + x^-P^+)  \exp(x^+P^-) \,, \ee and
$\epsilon_{0\bQ}$ and $\epsilon_{0\bS}$ are constant sixteen
component  fermionic parameters. Note that the Killing spinors
above defined satisfy the equations \be\label{kilspiequ} \DL{\mun}
\epsilon_{_\bQ} = \gamma_\mun \epsilon_{_\bS}\,, \qquad \DL{\mun}
\epsilon_{_\bS} = - \frac{\f^2}{2} \delta_\mun^+
\bar{\gamma}^+\epsilon_{_\bQ }\,, \ee which should be supplemented
by initial conditions \be
\epsilon_{_\bQ}|_{x^\mun\,\,=0}^{\vspace{7pt}} =
\epsilon_{_{0\bQ}}\,, \qquad
\epsilon_{_\bS}|_{x^\mun\,\,=0}^{\vspace{7pt}} =
\epsilon_{_{0\bS}}\,. \ee Explicit representation  for Killing
spinors can be derived then either from these equations or from
formula \rf{kilspifor}\footnote{The formula  \rf{kilspifor} can be
used by exploiting the relations \rf{pwlor1},\rf{pwlor7} and
commutation relations given in \rf{comrel4}-\rf{comrel8}.}
\begin{eqnarray}
\label{kilexpsol1} && \epsilon_{_\bQ} =
\frac{1}{2}(\gamma^+\bar{\gamma}^- + \gamma^-\bar{\gamma}^+\cos\f
x^+)\epsilon_{_{0\bQ}} + (x^-\bar{\gamma}^+ +
x^\hi\bar{\gamma}^\hi)\epsilon_{_\bS} + \bar{\gamma}^-\frac{\sin
\f x^+}{\f}\epsilon_{_{0\bS}}\,,
\\
\label{kilexpsol2}&& \epsilon_{_\bS} =
\frac{1}{2}(\bar{\gamma}^-\gamma^+ + \bar{\gamma}^+\gamma^-\cos\f
x^+ )\epsilon_{_{0\bS}} - \frac{\f}{2}\bar{\gamma}^+ \sin \f x^+
\epsilon_{_{0\bQ}}\,.
\end{eqnarray}
Making use equations for Killing spinors \rf{kilspiequ} one can
make sure that the action \rf{covlag} is indeed invariant with
respect to transformations given in \rf{lorsup1}. Thus there are
32 Killing spinors which are responsible for 32 super(conformal)
symmetries.

By using interrelation between the Lorentz basis supercharges
$\bQ$, $\bS$ and plane wave basis supercharges $\Omega$, $S^-$
\rf{omqs} we can bring Lorentz basis supersymmetry transformations
\rf{lorsup1} to the plane wave basis supersymmetry transformations

\be\label{pwsusytra1} \delta_{_\Omega}^{cov} A^\mu ={\rm i}\psi
\bar{\gamma}^\mu \epsilon_{_\Omega}\,, \qquad
\delta_{_{S^-}}^{cov} A^\mu ={\rm i}\psi \bar{\gamma}^\mu
(x^-\gamma^+ + x^\hi \gamma^\hi)\epsilon_{_{0S^-}}\,, \ee
\begin{eqnarray}
\label{pwsusytra2} && \delta_{_\Omega}^{cov} \psi
=\frac{1}{2}F^{\mu\nu}\gamma^{\mu\nu} \epsilon_{_\Omega} + {\rm
i}\f \phi^M\gamma^M \bar{\gamma}^+\epsilon_{_\Omega}\,,
\\
\label{pwsusytra3}&& \delta_{_{S^-}}^{cov} \psi
=\frac{1}{2}F^{\mu\nu}\gamma^{\mu\nu}(x^-\gamma^+ + x^\hi
\gamma^\hi)\epsilon_{_{0S^-}}
-2\phi^M\gamma^M\epsilon_{_{0S^-}}\,,
\end{eqnarray}
where the plane wave basis Killing spinors $\epsilon_{_\Omega}$
corresponding to the supercharge $\Omega$ is given by \be
\label{epsOm} \epsilon_{_\Omega} = \exp(-\frac{{\rm i}\f}{2}x^\hi
\gamma^\hi \bar{\gamma}^+) \exp(-\frac{{\rm i}\f}{2} x^+ \gamma^-
\bar{\gamma}^+)\epsilon_{_{0\Omega}}\,. \ee The constant sixteen
component spinors $(\epsilon_{_{0\Omega}})^\alpha$, which is
complex-valued, and $(\epsilon_{_{0S^-}})_\alpha$, which
real-valued, subject to the constraints implemented by the ones
for the supercharges $\Omega$, $S^-$ \rf{supherrul} \be
(\epsilon_{_{0\Omega}}^-)^\dagger =\epsilon_{_{0\Omega}}^-\,,
\qquad \gamma^-\epsilon_{_{0S^-}}=0\,. \ee

\newsection{Hamiltonian light cone gauge dynamics of plane wave
SYM}\label{HAMFOR}

Following the standard procedure we impose light cone gauge

\be\label{lcgau} A_- =0\,, \ee and plug this gauge into the
covariant Lagrangian \rf{covlag}. Because such gauge fixed
Lagrangian describes non-propagating field $A_+$ we find that the
equation of motions for $A_+$ leads to the constraint

\be\label{solam}
\partial^+ A^- =
-\partial^I A^I - \frac{1}{\partial^+}[A^I,\partial^+A^I]
+\frac{\rm i}{\partial^+}(\psi^\oplus\bar{\gamma}^+\psi^\oplus)\,,
\ee which allows us to express non-physical field\footnote{Note
that after imposing light cone gauge \rf{lcgau} the relations
between covariant and contr-variant vectors are simplified $A_- =
A^+=0$, $A_+ = g_{+\nun} A^\nun =A^- $.} $A_+= A^-$ in terms of
physical bosonic field $A^I=(A^\hi,A^M=\phi^M)$ and physical
fermionic field $\psi^\oplus$, where

\be \psi= \psi^\oplus+\psi^\ominus\,,\qquad \psi^\oplus
=\frac{1}{2}\gamma^-\bar{\gamma}^+\psi\,, \qquad \psi^\ominus
=\frac{1}{2}\gamma^+\bar{\gamma}^-\psi\,. \ee Fermionic field
$\psi^\ominus$ also turns out to be non-propagating and therefore
we find that equation of motion for $\psi^\ominus$ leads to the
constraint which allows us to express the $\psi^\ominus$ in terms
of physical field $\psi^\oplus$

\be\label{solpsim} \psi^\ominus =
-\frac{1}{2\partial^+}\gamma^+\bar{\gamma}^I D^I \psi^\oplus\,,
\qquad D^I\psi^\oplus \equiv \partial^I\psi^\oplus
+[A^I,\psi^\oplus]\,. \ee Plugging solution for $A^-$ and
$\psi^\ominus$ into Lagrangian \rf{covlag} we get light cone gauge
Lagrangian which can be presented as follows

\be {\cal L}= {\cal L}_2+{\cal L}_3+{\cal L}_4\,, \ee where ${\cal
L}_2$ describes quadratic part of Lagrangian while ${\cal L}_3$
and ${\cal L}_4$ describe 3-point and 4-point interaction vertices
respectively. The quadratic part ${\cal L}_2$ is given by

\be\label{kinpw} {\cal L}_2 =\rTrpr\,\, \frac{1}{2}A^I\Box A^I
-\frac{\rm i}{4}\psi^\oplus \frac{\bar{\gamma}^+}{\partial^+}\Box
\psi^\oplus\,, \ee where the covariant D'Alembertian operator in
$4d$ plane wave background \rf{4dpw} defined by standard relation
$\Box= \frac{1}{\sqrt{g}}\partial_\mun \sqrt{g} g^{\mun\,\nun}\,
\partial_\nun$ is given by

\be\label{box} \Box = 2\partial^+\partial^- +
\partial^\hi\partial^\hi + \f^2 x^\hi x^\hi\partial^+\partial^+\,. \ee The
3-point and 4-point interaction vertices are given by

\begin{eqnarray}\label{lag3}
{\cal L}_3 & = &\rTrpr\,\,
 -[A^I,A^J]\partial^I  A^J -
\partial^J A^J\frac{1}{\partial^+} [A^I,\partial^+A^I]
\\
& + &{\rm i} \partial^IA^I \frac{1}{\partial^+}
(\psi^\oplus\bar{\gamma}^+\psi^\oplus ) +\frac{\rm
i}{4}[A^I,\psi^\oplus]
\frac{\bar{\gamma}^+\gamma^I\bar{\gamma}^J}{\partial^+}
\partial^J\psi^\oplus
+ \frac{\rm i}{4}\partial^J\psi^\oplus
\frac{\bar{\gamma}^+\gamma^J\bar{\gamma}^I}{\partial^+}
[A^I,\psi^\oplus]\,, \nonumber
\\
&&\nonumber
\\
\label{lag4} {\cal L}_4 & =  &\rTrpr\,\, -\frac{1}{4}[A^I,A^J]^2
-\frac{1}{2} (\frac{1}{\partial^+}[A^J,\partial^+ A^J])^2
\nonumber\\
& + & \frac{{\rm
i}}{\partial^+}(\psi^\oplus\bar{\gamma}^+\psi^\oplus)
\frac{1}{\partial^+}[A^I,\partial^+ A^I] +\frac{\rm
i}{4}[A^I,\psi^\oplus]\frac{\bar{\gamma}^+\gamma^I
\bar{\gamma}^J}{\partial^+}
[A^J,\psi^\oplus]
\nonumber\\
& + &
\frac{1}{2}(\frac{1}{\partial^+}(\psi^\oplus\bar{\gamma}^+
\psi^\oplus))^2\,.
\end{eqnarray}
{}From these expressions it is easily seen that dimensionfull
constant $\f$ does not appear in the  interaction vertices ${\cal
L}_3$ and ${\cal L}_4$. In other words, the light cone gauge
vertices of SYM in plane wave background take the same form as the
ones in flat Minkowski space. This implies that many interesting
investigations carried out previously in the literature for the
case of light cone gauge SYM theory in flat space can be extended
in relatively straightforward way to the case of SYM theory in
plane wave background. All that is required is to take into
account the new modified kinetic term ${\cal L}_2$ \rf{kinpw}.
Note that this extension is not straightforward because there are
subtleties related to the mass-like terms proportional in the
covariant D'Alembertian operator (see $\f^2$-term in \rf{box}).
These mass terms lead to discretization of the spectrum of the
energy operator and this should be taken into account upon
quantization.

Light-cone gauge action of plane wave SYM theory

\be\label{actlc}
S = \int dx^+d^3 x ({\cal L}_2 +{\cal L}_3 + {\cal L}_4)\,,\qquad
d^3x \equiv dx^-dx^1dx^2\,,
\ee
can be brought into the Hamiltonian form

\be S = \int dx^+ d^3 x\, \rTrpr(- \partial^+ A^I\partial^- A^I -
\frac{\rm i}{2}\psi^\oplus \bar{\gamma}^+ \partial^- \psi^\oplus)
+\int dx^+ P^-\,, \ee with  the Hamiltonian $H= -P^-$

\be
P^- = \int d^3x\, \PP^-\,,
\ee
where the Hamiltonian density $\PP^-$ is given by

\be\label{hamden} \PP^- =  \rTrpr\Bigl(- \frac{1}{2}
\partial_\hi A^J\partial_\hi A^J
-\frac{\f^2}{2}x^2(\partial^+A^I)^2 - \frac{\rm i}{4}\psi^\oplus
\frac{\bar{\gamma}^+}{\partial^+} (\partial^2+\f^2x^2\partial^{+2}
) \psi^\oplus\Bigr) +{\cal L}_3 + {\cal L}_4\,. \ee Applying
standard methods to the action \rf{actlc} we find the well known
canonical Poisson-Dirac brackets

\begin{eqnarray}
&& \label{aacom} [A^I(x)^{\sf a}, A^J(x')^{\sf
a'}]\left|_{P.D.\,\,equal\, x^+}\right.
 = -\frac{1}{2\partial^+}\delta(x^- - x^{-\prime})
\delta^{(2)}(x-x') \delta^{IJ}\,\hbox{\'I}{}^{\,\sf a\, a'}\,,
\\
&&\label{ppcom} \{\psi^{\oplus\alpha}(x)^{\sf a}
,\psi^{\oplus\beta}(x')^{\sf a'}\} \left|_{P.D.\,\,equal\,
x^+}\right. = - \frac{\rm i}{2}(\gamma^-)^{\alpha\beta} \delta(x^-
- x^{-\prime})\delta^{(2)}(x-x')\,\hbox{\'I}{}^{\,\sf a\, a'}\,,\
\ \ \
\end{eqnarray} where $\hbox{\'I}{}^{\,\sf a\, a'}$ is a minus projector
operator which we insert to respect the Lie algebra indices of the
physical fields $A^I$ and $\psi^\oplus$. All that is required this
operator should satisfy the relation

\be  \hbox{\'I}{}^{\,\sf a\, c}\, \rTrpr (t_{\sf c} t_{\sf b}) =
\delta_{\sf b}^{\sf a}\,. \ee Equation of motions for the physical
fields $A^I$ and $\psi^\oplus$ takes then the standard Hamiltonian
form

\be\label{hamequ}
\partial^ - A^I =[A^I, P^-]_{P.D.}\,,
\qquad
\partial^ - \psi^\oplus =[\psi^\oplus, P^-]_{P.D.}\,.
\ee Here and below brackets $[..,..]_{P.D.}$ stand for
Poisson-Dirac brackets evaluated for equal values of evolution
parameter $x^+$. In \rf{hamequ} and below the suffix `equal $x^+$'
is implicit.

\newsection{Global symmetries of $\cN=4$ SYM in
plane wave background}\label{GLOSYM}

In this section we discuss field theoretical realization of global
supersymmetries of $\cN=4$ SYM theory in plane wave background which
are generated by $psu(2,2|4)$ superalgebra. To do that we use the
framework of Noether charges. The Noether charges play an
important role in analysis of the symmetries of dynamical systems.
The choice of the light cone gauge spoils manifest global
symmetries,  and  in order to demonstrate that these global
invariances are still present, one needs to  find the Noether
charges that generate them.

\subsection{Bosonic Noether charges as generators of the
$psu(2,2|4)$ \\ superalgebra}

We start our discussion with bosonic Noether charges. These
charges can be found following the  standard procedure. Let
$T^{\mun\,\nun}$ be symmetric, conserved, and traceless
energy-momentum tensor

\be\label{enecon} D_\mun T^{\mun\,\nun} =0\,,\qquad
T^\mun{}_\mun=0\,,\qquad T^{\mun\,\nun} = T^{\nun\,\mun}\,. \ee
For each Killing vector $\xi^{G\mun}$ satisfying the equations
\rf{isokil} (or \rf{conkil}) we can construct conserved current
${\cal G}^\mun $:

\be\label{concur} {\cal G}^\mun =  T^\mun{}_\nun \xi^{G
\nun}\,,\qquad
\partial_\mun (\sqrt{g} {\cal G}^\mun) =0\,.
\ee Making use of equations for Killing vectors
\rf{isokil},\rf{conkil} and relations for energy - momentum tensor
given in \rf{enecon} one can make sure that the currents
\rf{concur} satisfy the conservation law. Note that we use the
coordinates given by \rf{4dpw} in which $\sqrt{g} =1$ and
therefore the conservation law takes simplified form
$\partial_\mun {\cal G}^\mun =0$. Appropriate bosonic charges $G$
take then the standard form

\be\label{gint} G = \int  d^3x \, T^+{}_\nun\xi^{G \nun}\,, \ee
where the measure $d^3x$ is defined in \rf{actlc}. The energy
momentum tensor for $\cN=4$, $4d$ SYM \rf{covlag} satisfying the
requirements \rf{enecon} is well known and is given by

\be\label{tmn} T^{\muun\,\nuun} = \rTrpr\Bigl(F^{\muun\,\rhoun}\,
F^\nuun{}_\rhoun + \frac{\rm i}{4}\psi(\gamma^\muun\, D^\nuun
+\gamma^\nuun\, D^\muun)\psi\Bigr) + \Delta T^{\muun\,\nuun} +
g^{\muun\,\nuun}\,{\cal L}\,, \ee where `improving' contribution
given by

\be\label{delt} \Delta T^{\mun\,\nun} =
\frac{1}{6}(g^{\mun\,\nun}D_\kun D^\kun -D^\mun D^\nun +
R^{\mun\,\nun})\rTrpr\,\phi^2\,,\qquad \Delta T^{M\,\mun}=
\Delta T^{MN}=0\,,\ee
is to respect traceless condition \rf{enecon}. While writing the
expression for $T^{\muun\,\nuun}$ \rf{tmn} we take into account
that Richi curvature scalar in plane wave background \rf{4dpw} is
equal to zero $R=0$. Note that the only non-zero component of
Richi tensor is $R^{--}=2\f^2$.

To evaluate the charges \rf{gint} we have to determine expressions
for $T^+{}_-$, $T^+{}_I$, $T^+{}_+$. To this end we plug light
cone gauge \rf{lcgau} and solution for non-physical fields
\rf{solam}, \rf{solpsim} into expressions for energy-momentum
tensor \rf{tmn} and find\footnote{In formulas \rf{tpm}-\rf{RmI} we
keep dependence on six coordinates $x^M$. To get expressions
corresponding to $4d$ SYM one needs to apply rules \rf{redrul}.}

\begin{eqnarray}
\label{tpm} T^+{}_-  & = &\PP^+ -\frac{1}{6}\partial^+\partial^+
\rTrpr\, \phi^2\,,
\\
\label{tpi} T^+{}_I & = & \PP^I -\frac{1}{2}\partial^J{\cal
M}^{JI} - \frac{1}{2} \partial^+({\cal M}^{-I} -{\cal R}^{-I})
-\frac{1}{6}\delta_\hi^I\partial^\hi \partial^+ \rTrpr\,\phi^2\,,
\\
\label{tpp} T^+{}_+ & = &  \PP^- +\frac{1}{2}
\partial^I (
{\cal M}^{-I} + {\cal R}^{-I}) +\frac{1}{6}(\Box -
\partial^+\partial^-)\rTrpr\, \phi^2\,,
\end{eqnarray}
where D'Alembertian operator
$\Box$ is defined by \rf{box} and we introduce
momentum densities

\begin{eqnarray}
&&\label{ppp} \PP^+ \equiv \rTrpr\,\, \partial^+ A^J\partial^+ A^J
+ \frac{\rm i}{2}\psi^\oplus\gamma^+
\partial^+\psi^\oplus\,,
\\
&&\label{ppi} \PP^I \equiv \rTrpr\,\, \partial^+ A^J\partial^I A^J
+ \frac{\rm i}{2}\psi^\oplus\gamma^+ \partial^I\psi^\oplus\,,
\\
&& \PP^- = \rTrpr\Bigl( - \frac{1}{2}
\partial^I A^J\partial^I A^J
-\frac{\f^2}{2}x^2(\partial^+A^I)^2 - \frac{\rm i}{4}\psi^\oplus
\frac{\gamma^+}{\partial^+} (\partial_I^2+\f^2x^2\partial^{+2} )
\psi^\oplus\Bigr)
\nonumber\\
\label{pmin}
&&\hspace{1cm} + \,\, {\cal L}_3 + {\cal L}_4\,,
\end{eqnarray}
and spin densities\footnote{The matrix ${\cal M}^{IJ}$ should not
be confused with the one given given in \rf{exprep1},\rf{comM}.}

\begin{eqnarray}
&& {\cal M}^{IJ}\equiv\rTrpr\,\,
\partial^+ A^I A^J - \partial^+ A^J A^I
+ \frac{\rm i}{4}\psi^\oplus \gamma^{+IJ}\psi^\oplus\,,
\\
&&\label{mmi} {\cal M}^{-I} \equiv\rTrpr\,\,
 - A^I\partial^J A^J +  A^J\partial^J A^I
+ \frac{\rm i}{4}\psi^\oplus \gamma^+\gamma^I\gamma^J
\frac{\partial^J}{\partial^+}\psi^\oplus
\nonumber\\
&& \hspace{1cm} - A^I\frac{1}{\partial^+}([A^J,\partial^+ A^J]
-{\rm i}\psi^\oplus \gamma^+\psi^\oplus)\,.
\end{eqnarray}
The expressions for 3-point and 4-point vertices  ${\cal L}_3$,
${\cal L}_3$ which enter definition of Hamiltonian density $\PP^-$
\rf{pmin},\rf{hamden} are given in \rf{lag3},\rf{lag4}. The
density ${\cal R}^{-I}$ which appears in \rf{tpi},\rf{tpp} is
given by

\be\label{RmI} {\cal R}^{-I}\equiv\rTrpr\,\,
A^I\frac{1}{\partial^+}([A^J,\partial^+ A^J] -{\rm i}\psi^\oplus
\gamma^+\psi^\oplus) -\frac{\rm
i}{4}\psi^\oplus\frac{\gamma^+\gamma^I\gamma^J}{
\partial^+}[A^J,\psi^\oplus]\,.
\ee
In contrast to the momentum and spin densities \rf{ppp}-\rf{mmi}
the density ${\cal R}^{-I}$ does not contribute to charges.

Making use of general formula for charges \rf{gint}, explicit
representation for components of energy-momentum tensor
\rf{tpm}-\rf{tpp}, and for Killing vectors \rf{isokil1},
\rf{dkil}-\rf{kmkil} we can get explicit representation for
charges in a rather straightforward way. For instance field
theoretical  representation for kinematical generators of isometry
transformations, i.e. $P^+$, $T^\hi$, and $J^{\hi\hj}$ is given by

\begin{eqnarray}
\label{genpp}
P^+ & = &\int d^3x\, \PP^+\,,
\\
T^\hi & = & \int\! d^3x\, e^{-{\rm i}\f x^+}\,( \PP^\hi+ {\rm i}\f
x^\hi \PP^+)\,,
\\
\label{genjij}J^{\hi\hj}
& = & \int\!  d^3x (x^\hi \PP^\hj - x^\hj \PP^\hi +{\cal M}^{\hi\hj})\,.
\end{eqnarray}
{}From these expressions it is easily seen that expressions for
generators $P^+$, $J^{\hi\hj}$ coincide with the ones in flat
Minkowski space-time. Field theoretical representation for
generators of $R$-symmetries $so(6)$ is given by

\be\label{genso6} J^{MN} = \int\!  d^3x\,\rTrpr\,\,\partial^+
\phi^M \phi^N -
\partial^+ \phi^N \phi^M +\frac{\rm i}{4}\psi^\oplus
\gamma^{+MN}\psi^\oplus\,. \ee This expression for $J^{MN}$
obviously coincides with the one in flat space-time.

Field theoretical representation for generators of conformal
transformations, i.e. $D$, $C$, $C^\hi$ and $K^-$, is given by

\begin{eqnarray}
\label{gend} && D = \int\! d^3x\,\, (2 x^-\PP^+ + x^\hi
\PP^\hi)\,,
\\[8pt]
&&
C =  \int\! d^3x\,   e^{-2{\rm i}\f x^+}(\PP^-
-{\rm i}\f x^\hi \PP^\hi + \f^2 x^2\PP^+)\,,
\\[8pt]
&&
C^\hi =  \int\! d^3x\,
e^{-{\rm i}\f x^+}\Bigl(x^- \PP^\hi - x^\hi\PP^-
+{\rm i}\f(
-\frac{1}{2}x^2\PP^\hi +x^\hi( x^-\PP^+ + x^\hj\PP^\hj))
\\
&&\hspace{3cm}  -\frac{\f^2}{2}x^2 x^\hi \PP^+ +{\cal M}^{-\hi}
+{\rm i}\f {\cal M}^{\hi\hj} x^\hj \Bigr)\,, \nonumber
\\[8pt]
&& \label{genkm} K^- = \int\! d^3x\,\Bigl( -\frac{1}{2}x^2\PP^- +
x^-( x^-\PP^+ + x^\hi\PP^\hi) -\frac{\f^2}{4}(x^2)^2\PP^+ +{\cal
M}^{-\hi} x^\hi - \rTrpr\,\phi^2\Bigr)\,. \qquad
\end{eqnarray}
Taking into account expressions for momentum densities
\rf{ppp},\rf{ppi} it is easy to see that dilatation generator $D$
\rf{gend} is quadratic in fields, i.e. the $D$ is a kinematical
generator. The remaining proper conformal generators $C$, $C^\hi$,
and $K^-$ are realized non-linearly, i.e. they can be considered
as dynamical generators.

With the definition of charges given in \rf{genpp}-\rf{genkm} and
commutation relations for fields given in \rf{aacom},\rf{ppcom}
the transformation rules of the physical fields
$A^I=(A^\hi,A^M=\phi^M)$ and $\psi^\oplus$ under action the
conformal algebra $so(4,2)$ and $so(6)$ algebra are given by

\be\label{fietra} \delta_G A^I = [A^I,G]_{P.D.}\,, \qquad \delta_G
\psi^\oplus = [\psi^\oplus,G]_{P.D.}\,. \ee  Discussion of some
details of these transformations may be found in Appendix D.

\subsection{Noether supercharges as generators of \\ the superalgebra
$psu(2,2|4)$}

In this section we describe field theoretical realization of
supercharges of $\cN=4$ plane wave SYM theory. We start our
discussion with description of supercurrents. As before it is
convenient to start with study of supercurrents taken to be in
Lorentz basis. Because in the Lorentz basis we deal with the
supercharges $\bQ$, $\bS$ we introduce corresponding supercurrents
denoted by $\bQ^\mun$ and $\bS^\mun$. These supercurrents can be
found by using standard procedure and explicit expression for them
can be fixed by using the formula

\be\label{supcur} \epsilon_{_{0\bQ}} \bQ^\mun
+\epsilon_{_{0\bS}}\bS^\mun \,=\, \rTrpr\,\,\frac{1}{2}
\epsilon_{_\bQ} \gamma^{\mu\nu}F^{\mu\nu}\gamma^\mun\, \psi
 + 2\epsilon_{_\bS}\phi^M\gamma^M\gamma^\mun\, \psi\,,
\ee where $\epsilon_{_{0\bQ}}$ and $\epsilon_{_{0\bS}}$ are
constant spinor while $\epsilon_{_\bQ}$ and $\epsilon_{_\bS}$ are
the Killing spinor in $4d$ plane wave background. These Killing
spinors satisfy the defining equations \rf{kilspiequ} and explicit
solution to these equations is given in
\rf{kilexpsol1},\rf{kilexpsol2}. Making use of defining equations
for Killing spinors \rf{kilspiequ} and equations of motion for
fields of SYM one can check that super-currents \rf{supcur} are
indeed conserved

\be
\partial_\mun(\sqrt{g}\, \bQ^\mun) =0\,,\qquad
\partial_\mun(\sqrt{g}\, \bS^\mun) =0\,.
\ee Explicit expressions for supercurrents $\bQ^\mun$ and
$\bS^\mun$ in terms of covariant fields can be obtained by
plugging solution to Killing spinors into \rf{supcur}

\begin{eqnarray}
\label{qsup1}\bQ^\mun  &  = & \rTrpr\,
\frac{1}{4}\Bigl(\gamma^-\gamma^+ +\gamma^+\gamma^-\cos\f x^+
-\f\gamma^{+\hi}x^\hi\sin\f x^+\Bigr)F^{\mu\nu} \gamma^{\mu\nu}
\gamma^\mun\,\psi
\nonumber\\
& - & \f\sin\f x^+ \gamma^+\phi^M\gamma^M \gamma^\mun\,\psi\,,
\\[8pt]
\label{ssup1}\bS^\mun  &  = &\rTrpr\, \frac{1}{4}(\gamma^+\gamma^-
+ \gamma^-\gamma^+\cos\f x^+) (x^-\gamma^+ + x^\hi\gamma^\hi)
F^{\mu\nu} \gamma^{\mu\nu} \gamma^\mun\,\psi
\nonumber\\
& +  &  \gamma^-\frac{\sin\f x^+}{2\f} F^{\mu\nu} \gamma^{\mu\nu}
\gamma^m\psi +(\gamma^+\gamma^-  + \gamma^-\gamma^+\cos\f x^+
)\phi^M\gamma^M \gamma^\mun\,\psi\,.
\end{eqnarray}
The expressions for supercharges $\bQ$ and $\bS$ can be found then
from $\mun=+$ components of the supercurrents
\rf{qsup1},\rf{ssup1}

\be \bQ =\int d^3x\, \bQ^\mun\,|_{\mun=+}\,,\qquad \bS =\int
d^3x\, \bS^\mun\,|_{\mun=+}\,.\ee Taking into account the
interrelation of Lorentz basis supercharges and the ones of plane
wave basis \rf{omqs} we find plane wave Noether supercharge

\be \Omega =  \int d^3x\, \rTrpr\, \frac{1}{2}\exp(- \frac{{\rm
i}\f}{2} x^+ \gamma^+\gamma^-) \exp(- \frac{{\rm i}\f}{2}x^\hi
\gamma^+\gamma^\hi) F^{\mu\nu} \gamma^{\mu\nu} \gamma^+\psi\,. \ee
Dividing the supercharge $\Omega$ into $\Omega^+$ and $\Omega^-$
parts and working out an expression for the remaining supercharge
$S^-$ gives the following representation for the plane wave basis
Noether supercharges in terms of the covariant fields

\begin{eqnarray}
&& \Omega^+ =\int d^3x\,\rTrpr\, \frac{1}{4}\gamma^-\gamma^+
F_{\mu\nu}\gamma^{\mu\nu}\gamma^+\psi\,,
\\
&& \Omega^- = \int d^3x\,\rTrpr\, \frac{1}{4}e^{-{\rm i}\f x^+}
(\gamma^+\gamma^- -{\rm i}\f \gamma^{+\hi} x^\hi)
F_{\mu\nu}\gamma^{\mu\nu}\gamma^+\psi\,,
\\
&& S^- = \int d^3x\,\rTrpr\, \frac{1}{4}\gamma^+\gamma^-
(x^-\gamma^+  + x^\hi \gamma^\hi)
F_{\mu\nu}\gamma^{\mu\nu}\gamma^+\psi
+2\phi^M\gamma^M\gamma^+\psi\,.
\end{eqnarray}
In light cone frame these expressions take the form

\begin{eqnarray}
&& \Omega^+ = \int d^3x\,\rTrpr\, (-2 F^{+I}
\gamma^I\psi^\oplus)\,,
\\
&& \Omega^- =\int d^3x\,\rTrpr\,
 e^{-{\rm i}\f x^+}\Bigl(-F^{+-}
+\frac{1}{2}\gamma^{IJ}F^{IJ} +{\rm i}\f x^\hi \gamma^\hi \gamma^J
F^{+J} \Bigr)\gamma^+\psi^\oplus\,,
\\
&& S^- = \int d^3x\,\rTrpr\,
\Bigl(2x^-\gamma^IF^{+I}+x^\hi\gamma^\hi(- F^{+-}
+\frac{1}{2}\gamma^{IJ}F^{IJ}) +2
\gamma^M\phi^M\Bigr)\gamma^+\psi^\oplus\,.
\end{eqnarray}
Note that while writing these expressions we have not
used light cone gauge.

Light cone gauge representation for above-given supercharges in
terms of physical fields is obtainable then by using light cone
gauge fixed field strengths. Such field strengths can be obtained
by plugging light cone gauge \rf{lcgau} and solution to
constraints \rf{solam} into covariant field strengths given in
\rf{fiestr}. Doing this we get the following light cone gauge
fixed field strengths

\begin{eqnarray}
\label{fiestr1}
&& F^{+I} =\partial^+ A^I\,, \qquad F^{IJ} =
\partial^I A^J -\partial^J A^I +[A^I,A^J]\,,
\\
\label{fiestr2}&& F^{+-} =  - \partial^I A^I
-\frac{1}{\partial^+}[A^I,\partial^+A^I] +\frac{\rm
i}{\partial^+}(\psi^\oplus\bar{\gamma}^+\psi^\oplus)\,.
\end{eqnarray}
Making use of these formulas we get the following useful relation
to be inserted into expression for supercharges $\Omega^-$, $S^-$
\be - F^{+-}  + \frac{1}{2}\gamma^{IJ}F^{IJ} =\Bigl(\partial^I A^J
+\frac{1}{\partial^+}[A^I,\partial^+ A^J]\Bigr) \gamma^I\gamma^J
-\frac{\rm i}{\partial^+}(\psi^\oplus\bar{\gamma}^+\psi^\oplus)\,.
\ee In Section \ref{HAMFOR} we demonstrated that the light cone
gauge vertices of SYM in plane wave background take the same form
as the ones in flat Minkowski space-time. Now it can easily be
checked that on the surface of initial data $x^+=0$ all bosonic
and fermionic Noether charges of plane wave SYM coincide with the
ones of flat Minkowski space-time. It is clear that underlying
reason for these coincides is related with conformal invariance of
these SYM theories.

\newsection {Transformations of physical fields under
\\ action of $psu(2,2|4)$ superalgebra}\label{TRARUL}

Because the expressions for Noether charges are given entirely in
terms of physical field $A^\hi$, $A^M=\phi^M$ and $\psi^\oplus$
transformations of these fields under action of $psu(2,2|4)$
superalgebra could be found in principle by using formulas
\rf{fietra} and Poisson-Dirac brackets for physical fields
\rf{aacom}, \rf{ppcom}. Here we discuss simpler method of deriving
field transformations generated by $psu(2,2|4)$ algebra. Let us
start our discussion with transformations generated by the bosonic
symmetries which are $so(4,2)\oplus so(6)$. Because realization of
$R$-symmetries of $so(6)$ takes the standard form we proceed to
discussion of transformations generated by the conformal $so(4,2)$
algebra.

In order to find global transformations of physical fields we
start with corresponding global transformations of covariant
fields supplemented with local gauge transformations \be
\label{exttra1} \delta_G A^\muun = \delta^{cov} A^\muun +
D^\muun\, \alpha^G\,, \qquad\quad \delta_G \psi = \delta^{cov}
\psi + [\psi,\alpha^G]\,,\ee where global transformations of the
covariant fields $A^\muun =(A^\mun, \phi^M)$ and $\psi$ are given
in \rf{delamu1}-\rf{delamu3}. As usual the parameter of gauge
transformation $\alpha^G$ corresponding to global transformation
generated by Noether charge $G$ is fixed by the requirement that the
transformation \rf{exttra1} maintains the light cone gauge
\be\label{DGA} \delta_G A^+ =0\,. \ee From this equation we can
get solution to parameters of compensating gauge transformations.
It turns out that parameters of compensating gauge transformations
for the isometry transformations generated by $P^+$, $T^\hi$,
$J^{\hi\hj}$ and three proper conformal transformations generated
by $D$ and $C$ are equal to zero \be\label{alpsol}
\alpha^{P^+}=0\,,\quad \alpha^{T^\hi}=0\,,\quad
\alpha^{J^{\hi\hj}}=0\,,\quad \alpha^D=0\,,\qquad \alpha^C=0\,,
\ee while the parameters corresponding to $C^\hi$ and $K^-$ are
given by \be \qquad \alpha^{C^\hi} = -e^{-{\rm i}\f
x^+}\frac{1}{\partial^+} A^\hi\,, \qquad \alpha^{K^-} =
-\frac{1}{\partial^+} x^\hi A^\hi\,. \ee Plugging the solutions
parameters into \rf{exttra1} we can get desired transformations
for physical fields. The transformations corresponding to isometry
generators $P^+$, $T^\hi$, $J^{\hi\hj}$ are given in
\rf{pwpw01}-\rf{pwpw04} while the transformations corresponding to
proper conformal generators $D$, $C$, $C^\hi$, $K^-$ are listed
below.

{\it Proper conformal transformation of the physical spin 1 field
$A^\hi$:}
\begin{eqnarray}
\label{comtrarul1} && \delta_D A^\hi =(\xi^{D}\partial +1
)A^\hi\,,
\\
\label{comtrarul2}&& \delta_C A^\hi =(\xi^{C}\partial - {\rm i}\f
e^{-2{\rm i}\f x^+})A^\hi\,,
\\
\label{comtrarul3}&& \delta_{C^\hj} A^\hi =(\xi^{C^\hj}\partial
+{\rm i}\f x^je^{-{\rm i}\f x^+})A^\hi -e^{-{\rm i}\f
x^+}\delta^{\hi\hj} A^-
\nonumber\\
&&\hspace{1.5cm} +{\rm i}\f e^{-{\rm i}\f x^+}(x^\hk A^\hk
\delta^{\hi\hj} -x^\hi A^\hj) - e^{-{\rm i}\f x^+}D^\hi
\frac{1}{\partial^+}A^\hj\,,
\\
\label{comtrarul4}&& \delta_{K^-} A^\hi =(\xi^{K^-}\partial +
x^-)A^\hi - x^\hi A^- - D^\hi \frac{1}{\partial^+} (x^\hj
A^\hj)\,.
\end{eqnarray}

{\it Proper conformal transformation of the scalar fields
$A^M=\phi^M$:}
\begin{eqnarray}
\label{comtrarul5}&& \delta_D\phi^M =(\xi^{D}\partial +
1)\phi^M\,,
\\
\label{comtrarul6}&& \delta_C \phi^M =(\xi^C\partial -{\rm i}\f
e^{-2{\rm i}\f x^+} )\phi^M\,,
\\
\label{comtrarul7}&& \delta_{C^\hi} \phi^M = (\xi^{C^i}\partial +
{\rm i}\f e^{-{\rm i}\f x^+}x^\hi )\phi^M
-e^{-{\rm i}\f x^+}[\phi^M,\frac{1}{\partial^+}A^\hi]\,,
\\
\label{comtrarul8}&& \delta_{K^-} \phi^M =(\xi^{K^-}\partial +x^-
)\phi^M - [\phi^M,\frac{1}{\partial^+} x^\hi A^\hi]\,.
\end{eqnarray}

{\it Proper conformal transformation of the physical spin 1/2
field $\psi^\oplus$:}
\begin{eqnarray}
\label{comtrarul9}&& \delta_D\psi^\oplus =(\xi^D\partial +
2)\psi^\oplus\,,
\\
\label{comtrarul10}&& \delta_C \psi^\oplus =(\xi^C\partial -{\rm
i}\f e^{-2{\rm i}\f x^+} )\psi^\oplus\,,
\\
\label{comtrarul11}&& \delta_{C^\hi} \psi^\oplus
=\Bigl(\xi^{C^\hi}\partial + e^{-{\rm i}\f x^+}(\frac{3{\rm i}
}{2}\f x^\hi +\frac{\rm i}{2}\f \gamma^{\hi\hj}
x^\hj)\Bigr)\psi^\oplus +\frac{1}{2} e^{-{\rm i}\f
x^+}\gamma^{-\hi}\psi^\ominus +[\psi^\oplus,\alpha^{C^\hi}]\,, \ \
\ \ \
\\
\label{comtrarul12}&& \delta_{K^-} \psi^\oplus =(\xi^{K^-}\partial
+2x^-)\psi^\oplus + \frac{1}{2} \gamma^{- \hi} x^\hi \psi^\ominus
+[\psi^\oplus,\alpha^{K^-}]\,.
\end{eqnarray}
Solutions to the non-physical fields $A^-$, $\psi^\ominus$ to be
inserted in these expressions are given by \rf{solam},
\rf{solpsim}, while the expressions for Killing vectors in terms
of differential operators $\xi^G\partial\equiv
\xi^{G\mun}\partial_\mun$ are given in \rf{dkil}-\rf{kmkil}.
Replacing in above given transformations rules the time
derivatives of physical fields by the commutator with Hamiltonian
\rf{hamequ} we get famous {\it off-shell} light cone realization
of global symmetries.

\subsection{Supersymmetry transformations of physical fields}

Supersymmetry transformations of physical fields can be fixed by
using the same procedure as in previous section. As before we
start with general transformation rules \rf{exttra1} where the
supersymmetry transformations of covariant fields on r.h.s. are
given by \rf{pwsusytra1}-\rf{pwsusytra3}. Before to proceed to
transformations of physical fields let us cast the transformations
rules of covariant fields into more convenient form. In light cone
formalism it is convenient to divide explicitly the supersymmetry
transformations into the ones generated by supercharges $\Omega^+$
and $\Omega^-$. This can be made in a straightforward way by using
formulas for $\Omega$ transformations given in
\rf{pwsusytra1}-\rf{pwsusytra3}. In transparent form supersymmetry
transformations in basis formed by supercharges $\Omega^\pm$,
$S^-$ take then the form
\begin{eqnarray}
&& \delta_{_{\Omega^+}}^{cov}A^\mu
 = {\rm i}\psi\bar{\gamma}^\mu \epsilon_{_{0Q^+}}\,,
\\
&& \delta_{_{\Omega^-}}^{cov} A^\mu ={\rm i} e^{ -{\rm i}\f
x^+}\psi \bar{\gamma}^\mu (1+\frac{\rm i\f}{2}x^\hi\gamma^{+\hi})
\epsilon_{_{0\Omega^-}}\,,
\\
&& \delta_{_{S^-}}^{cov} A^\mu ={\rm i}\psi \bar{\gamma}^\mu
(x^-\gamma^+ + x^\hi \gamma^\hi) \epsilon_{_{0S^-}}\,,
\end{eqnarray}
\begin{eqnarray}
&& \delta_{_{\Omega^+}}^{cov} \psi
=\frac{1}{2}F^{\mu\nu}\gamma^{\mu\nu} \epsilon_{_{0\Omega^+}}\,,
\\
&& \delta_{_{\Omega^-}}^{cov} \psi = \frac{1}{2} e^{ -{\rm i}\f
x^+} F^{\mu\nu}\gamma^{\mu\nu} (1+\frac{\rm
i\f}{2}x^\hi\gamma^{+\hi}) \epsilon_{_{0\Omega^-}} +{\rm i}\f
e^{-{\rm i}\f x^+}\phi^M\gamma^M\bar{\gamma}^+
\epsilon_{_{0\Omega^-}}\,,
\\
&& \delta_{_{S^-}}^{cov} \psi
=\frac{1}{2}F^{\mu\nu}\gamma^{\mu\nu}(x^-\gamma^+ + x^\hi
\gamma^\hi) \epsilon_{_{0S^-}}
-2\phi^M\gamma^M\epsilon_{_{0S^-}}\,,
\end{eqnarray}
where the constant parameters of transformations satisfy the
constraints \be \bar{\gamma}^+\epsilon_{_{0\Omega^+}} =0\,, \qquad
\bar{\gamma}^-\epsilon_{_{0\Omega^-}}=0\,, \qquad
\gamma^-\epsilon_{_{0S^-}}=0\,. \ee Now we plug these
transformations rules on r.h.s. of \rf{exttra1} and from the
requirement \rf{DGA} we find solution to the parameters of
compensating gauge transformations \be \alpha^{\Omega^+}=0\,,
\qquad \alpha^{\Omega^-} = -\frac{\rm i}{\partial^+} e^{-{\rm i}\f
x^+}\psi^\oplus\bar{\gamma}^+ \epsilon_{_{0\Omega^-}}\,, \qquad
\alpha^{S^-} = -\frac{\rm
i}{\partial^+}\psi^\oplus\bar{\gamma}^{+\hi} x^\hi
\epsilon_{_{0S^-}}\,. \ee Plugging these compensating parameters
into \rf{exttra1} we get the desired supersymmetry transformations
of the physical fields $A^I=(A^\hi,A^M=\phi^M)$ and $\psi^\oplus$:
\begin{eqnarray}
&& \delta_{_{\Omega^+}} A^I = {\rm i}\psi^\oplus \bar{\gamma}^I
\epsilon_{_{0Q^+}}\,,
\\
&& \delta_{_{\Omega^-}} A^I = {\rm i}e^{-{\rm i}\f x^+}(
\psi^\ominus \bar{\gamma}^I +\frac{{\rm i}\f}{2} x^\hj \psi^\oplus
\bar{\gamma}^I \gamma^{+\hj})\epsilon_{_{0\Omega^-}}
+D^I\alpha^{\Omega^-}\,,
\\
&& \delta_{_{S^-}} A^I = {\rm
i}(x^-\psi^\oplus\bar{\gamma}^I\gamma^+ + x^\hj \psi^\ominus
\bar{\gamma}^I \gamma^\hj)\epsilon_{_{0S^-}} +D^I\alpha^{S^-}\,,
\end{eqnarray}
\begin{eqnarray}
&& \delta_{_{\Omega^+}}\psi^\oplus =\gamma^{-I}F^{+I}
\epsilon_{_{0\Omega^+}}\,,
\\
&& \delta_{_{\Omega^-}}\psi^\oplus =e^{-{\rm i}\f x^+}(F^{+-}
+\frac{1}{2}\gamma^{IJ}F^{IJ} -{\rm i}\f
F^{+J}\gamma^J\bar{\gamma}^\hi x^\hi) \epsilon_{_{0\Omega^-}}
+[\psi^\oplus,\alpha^{\Omega^-}]\,,
\\
&& \delta_{_{S^-}}\psi^\oplus =\Bigl((F^{+-}
+\frac{1}{2}\gamma^{IJ}F^{IJ})x^\hk\gamma^\hk -2x^-F^{+J}\gamma^J
-2\gamma^M\phi^M\Bigr) \epsilon_{_{0S^-}}
+[\psi^\oplus,\alpha^{S^-}]\,. \ \ \ \ \ \ \
\end{eqnarray}
The expressions for the
non-physical field $\psi^\ominus$ and gauge fixed field
strengths to be inserted in these formulas may be found in
\rf{solpsim},\rf{fiestr1},\rf{fiestr2}.

\newsection{Conclusions}\label{CONCLU}

To summarize, we have found  the supersymmetric  action for a
$D3$ brane probe propagating in plane wave Ramond-Ramond background.
The action is   given by \rf{action}--\rf{fexp} with the closed
5-form  defining the WZ term  given in \rf{h5}. This action is
world-volume reparametrisation invariant and $\kappa$-invariant.
Its advantage is that it is manifestly invariant under the
symmetries of plane wave vacuum: 30 bosonic isometries and 32
supersymmetries. It  does not have a particularly  simple form
when written in terms of the coordinates $(x,\theta)$,  even using
the closed expressions   for the Cartan 1-forms in terms of
$\theta$ (see Appendix C).

The action can be put in a more explicit form by imposing various
kappa symmetry gauges. For instance to establish a connection to
the abelian $\cN=4$ SYM  theory as discussed in the Introduction we

(i) fix  the $\kappa$-symmetry gauge in a way that simplifies the
fermionic part of the action.

(ii)  fix the static gauge  so that the $D3$ probe is oriented
parallel to the $D3$ source.

After fixing the  local symmetry gauges only the bosonic seven
isometry symmetries of $4d$ plane wave background, the seven
R-symmetries $SO(2)\times SO'(4)$, and 16 supersymmetries of
the original symmetries remain unbroken, while the remaining 16
bosonic and 16 supersymmetries are broken and realized
non-linearly. Interesting fact is that the number of unbroken
space-time symmetries (seven isometry symmetries of $4d$ plane wave
background) is equal to the number of unbroken R-symmetries (seven
symmetries of $SO(2)\times SO'(4)$). Note also that the number of broken
bosonic symmetries is equal to the number of broken supersymmetries.

We developed the Hamiltonian light cone gauge formulation of plane
wave SYM and demonstrated that in contrast to covariant and gauge
invariant formulation the 3-point and 4-point light cone gauge
vertices of plane wave SYM theory takes exactly the same form as
the ones of SYM in flat Minkowski space-time.

The results presented here should have a number of interesting
applications and generalizations, some of which are:

1. In Refs. \cite{Kristjansen:2002bb,Bianchi:2002rw} various
improvements of the original BMN operators \cite{beren} were
suggested. We expect that study based on analysis of conformal
properties of plane wave SYM which we performed in this papers
together with ideas and approaches of Refs.
\cite{Gopakumar:1994iq,pin} should allow us to derive BMN
operators from first principles and fix their precise form.

2. In this paper we develop light cone gauge formulation of plane
wave SYM using field component formulation. As in flat space
\cite{brink} the plane wave SYM could also be reformulated in
terms of unconstrained light cone superfield. It would be
interesting then to apply  such superfield formulation to the
study of BMN conjecture \cite{beren} along the line of Ref.
\cite{Santambrogio:2002sb}.

3. The plane wave background \rf{4dpw} can be obtained from the
$R\times S^3$ space-time via the Penrose limit. Indeed consider
the line element of $R\times S^3$ \be ds^2 = -dt^2 +R^2
(\cos^2\theta d\psi^2 + d\psi^2 +\sin^2\theta d\varphi^2)\,, \ee
where $R$ is a radius of $S^3$. Introducing the coordinates
$x^\pm$, $x^\hi$ \be t
= \sqrt{2}\,\f R x^+ -\frac{x^-}{2\sqrt{2}\,\f R}\,, \qquad \psi =
\sqrt{2}\,\f  x^+ + \frac{x^-}{2\sqrt{2}\,\f R^2}\,,\qquad \theta
= \frac{|x|}{R}\,, \ee $|x|=(x^\hi x^\hi)^{1/2}$, and taking the
Penrose limit $R\rightarrow \infty $ we get the line element of
plane wave background \rf{4dpw}. Making use of relations \be
\partial_t = -{\rm i}H = -{\rm i}\frac{\Delta}{R}\,,\qquad
\partial_{x^+}=-{\rm i}H_{l.c.}=-{\rm i}E_{l.c.}\,,\qquad
\partial_{x^-} = {\rm i} p^+\,,\qquad
\partial_\psi = {\rm i}J\,,
\ee we find then the relations between light cone energy
$E_{l.c.}$ and momentum $p^+$ of (states) operators of plane wave
SYM and conformal dimension $\Delta$ and angular momentum $J$ of
(states)operators of SYM in $R\times S^3$ space-time \be E_{l.c.}
= \sqrt{2}\,\f (\Delta -J)\,, \qquad p^+ =
\frac{\Delta+J}{2\sqrt{2}\, \f R^2}\,.\ee Taking into account
these relations and interrelations of $\Delta$ and $J$ with plane
wave superstring energy and momentum $p^+$ found in \cite{beren}
it seems highly likely that new duality \cite{beren} can be
re-formulated as interrelations between light cone energies and
$p^+$ momenta of (states) operators of $4d$ plane wave SYM and
(states) operators of $10d$ plane wave superstring theory \be
E_{l.c.}(\hbox{pw SYM})  \sim E_{l.c.}(\hbox{pw string})\,, \qquad
p^+(\hbox{pw SYM})  \sim p^+(\hbox{pw string})\,. \ee It is
desirable to understand these interrelations better.

\setcounter{section}{0}
\setcounter{subsection}{0}
\begin{center}
{\bf Acknowledgments}
\end{center}

At the initial states of this work we were fortunate to
collaborate with A.A. Tseytlin, and we are grateful for his help.
Conversations with E.A. Ivanov are happily acknowledged. Author
would like to express his gratitude to Prof. A. Sagnotti for the
hospitality at the University Tor Vergata where some part of this
work was done. This work was supported by the INTAS project
00-00254, by the MURST-COFIN contract 2001-025492,
by the RFBR Grant 02-02-16944 and RFBR Grant 01-02-30024
for Leading Scientific Schools.

\appendix{Notation and conventions}

We use the following  conventions for the indices:
\begin{eqnarray*}
\mu,\nu,\rho = 0,1,\ldots, 9 && \quad  so(9,1) \  \hbox{ vector
indices (tangent space indices) }
\\
\muun,\nuun,\rhoun = 0,1,\ldots, 9 && \quad  \hbox{Sects.
\ref{2}-\ref{LCD3}: coordinate indices of $10d$ plane wave
background}
\\
\muun,\nuun,\rhoun = 0,1,\ldots, 9 && \quad  \hbox{Sects.
$\!\!$\ref{PSUVARBAS}-\ref{TRARUL}:$\!$ coordinate indices of $4d$
plane wave $\!\!\times R^6$ background}
\\
I,J,K,L = 1,\ldots, 8 && \quad  so(8) \  \hbox{ vector indices
(tangent space indices) }
\\
M,N, = 3,\ldots, 8 && \quad  so(6) \  \hbox{ vector indices
(tangent space indices) }
\\
i,j,k,l = 1,\ldots, 4 && \quad  so(4) \  \hbox{ vector indices
(tangent space indices) }
\\
\ipr,\jpr,\kpr,\lpr = 5,\ldots, 8 && \quad  so^\prime(4) \ \hbox{
vector indices (tangent space indices) }
\\
\hi,\hj,\hk, = 1,2 && \quad \ so(2) \  \hbox{ vector indices
(tangent space indices) }
\\
a,b = 0,1,2,9  && \quad  \hbox{ 4-d $D3$ brane world-volume
coordinate indices}
\\
\mun,\nun = 0,1,2,9  && \quad  \hbox{ $4d$ plane wave space-time
coordinate indices}
\\
m,n = 0,1,2,9  && \quad  \hbox{ $so(3,1)$ tangent space indices}
\\
\alpha,\beta,\gamma = 1,\ldots, 16 && \quad \ so(9,1) \  \hbox{
spinor indices in chiral representation}
\\
\cI,\cJ = 1,2 && \quad  \hbox{ labels of the two real MW spinors}
\end{eqnarray*}
We suppress the flat space metric tensor
$\eta_{\mu\nu}=(-,+,\ldots, +)$ in scalar products, i.e. $ X^\mu
Y^\mu\equiv \eta_{\mu\nu}X^\mu Y^\nu.$ We decompose
$x^{\underline{\mu}}$ into the light cone and transverse
coordinates: $x^{\underline{\mu}}= (x^+,x^-, x^I)$,
$x^I=(x^i,x^\ipr)$, where $x^\pm\equiv \frac{1}{\sqrt{2}}(x^9\pm
x^0)$. The scalar products of tangent space vectors are decomposed
as \be X^\mu Y^\mu = X^+Y^- + X^-Y^+ +X^IY^I\,, \qquad
X^IY^I=X^iY^i+X^\ipr Y^\ipr\,. \ee

We use the chiral representation for the $32\times 32$ Dirac
matrices $\Gamma^\mu$ in terms of the $16\times 16 $
matrices $\gamma^\mu$
\be \Gamma^\mu =\left(\begin{array}{cc} 0  & \gamma^\mu \\
\bar{\gamma}^\mu & 0
\end{array}\right) \,,\ee
\be\label{gammaspl0} \gamma^\mu\bar{\gamma}^\nu +
\gamma^\nu\bar{\gamma}^\mu =2\eta^{\mu\nu}\,,\qquad \gamma^\mu =
(\gamma^\mu)^{\alpha\beta}\,, \qquad  \bar{\gamma}^\mu
=\gamma^\mu_{\alpha\beta}\,, \ee \be
\label{gammaspl}\gamma^\mu=(1,\gamma^I,\gamma^9)\,,\qquad
\bar{\gamma}^\mu=(-1,\gamma^I,\gamma^9)\,,\qquad
\alpha,\beta=1,\ldots 16\,.\ee We adopt the Majorana
representation for $\Gamma$-matrices, $ C= \Gamma^0$, which
implies that all $\gamma^\mu$ matrices are real and symmetric,
$\gamma^\mu_{\alpha\beta} = \gamma^\mu_{\beta\alpha}$,
$(\gamma^\mu_{\alpha\beta })^* = \gamma^\mu_{\alpha\beta}$. As in
\cite{rrm0112} $\gamma^{\mu_1\ldots \mu_k}$ are the
antisymmetrized products of $k$ gamma matrices, e.g.,
$(\gamma^{\mu\nu})^\alpha{}_\beta \equiv
\frac{1}{2}(\gamma^\mu\bar{\gamma}^\nu)^\alpha{}_\beta -(\mu
\leftrightarrow \nu)$,
$(\bar{\gamma}^{\mu\nu})_\alpha{}^\beta \equiv
\frac{1}{2}(\bar{\gamma}^\mu\gamma^\nu)_\alpha{}^\beta -(\mu
\leftrightarrow \nu)$,
$ (\gamma^{\mu\nu\rho})^{\alpha\beta}
\equiv
\frac{1}{6}(\gamma^\mu\bar{\gamma}^\nu\gamma^\rho)^{\alpha\beta}
\pm 5 \hbox{ terms}$. Note that
$(\gamma^{\mu\nu\rho})^{\alpha\beta}$ are antisymmetric in
$\alpha$, $\beta$ while
$(\gamma^{\mu_1\ldots\mu_5})^{\alpha\beta}$ is symmetric. We
assume the normalization \be\label{g11} \Gamma_{11} \equiv
\Gamma^0\ldots \Gamma^9 =\left(\begin{array}{cc}
I_{16} & 0\\
0 & - I_{16}
\end{array}\right)\ , \ \ \ \ \ \ \ \ \
 \gamma^0\bar{\gamma}^1
\ldots \gamma^8\bar{\gamma}^9=I_{16} \ . \ee We use the following
definitions $\Pi^\alpha{}_\beta \equiv
(\gamma^1\bar{\gamma}^2\gamma^3\bar{\gamma}^4)^\alpha{}_\beta$, $
(\Pi^\prime)^\alpha{}_\beta \equiv
(\gamma^5\bar{\gamma}^6\gamma^7\bar{\gamma}^8)^\alpha{}_\beta$.
Because of the relation $\gamma^0\bar{\gamma}^9 =\gamma^{+-}$ the
normalization condition \rf{g11} takes the form $
\gamma^{+-}\Pi\Pi^\prime = 1$. The 32-component positive chirality
spinor $\theta$ and  the negative chirality spinor $Q$  are
decomposed in terms of the 16-component spinors as
\be\label{3216rel} \theta = \left( \begin{array}{c} \theta^\alpha \\
0\end{array}\right)\,, \qquad\quad
Q = \left( \begin{array}{c} 0 \\
Q_\alpha\end{array}\right)\,.\ee
The   complex
Weyl spinor $\theta$ is related to the  two real
Majorana-Weyl spinors $\theta^1$ and $\theta^2$  by
\be\label{comrea} \theta = \frac{1}{\sqrt{2}}(\theta^1 +{\rm
i}\theta^2)\,,\qquad \bar{\theta} = \frac{1}{\sqrt{2}}(\theta^1 -
{\rm i}\theta^2)\,.\ee
The short-hand notation like
$\bar{\theta}\bar{\gamma}^\mu\theta$ and $\bar{\gamma}^\mu\theta$
 stand for
$\bar{\theta}{}^\alpha\gamma_{\alpha\beta}^\mu\theta^\beta$ and
$\gamma_{\alpha\beta}^\mu \theta^\beta$ respectively.

\appendix{Plane wave Ramond-Ramond superalgebra}

Plane wave Ramond-Ramond superalgebra contains ten translation
generators $P^\mu$ (in light cone frame $P^+$, $P^-$, $P^I$,
$I=1,\ldots 8)$, eight Lorentz boosts $J^{+I}$, six generators of
$SO(4)$ rotations, $J^{ij}$, $i,j=1,\ldots, 4$, six generators of
$SO'(4)$ rotations, $J^{i'j'}$, $i',j'=5,\ldots, 8$ and two
sixteen component real-valued spinor $Q_\alpha^\cI$, $\cI=1,2$,
$\alpha=1,\ldots, 16$. In the 32 spinor component notation the
$Q_\alpha^\cI$ correspond to two negative chirality Majorana-Weyl
spinors (see \rf{3216rel}). Commutation relations between the even
generators are given by\footnote{This algebra was found in
\cite{blau} as an algebra of isometry symmetries of plane wave
Ramond-Ramond solution of IIB supergravity. Derivation of this
superalgebra from the super-$AdS_5\times S^5$ algebra by the
Inonu-Wigner contractions may be found in \cite{Hatsuda:2002xp}.
Discussion of the corresponding plane wave solution of $11d$
supergravity is given in
\cite{Kowalski-Glikman:ux,Gauntlett:2002cs}. Plane wave
supermembrane and matrix theories were investigated in
\cite{Sugiyama:2002rs} and \cite{Dasgupta:2002hx} (see also
\cite{Sugiyama:2002bw} for some related studies).}

\be\label{pmpi1} [P^-,P^I] = -\f^2J^{+I}\,,\ee \be [P^I,J^{+J}] =
-\delta^{IJ}P^+\,, \qquad [P^-,J^{+I}]=P^I\,,\ee \be [P^i,J^{jk}]
= \delta^{ij}P^k -\delta^{ik}P^j\,, \qquad [P^\ipr,J^{\jpr\kpr}] =
\delta^{\ipr\jpr}P^\kpr -\delta^{\ipr\kpr}P^\jpr\,,\ee \be
[J^{+i},J^{jk}] = \delta^{ij}J^{+k} -\delta^{ik}J^{+j}\,, \qquad
[J^{+\ipr},J^{\jpr\kpr}] = \delta^{\ipr\jpr}J^{+\kpr}
-\delta^{\ipr\kpr}J^{+\jpr}\,,\ee \be [J^{ij},J^{kl}] =
\delta^{jk}J^{il} + 3\hbox{ terms}\,, \qquad
[J^{\ipr\jpr},J^{\kpr\lpr}] = \delta^{\jpr\kpr}J^{\ipr\lpr} +
3\hbox{ terms}\,, \ee where $\f$ is a dimensionful parameter.

Commutation relations  between the even and odd parts are
 \be\label{jq1} [J^{ij},Q_\alpha^\cI ] =
 \frac{1}{2}Q_\beta^\cI (\gamma^{ij})^\beta{}_\alpha
\,, \qquad [J^{\ipr\jpr},Q_\alpha^\cI] = \frac{1}{2}Q_\beta^\cI
(\gamma^{\ipr\jpr})^\beta{}_\alpha \,, \ee \be {}
[J^{+I},Q_\alpha^\cI] =
 \frac{1}{2}Q_\beta^\cI (\gamma^{+I})^\beta{}_\alpha\,,
\qquad
[P^\mu,Q_\alpha^\cI] =  -\frac{\f}{2}\tau_2^{\cI\cJ} Q_\beta^\cJ ( \Pi
\gamma^+\bar{\gamma}^\mu)^\beta{}_\alpha\,,
 \ee

The anticommutator takes the form
\begin{eqnarray} \{Q_\alpha^\cI, Q_\beta^\cJ \}
&=& -2{\rm i}\delta^{\cI\cJ}\gamma^\mu_{\alpha\beta} P^\mu - 2{\rm i}\f
\tau_2^{\cI\cJ}
\Bigl((\bar{\gamma}^i\Pi)_{\alpha\beta}J^{+i} +(\bar{\gamma}^\ipr
\Pi^\prime)_{\alpha\beta}J^{+\ipr}\Bigr)
\nonumber\\[7pt]
\label{qq} &+&{\rm i}\f\tau_2^{\cI\cJ}\Bigl(
(\bar{\gamma}^+\gamma^{ij}\Pi)_{\alpha\beta}J^{ij}
+(\bar{\gamma}^+
\gamma^{\ipr\jpr}\Pi^\prime)_{\alpha\beta}J^{\ipr\jpr}\Bigr)\,,
\end{eqnarray}
where $\tau_2$ is defined in \rf{taumat}. All the other
commutators and anticommutators vanish. The bosonic generators are
assumed to be anti-hermitian while the fermionic generators are
hermitian, $ (Q_\alpha^\cI)^\dagger=Q_\alpha^\cI$. The generators
of the full transformation group $G$ of the plane wave RR
superspace  are $P^\mu$, $Q_\alpha^\cI$, $J^{ij}$, $J^{\ipr\jpr}$,
$J^{+I}$. The generators of the stability subgroup $H$ are
$J^{ij}$, $J^{\ipr\jpr}$, $J^{+I}$. The plane wave RR superspace
is defined then as coset superspace $G/H$. The specific choice of
the representative of supercoset which we use frequently in this
paper is given by \be\label{G1} G(x,\theta)= e^{x^+P^-}e^{x^-P^+
+x^IP^I}e^{\theta^{\alpha\cI} Q_\alpha^\cI}\,. \ee Representation
of the bosonic generators of plane wave superalgebra in terms of
Killing vectors acting on the superspace
$(x^+,x^-,x^I,\theta^{\alpha \cI})$ defined by \rf{G1} takes the
form
\begin{eqnarray}
\label{PWK1}&& P^\pm =\partial^\pm\,,
\\
&& P^I = \cos\f x^+\partial^I +\f\sin\f x^+ x^I\partial^+
-\frac{\f}{2}\sin\f x^+
\partial_{\theta^{\alpha\cI}}(\gamma^{+I})^\alpha{}_\beta
\theta^{\beta\cI}\,,
\\
&&
J^{+I} = \frac{\sin \f x^+}{\f}\partial^I - \cos \f x^+
x^I\partial^+ +\frac{1}{2}\cos\f x^+
\partial_{\theta^{\alpha\cI}}(\gamma^{+I})^\alpha{}_\beta
\theta^{\beta\cI}\,,
\\
\label{PWK4} && J^{IJ} = x^I \partial^J - x^J\partial^I
+\frac{1}{2}
\partial_{\theta^{\alpha \cI}}
(\gamma^{IJ})^\alpha{}_\beta \theta^{\beta \cI}\,,
\end{eqnarray}
where just the $so(4)\oplus so'(4)$ part of  $J^{IJ}$ which is
given by $J^{ij}$, $J^{i'j'}$ enters the plane wave superalgebra.
Complete expressions for fermionic generators $Q_\alpha^\cI$ are
complicated and not illuminating. Leading terms of their expansion
in fermionic coordinates  are given by \be\label{Q1} \epsilon_0 Q
=\epsilon^{\alpha\cI}(x)
\partial_{\theta^{\alpha\cI}} -{\rm i}e^{\underline{\nu}}_\nu \,\,
\epsilon(x)\gamma^\nu \Theta
\partial_{x^{\underline{\nu}}}
+O(\Theta^2\partial_\Theta,\Theta^3)\,, \ee where the Killing
spinor $\epsilon(x)$ is defined by relation \rf{epsU}. In light
cone gauge only the terms shown explicitly in \rf{Q1} gives
non-zero contribution to supercharge $Q_\alpha^\cI$. In above
expressions the fermionic partial derivatives are defined to be
$\partial_{\theta^{\alpha \cI}} =\partial/\partial
\theta^{\alpha\cI}$ and we use the following conventions for
bosonic partial derivatives \be
\partial^+ \equiv \partial_-=\frac{\partial}{\partial x^-}\,,\qquad
\partial^-\equiv\partial_+=\frac{\partial}{\partial x^+}\,,\qquad
\partial^I\equiv \partial_I=\frac{\partial}{\partial x^I}\,.\ee

\appendix{Basic relations for Cartan forms on  coset superspace}

The left-invariant Cartan  1-forms $L^{^{...}} =
dX^{\underline{A}}\, L^{^{...}}_{\underline{A}}$,
$X^{\underline{A}}= (x^{\underline{\nu}}\,,
\theta^{\alpha\cI})$ are defined by \be\label{def1} G^{-1}dG =L^\mu
P^\mu+\frac{1}{2}L^{\mu \nu}J^{\mu\nu} +L^{\alpha
\cI}Q_{\alpha}^\cI\ , \ee where $G= G({x,\theta})$ is a coset
representative plane wave supergroup. $L^\mu$ are the 10-beins,
$L^{\alpha \cI}$, $(L^{\alpha \cI})^\dagger=L^{\alpha \cI}$, are
the two spinor 16-beins and  $L^{\mu\nu}$ ($L^{+\mu}=0$,
$L^{ij'}=0$) are the Cartan connections. The two 16 component
Cartan 1-forms $L^{\cI \alpha}$ and fermionic coordinates
$\theta^{\cI\alpha}$ are combined into 2-vectors
\be\label{vec2com} {\bf
L}^\alpha=\left(\begin{array}{c}L^{1\alpha}
\\ L^{2\alpha}\end{array} \right)\,,
\qquad
\Theta^\alpha=\left(\begin{array}{c}\theta^{1\alpha}
\\ \theta^{2\alpha}\end{array} \right)\,.
\ee Throughout this paper we use $2\times 2$ matrices $\tau_1$,
$\tau_2$, $\tau_3$ defined to be \be\label{taumat}\tau_1= \left(
\begin{array}{cc}
0 & 1
\\
1 & 0
\end{array}\right)\,,
\qquad \tau_2= \left(
\begin{array}{cc}
0 & 1
\\
-1 & 0
\end{array}\right)
\qquad \tau_3= \left(
\begin{array}{cc}
1 & 0
\\
0 & -1
\end{array}\right)\,.
\ee
Complex notation fermionic Cartan 1-forms $L^\alpha$, $\bar{L}^\alpha$
are defined as in \rf{comrea}.
The Cartan 1-forms satisfy the Maurer-Cartan equations
implied by  the  basic symmetry superalgebra
\begin{eqnarray}
\label{mc1} && dL^\mu = -L^{\mu\nu}\wedge L^\nu -{\rm i} {\bf L}
\bar{\gamma}^\mu \wedge {\bf L}\,,
\\
\label{mc2} && d{\bf L} =
-\frac{1}{4}L^{\mu\nu} \gamma^{\mu\nu} \wedge {\bf L}
-\frac{\f}{2}L^\mu  \Pi\gamma^+\bar{\gamma}^\mu
\tau_2 \wedge {\bf L}\,.
 \end{eqnarray}
We note that we use the following sign conventions under
permutations of Cartan 1-forms: \be \label{sigc} L^\mu\wedge L^\nu
= - L^\nu \wedge L^\mu\,, \qquad L^\mu \wedge L^\alpha = -
L^\alpha\wedge L^\mu\,,\qquad L^\alpha \wedge
L^\beta=L^\beta\wedge L^\alpha\,.\ee It is often (while analysis
of $kappa$-invariance and etc) useful to use the following
expressions for the  variations of Cartan 1-forms which  are also
implied by the structure of the  basic symmetry superalgebra
\begin{eqnarray}
\delta L^\mu  &=& d\widehat{\delta x}{}^\mu + L^\nu
\widehat{\delta x}{}^{\nu\mu} + L^{\mu\nu} \widehat{\delta
x}{}^\nu
 +  2{\rm i}  {\bf L} \bar{\gamma}^\mu \widehat{\delta \Theta} \,,
\\
\label{delL}
\delta {\bf L} & = & d \widehat{\delta \Theta}
+ \frac{\f}{2}L^\mu
\Pi \gamma^+\bar{\gamma}^\mu \tau_2 \widehat{\delta \Theta}
+\frac{1}{4}L^{\mu\nu} \gamma^{\mu\nu} \widehat{\delta \Theta}
\nonumber\\
&-&\frac{\f}{2}\widehat{\delta x}{}^\mu  \Pi \gamma^+
\bar{\gamma}^\mu\tau_2 {\bf L} -\frac{1}{4}\widehat{\delta x}{}^{\mu\nu}
\gamma^{\mu\nu} {\bf L}\,.
\end{eqnarray}
where \be\label{hatdel} \widehat{\delta x}{}^\mu \equiv
{\delta X}^{\underline{A}}\,
L_{\underline{A}}^\mu\,, \qquad \widehat{\delta x}{}^{\mu\nu} \equiv
{\delta X}^{\underline{A}}\,
L_{\underline{A}}^{\mu\nu}\,, \qquad \widehat{\delta
\Theta}{}^\alpha \equiv {\delta X}^{\underline{A}}\,
L_{\underline{A}}^\alpha\,. \ee

A specific choice of $G(x,\theta)$ which we use in this paper is
\be\label{wzpar}
G=g(x)e^{\theta^{\alpha \cI} Q_\alpha^\cI }\,,\qquad
\ee
where $g(x)$ is a bosonic body of coset representative of, i.e.
 $x=(x^\mu)$ provides a
certain parametrization of plane wave background which may be kept
arbitrary.\footnote{The  use of  a
 concrete parametrization for $\theta$ is needed, however,
to find  the representation for the 2-form ${\FF}$  which enters
the BI action (see below). As in the flat space case  \cite{aps},
${\FF}$ cannot be expressed in terms of the  Cartan forms  only.}
Let us make  the rescaling $\theta\rightarrow t\theta$ and
introduce \be L_t^\mu(x,\Theta)\equiv L^\mu(x,t\Theta)\,, \qquad
L_t^{\mu\nu}(x,t\Theta)\equiv L^{\mu\nu}(x,t\Theta)\,, \qquad {\bf
L}_t(x,\Theta)\equiv {\bf L}(x,t\Theta)\,, \label{noot}\ee with
the initial condition \be\label{inicon} L_{t=0}^\mu =
e^\mu\,,\qquad L_{t=0}^{\mu\nu}= \omega^{\mu\nu}\,,\qquad {\bf
L}_{t=0}=0\,,\ee where $e^\mu$, $\omega^{\mu\nu}$,
$\omega^{+\mu}=0$, $\omega^{i\jpr}$=0, are the vielbeins and the
Lorentz connections for plane wave RR background. Then the
defining equations for the  Cartan 1-forms are
\begin{eqnarray}
\label{t1} &&
\partial_t {\bf L}_t =  d\Theta +
\frac{1}{4}L_t^{\mu\nu}\gamma^{\mu\nu} \Theta + \frac{\f
}{2}L_t^\mu \Pi\gamma^+\bar{\gamma}^\mu \tau_2 \Theta\,,
\\[3pt]
\label{t2} &&
\partial_t L_t^\mu
=   - 2{\rm i}\Theta  \bar{\gamma}^\mu {\bf L}_t\,,
\\[3pt]
\label{t3} &&
\partial_t L_t^{-i} = - 2{\rm i}\f \Theta \bar{\gamma}^i\Pi
\tau_2{\bf L}_t\,, \qquad\ \
\partial_t L_t^{-\ipr} =
- 2{\rm i}\f \Theta  \bar{\gamma}^\ipr\Pi^\prime
\tau_2{\bf L}_t\,,
\\[3pt]
\label{t6} &&
\partial_t L_t^{ij} =   2{\rm i}\f \Theta\bar{\gamma}^+
\gamma^{ij}\Pi \tau_2{\bf L}_t\,, \qquad \partial_t L_t^{\ipr\jpr}
=    2{\rm i}\f\Theta \bar{\gamma}^+ \gamma^{\ipr\jpr}\Pi^\prime
\tau_2{\bf L}_t\,.
\end{eqnarray}
While the relations \rf{def1}--\rf{delL} are valid in  an
arbitrary parametrization of the coset superspace, the relations
\rf{t1}--\rf{t6} apply  only in the parametrization of \rf{wzpar}.
The equations \rf{t1}--\rf{t6} can be solved in a rather
straightforward way \be\label{exprep1} {\bf L} = \frac{\hbox{sinh}
{\cal M}}{\cal M}{\cal D}\Theta\,,\qquad L^\mu = e^\mu -2{\rm
i}\Theta \,\bar{\gamma}^\mu \frac{\hbox{cosh}{\cal M}-1}{{\cal
M}^2}{\cal D}\Theta\,,\ee where covariant differential ${\cal
D}\Theta$ and matrix ${\cal M}$ are defined to be
\be\label{comder} {\cal D}\Theta  = \Bigl(d +\frac{1}{4}
\omega^{\mu\nu} \gamma^{\mu\nu} +\frac{\f}{2} e^\mu \Pi \gamma^+
\bar{\gamma}^\mu \tau_2 \Bigr) \Theta\,, \ee
$$
({\cal M}^2)^{\cI\cJ}{}^\alpha{}_\beta = -{\rm i}\f
\Bigl((\Pi\gamma^+\bar{\gamma}^\mu\tau_2\theta)^{\alpha\cI}
(\theta\bar{\gamma}^\mu)_\beta^\cJ
+(\gamma^{+i}\theta)^{\alpha\cI}
(\theta\tau_2\bar{\gamma}^i\Pi)_\beta^\cJ +
(\gamma^{+\ipr}\theta)^{\alpha\cI}
(\theta\tau_2\bar{\gamma}^\ipr\Pi^\prime)_\beta^{\cJ}\Bigr)
$$
\be\label{comM} + \frac{{\rm
i}\f}{2}\Bigl((\gamma^{ij}\theta^\cI)^\alpha
(\theta\tau_2\bar{\gamma}^+\gamma^{ij}\Pi)_\beta^\cJ +
(\gamma^{\ipr\jpr}\theta^\cI)^\alpha (\theta\tau_2\bar{\gamma}^+
\gamma^{\ipr\jpr}\Pi^\prime)_\beta^\cJ\Bigr)\,, \ee

Note that in  many formal  calculations  it is more convenient
to use directly the defining equations \rf{t1}--\rf{t6}  rather
then the explicit solution above given (in complex parametrization
this solution was given in  \cite{rrm0112}).

\appendix{Action of conformal algebra symmetries on physical fields}

In this Appendix we give more details about field theoretical
realization of the conformal algebra generators and
transformations rules of the SYM physical fields and extend these
results to the case of plane wave massless arbitrary spin fields.
Let us start our discussion with generators of isometry symmetries
which are $P^+$, $T^\hi$, $J^{\hi\hj}$. Plugging momentum and spin
densities \rf{ppp}-\rf{mmi} into expressions
\rf{genpp}-\rf{genjij} we get the following explicit
representation for generators
\begin{eqnarray}
&& P^+ = \int d^3x\,\rTrpr\,\, \partial^+ A^I r_0(P^+) A^I
+\frac{\rm i}{2}\psi^\oplus \gamma^+ r_0(P^+) \psi^\oplus\,,
\\
&& T^\hi = \int\!  d^3x\, \rTrpr\,\,\partial^+ A^I r_0(T^\hi) A^I
+\frac{\rm i}{2}\psi^\oplus \gamma^+ r_0(T^\hi) \psi^\oplus\,,
\\
&&
J^{\hi\hj}
 = \int\!  d^3x\,\rTrpr\,\,\partial^+ A^L
r_0(J^{\hi\hj}) A^L
+\frac{\rm i}{2}\psi^\oplus \gamma^+ r_0(J^{\hi\hj})
\psi^\oplus
\nonumber\\
& &\hspace{1cm}+\,
\partial^+ A^\hi A^\hj - \partial^+ A^\hj A^\hi
+\frac{\rm i}{4}\psi^\oplus \gamma^{+\hi\hj}\psi^\oplus\,,
\end{eqnarray}
where we use the following conventions for differential operators
acting on the physical fields $A^I$, $\psi^\oplus$ \be
r_0(P^+)=\partial^+\,,\qquad r_0(T^\hi) = e^{-{\rm i}\f
x^+}(\partial^\hi + {\rm i}\f x^\hi\partial^+)\,, \qquad
r_0(J^{\hi\hj}) = x^\hi\partial^\hj - x^\hj\partial^\hi\,. \ee
Making use of commutation relations \rf{aacom},\rf{ppcom} we get
the following transformations of fields under action of isometry
symmetries generated by $P^+$ and $T^\hi$ \be\label{pwpw01}
[A^I,P^+] = r_0(P^+) A^I\,,\qquad [\psi^\oplus,P^+] =
r_0(P^+)\psi^\oplus\,, \ee \be\label{pwpw02} [A^I,T^\hi] =
r_0(T^\hi)A^I\,,\qquad [\psi^\oplus,T^\hi] =
r_0(T^\hi)\psi^\oplus\,. \ee Action of isometry $so(2)$ rotations
generated by $J^{\hi\hj}$ takes the form \be {}[A^\hk,
J^{\hi\hj}]= r_0(J^{\hi\hj}) A^\hk +\delta^{\hk\hi}A^\hj
-\delta^{\hk\hj}A^\hi\,, \qquad [\phi^M, J^{\hi\hj}]=
r_0(J^{\hi\hj}) \phi^M\,, \ee \be\label{pwpw04} {}[\psi^\oplus,
J^{\hi\hj}]= (x^\hi\partial^\hj -x^\hj\partial^\hi
+\frac{1}{2}\gamma^{\hi\hj} ) \psi^\oplus\,. \ee

Now we turn to discussion of the proper conformal generators and
the transformations rules of the physical fields. Complete
expressions for the generators and the transformations are given
in \rf{gend}-\rf{genkm} and \rf{comtrarul1}-\rf{comtrarul12}.
Because we are going to consider an arbitrary spin field we
restrict ourselves to the linear transformations, i.e. to the ones
generated by parts of the generators which are quadratic in
fields. Introducing notation $G_{(2)}$ for parts of generators
which are quadratic in physical fields and plugging momentum and
spin densities \rf{ppp}-\rf{mmi} into expressions
\rf{gend}-\rf{genkm} we get the following explicit representation
for generators
\begin{eqnarray}
&& D_{(2)} = \int\! d^3x\, \rTrpr\,\, \partial^+ A^I r_0(D) A^I
+\frac{\rm i}{2}\psi^\oplus \gamma^+ r_0(D) \psi^\oplus\,,
\quad\qquad D_{(2)} = D\,,
\\[6pt]
&& C_{(2)} =  \int\! d^3x\,\rTrpr\,\partial^+ A^I r_0(C) A^I
+\frac{\rm i}{2}\psi^\oplus \gamma^+ r_0(C)\psi^\oplus\,,
\\[6pt]
&& C_{(2)}^\hi =   \int\! d^3x\, \rTrpr\,\Bigl(\partial^+ A^I
r_0(C^\hi) A^I +\frac{\rm i}{2}\psi^\oplus \gamma^+
r_0(C^\hi)\psi^\oplus\Bigr) +  e^{-{\rm i}\f x^+}({\cal
M}_{(2)}^{-\hi} +{\rm i}\f {\cal M}^{\hi\hj} x^\hj)\,,
\\[6pt]
&& K_{(2)}^-
 =  \int\! d^3x\, \rTrpr\,\Bigl(\partial^+ A^I
(r_0(K^-) -\frac{1}{2\partial^+})A^I +\frac{\rm i}{2}\psi^\oplus
\gamma^+ r_0(K^-)\psi^\oplus - \phi^2 \Bigr) +
 {\cal M}_{(2)}^{-\hi}x^\hi\,,\ \ \ \ \ \ \ \ \ \ \
\end{eqnarray}
where we use the following notation for differential operators
acting on the physical fields
\begin{eqnarray}
\label{rdkil}
&&r_0(D) =2x^-\partial^+ + x\partial\,,
\\
&&
\label{rckil}
r_0(C) =
e^{-2{\rm i}\f x^+}(
-\frac{1}{2\partial^+}\partial^2
- {\rm i}\f x\partial + \frac{\f^2}{2} x^2\partial^+)\,,
\\
&&
r_0(C^\hi) =e^{-{\rm i}\f x^+}\Bigl(x^-\partial^\hi
+\frac{1}{2\partial^+}x^\hi\partial^2
+{\rm i}\f(
-\frac{1}{2}x^2\partial^\hi +x^\hi( x^-\partial^+ + x\partial))\Bigr)\,,
\\
&&
\label{rkmkil}
r_0(K^-) = \frac{1}{4\partial^+}x^2\partial^2 +x^-
( x^-\partial^+ + x\partial)\,.
\end{eqnarray}
These differential operators are obtainable from the ones given in
\rf{dkil}-\rf{kmkil} my making the following substitution there
\be
\partial^- \rightarrow -\frac{1}{2\partial^+}\partial^2
-\frac{\f^2}{2}x^2\partial^+\,. \ee This substitution reflects
simply the fact that we are using equations of motion in
expressions for generators. Making use of the commutation
relations \rf{aacom},\rf{ppcom} gives the following transformation
rules under action of plane wave dilatation operator
\be\label{pwpwd3} [A^I, D]=(r_0(D) +1)A^I\,, \qquad [\psi^\oplus,
D]=(r_0(D) + 2)\psi^\oplus\,. \ee Appearance of the unusual factor
2 in transformation of fermionic field under action plane wave
dilatation generator $D$ \rf{pwpwd3} can be understood from the
relation \rf{pwlor3}.

Now we consider action of the remaining conformal transformations
generated by $C$, $C^\hi$, and $K^-$. We find that the
transformations rules of the physical fields $A^I$, $\psi^\oplus$
under action of the generator $C_{(2)}$ take the same form:
\be\label{pwpw1} [A^I, C_{(2)}]= (r_0(C) -{\rm i}\f e^{-2{\rm i}\f
x^+})A^I\,,\qquad [\psi^\oplus,C_{(2)}]= (r_0(C) -{\rm i}\f
e^{-2{\rm i}\f x^+})\psi^\oplus\,. \ee As to the transformations
generated by $C_{(2)}^\hi$, and $K_{(2)}^-$ they take different
form
\begin{eqnarray}
\label{pwpw2}&& {} [A^\hi, C_{(2)}^\hj] =(r_0(C^\hj) +{\rm i}\f
x^\hj e^{-{\rm i}\f x^+})A^\hi +e^{-{\rm i}\f
x^+}(\frac{\partial^\hk}{\partial^+}+{\rm i}\f x^\hk)
(\delta^{\hi\hj} A^\hk -\delta^{\hi\hk}A^\hj)\,,
\\
\label{pwpw3}&& {}[\phi^M, C_{(2)}^\hi] =(r_0(C^\hi) +{\rm i}\f
x^\hi e^{-{\rm i}\f x^+})\phi^M\,,
\\
\label{pwpw4}&& {}[\psi^\oplus, C_{(2)}^\hi] = \Bigl(r_0(C^\hi)
+(\frac{\partial^\hi}{2\partial^+} +\frac{3\rm i}{2}\f x^\hi)
e^{-{\rm i}\f x^+} +\frac{1}{2} e^{-{\rm i}\f
x^+}(\frac{\partial^\hk}{\partial^+} +{\rm i}\f
x^\hk)\gamma^{\hi\hk} \Bigr) \psi^\oplus\,,\qquad\qquad
\end{eqnarray}
\begin{eqnarray}
\label{pwpw5}&& {}[A^\hi, K^-_{(2)}] = (r_0(K^-) + x^-
-\frac{1}{\partial^+})A^\hi +x^\hj \frac{\partial^\hk}{\partial^+}
(\delta^{\hi\hj} A^\hk -\delta^{\hi\hk} A^\hj)\,,
\\
\label{pwpw6}&& {}[\phi^M, K^-_{(2)}] = (r_0(K^-) +x^-)\phi^M\,,
\\
\label{pwpw7} && {}[\psi^\oplus , K^-_{(2)}] = \Bigl(r_0(K^-)
+2x^- +\frac{1}{2\partial^+}x\partial +
\frac{1}{2}\gamma^{\hi\hj}x^\hi\frac{\partial^\hj}{\partial^+}
\Bigr)\psi^\oplus\,.
\end{eqnarray}
These transformations can be cast however into a unifying form. To
this end we introduce a new fermionic field $ \psi^\oplus =
(\partial^+)^{1/2}\tilde{\psi}^\oplus$ and get the following
transformations
\begin{eqnarray}
&& {}[\tilde{\psi}{}^\oplus, C_{(2)}^\hi] = \Bigl(r_0(C^\hi) +
{\rm i}\f x^\hi e^{-{\rm i}\f x^+}  + \frac{1}{2} e^{-{\rm i}\f
x^+}(\frac{\partial^\hk}{\partial^+} +{\rm i}\f
x^\hk)\gamma^{\hi\hk} \Bigr) \tilde{\psi}^\oplus\,,
\\
&& [\tilde{\psi}^\oplus, K^-_{(2)}] =\Bigl(r_0(K^-) +x^-
-\frac{1}{4\partial^+} +
\frac{1}{2}\gamma^{\hi\hj}x^\hi\frac{\partial^\hj}{\partial^+}
\Bigr)\tilde{\psi}^\oplus\,, \end{eqnarray} which together with
the transformations given in
\rf{pwpwd3},\rf{pwpw2},\rf{pwpw3},\rf{pwpw5},\rf{pwpw6} can be
generalized to arbitrary spin $s$ plane wave massless field
$\Xi_s$ in a straightforward way
\begin{eqnarray}
&& {}[\Xi_s, D] = (r_0(D) +1)\Xi_s\,,
\\
&& {}[\Xi_s, C_{(2)}^\hi] = (r_0(C^\hi) + {\rm i}\f x^\hi e^{-{\rm
i}\f x^+})\Xi_s  +  e^{-{\rm i}\f
x^+}(\frac{\partial^\hk}{\partial^+} +{\rm i}\f
x^\hk)[\Xi_s,M^{\hi\hk}]\,,
\\
&&{}[\Xi_s, K^-_{(2)}] = (r_0(K^-) +x^-
-\frac{s^2}{\partial^+})\Xi_s +
x^\hi\frac{\partial^\hj}{\partial^+} \Bigr[\Xi_s,M^{\hi\hj}]\,.
\end{eqnarray} The transformations rules for the physical
fields of SYM theory $\phi^M$, $A^\hi$, $\tilde{\psi}^\oplus$ are
obtainable from these transformation by respective setting $s=0$,
$s=1$, $s=1/2$. The bracket $[\Xi_s,M^{ij}]$ denotes
transformation of spin $s$ field $\Xi_s$ under action of spin part
of generator of $so(2)$ algebra $J^{\hi\hj}$. For instance for
spin $s=0$ scalar field $[\phi,M^{\hi\hj}]=0$, while for spin
$s=1/2$ fermionic field
$[\tilde{\psi}{}^\oplus,M^{\hi\hj}]=\frac{1}{2}\gamma^{\hi\hj}
\tilde{\psi}{}^\oplus$. Light cone gauge transformations rules of
the spin $s$ field $\Xi_s$ under action of the generators $P^+$,
$T^\hi$, $C_{(2)}$ takes the same form as for the fields of SYM
theory (see \rf{pwpw01},\rf{pwpw02},\rf{pwpw1}). Light cone gauge
equations motion for the field $\Xi_s$ also take a simplified form
$\Box\, \Xi_s=0$ (with $\Box$ given in \rf{box}) .

\newpage

\end{document}